\documentclass[reqno,12pt]{article}
\usepackage{amsmath}
\usepackage{bm}
\usepackage{a4wide,amssymb}
\usepackage{graphicx,xcolor}
\usepackage{epstopdf}
\usepackage{bbm}
\usepackage{mathrsfs}

\usepackage[titletoc,toc,title]{appendix}

\usepackage[refpage]{nomencl}
\usepackage{etoolbox}

\newcommand{\qed}{ $\blacksquare$} \newcommand{\nn}{\nonumber}
         
\newcommand{\Z}{{\mathbb Z}} 
\newcommand{\E}{{\mathbb E}}

\renewcommand{\a}{\alpha}

\newcommand{\g}{\gamma}\renewcommand{\d}{\delta}
           
\newcommand\om{\omega}

\newcommand{\e}{\varepsilon}

\newcommand{\pa}{\partial}

\newcommand{\z}{\zeta}
\newcommand{\la}{\lambda}
\newcommand{\D}{\Delta}
\newcommand{\Lam}{\Lambda}
\newcommand{\supp}{\mathop{\mathrm{supp}}}
\def\G{\Gamma}
\def\s{\sigma}
\def\r{{\rho}}

\def\p{{\pi}}
\def\m{{\mu}}

\def\DD{\mathcal D}
\def\bz{{\bf z}}
\def\bx{{\bf x}}
\def\bv{{\bf v}}

\def\bW{{\bf W}}

\def\br{{\bm\r}}

\def\bo{{\bm \o}}

\def\bG{{\bm \G}}
\def\bt{{\bf t}}
\def\bs{{\bm \s}}
\def\bk{{\bf k}}
\def\bze{{\bm \zeta}}

\def\bxi{{\bm \xi}}
\def\bet{{\bm \eta}}

\def\T{{\mathcal T}}
\def\L{{\Lambda}}
\def\II{{\mathcal I}}
\def\EE{{\mathscr{E}}}
\def\JJ{{\mathcal J}}
\def\RRR{{\mathbb R}}
\def\EEE{{\mathbb{E}}}

\def\BB{{\cal B}}

\def\QQ{{\cal Q}}
\def\HH{{\cal H}}

\def\KK{{\cal K}}
\def\MM{{\cal M}}
\def\AA{{\cal A}}

\def\CC{{\cal C}}
\def\SS{{\cal S}}
\def\TT{\mathtt{T}}
\def\ZZ{{\cal Z}}

\newtheorem{thm}{Theorem}[section]
\newtheorem{prop}[thm]{Proposition}
\newtheorem{lem}[thm]{Lemma}
\newtheorem{defi}[thm]{Definition}
\newtheorem{cor}[thm]{Corollary}
\newtheorem{hyp}[thm]{Hypothesis}

\def\be{\begin{equation}}
\def\ee{\end{equation}}
\def\bea{\begin{eqnarray}}
\def\eea{\end{eqnarray}}
\def\ni{\noindent}
\def\nn{\nonumber}

\def\ol{\overline}

\def\d{\delta}
\def\o{\omega}
\def\b{\beta}
\def\t{\tau}

\DeclareMathOperator*{\esssup}{ess\,sup}

\usepackage{multicol}
\makenomenclature
\usepackage{xpatch}
\makeatletter
\xapptocmd\thenomenclature{\let\@item\nomencl@item\def\nomencl@width{0pt}}{}{}
\let\nomencl@item\@item
\xpretocmd\nomencl@item{\nomencl@measure{#1}}{}{}
\def\nomencl@measure#1{%
  \sbox0{#1}%
  \ifdim\wd0>\nomencl@width\relax
    \edef\nomencl@width{\the\wd0}%
  \fi
}

\xpatchcmd\thenomenclature
  {\section*{\nomname}}
  {\begin{multicols}{2}[\section*{\nomname}]}
  {}{}
\xpatchcmd\thenomenclature
  {\chapter*{\nomname}}
  {\begin{multicols}{2}[\chapter*{\nomname}]}
  {}{}

\xapptocmd\endthenomenclature{%
  \immediate\write\@mainaux{\global\nomlabelwidth\nomencl@width\relax}%
  \end{multicols}
}{}{}
\makeatother

\begin{document}

\begin{titlepage}

\begin{center} 
{\bf \Large{The Boltzmann--Grad Limit of a Hard Sphere System: Analysis of the Correlation Error}}

\vspace{1.5cm}
{\large M. Pulvirenti$^{1,2}$ and S. Simonella$^{3}$}

\vspace{0.5cm}
{$1.$\scshape {\small \ Dipartimento di Matematica, Universit\`a di Roma La Sapienza\\ 
Piazzale Aldo Moro 5, 00185 Roma -- Italy \\ \smallskip
$2.$\scshape \small \ International Research Center M\&MOCS, Universit\`{a} dell'Aquila\\ 
Palazzo Caetani, 04012 Cisterna di Latina -- Italy \\ \smallskip
$3.$\ Zentrum Mathematik, TU M\"{u}nchen \\ Boltzmannstrasse 3, 85748 Garching -- Germany}}
\end{center}

\vspace{1.5cm}
\hfill {\em In memory of Oscar Erasmus Lanford III}

\vspace{1.5cm}
\noindent
{\bf Abstract.} 
We present a quantitative analysis of the Boltzmann--Grad (low--density) limit of a hard sphere system.
We introduce and study a set of functions ({\em correlation errors}) measuring the deviations in time from the 
statistical independence of particles (propagation of chaos). In the context of the BBGKY hierarchy, a 
correlation error of order $k$ measures the event where $k$ particles are connected by a chain of 
interactions preventing the factorization.

\ni We show that, provided $k < \e^{-\a}$, such an error flows to zero with the average density $\e$, 
for short times, as $\e^{\g k}$, for some positive $\a,\g \in (0,1)$. This provides an information on the {\em size of
chaos}, namely, $j$ different particles behave as dictated by the Boltzmann equation 
even when $j$ diverges as a negative power of $\e$. 

\ni The result requires a rearrangement of Lanford perturbative series into a cumulant type expansion, 
and an analysis of many--recollision events.

\vspace{0.5cm}\noindent
{\bf Keywords.} Kinetic theory, low--density limit, BBGKY hierarchy, Boltzmann equation, propagation of chaos

\thispagestyle{empty}

\bigskip
\bigskip

\newpage

\tableofcontents

\thispagestyle{empty}

\end{titlepage}


\section{Introduction} \label{sec:intro}
\setcounter{equation}{0}    
\def\theequation{1.\arabic{equation}}

\smallskip
\ni {\bf 1.1\ \ \ Lanford's theorem and beyond} 
\bigskip

\ni In 1975  O. E. Lanford III presented his celebrated proof of the mathematical validity of the Boltzmann equation 
for hard spheres, in a time interval small enough \cite{La75}. To remind his result, let us consider a system of 
identical hard spheres of diameter $\e$ moving in the whole space $\RRR^3$  with collisions governed by the usual laws 
of elastic reflection. 

At a given time, say $t = 0$, the state of the system over the grand canonical phase space 
$\cup_{n\geq 0}(\RRR^3\times\RRR^3)^n$ is completely characterized via the symmetric 
(in the exchange of the particles) probability densities $(1/n!) W^\e_{0,n}$ 
of finding exactly $n$ spheres with given positions and velocities.
If $p_n =(1/n!) \int W^\e_{0,n}$, then $\sum_n p_n=1$ and 
the average number of particles is $\langle N \rangle=\sum_n n\,p_n$. 

We are interested in analyzing a low--density limit, namely the {\em Boltzmann--Grad limit}
\cite{Gr49,Gr58}, defined by
\be
\langle N \rangle   \to \infty, \quad \e \to 0 \quad \mbox{and} \quad \langle N \rangle \e^2 \to \la^{-1} >0\;,
\label{eq:BGlgc}
\ee
where $\la$ is a fixed constant proportional to the mean free path.
Since $\langle N \rangle $ and $\e$ are related in this limit, let us use a single parameter, 
say $\e$, and rescale properly $\langle N \rangle=\e^{-2} \la^{-1}$. 

From the densities $W^\e_{0,n}$ we construct the {\em rescaled correlation functions} 
(r.c.f. in the sequel)  $f^\e_{0,j} = \e^{2j}\r^{\e}_{0,j}$ for $j=1,2,3,\cdots$,
where $\r^\e_{0,j}$ are the correlation functions as usually introduced in statistical mechanics.
This is a way to encode all the statistical properties of the many-body system.
Let $N_\D$ be the number of particles in the subset $\D$ of $\RRR^3\times\RRR^3$.
Then the $j-$th correlation function is a function of $j$ positions and velocities such that
$\int_{\D^j}\r^\e_{0,j}$ expresses the mean value of the product
$N_\D (N_\D-1)\cdots (N_\D-j+1)$ (average number of unordered $j-$tuples in $\D$).
Note that the rescaling $\e^{2j}$ allows to see finite quantities as $\e\to 0$. 
The r.c.f. differ in fact from the {\em marginals} of the measure only for proper normalization factors.

We focus on the quantities $f^\e_{j}(t)$, namely the r.c.f.
of the system at time $t>0$, evolved deterministically (starting from $f^\e_{0,j}$)
in accord to the hard sphere dynamics.

Lanford proved that, if the initial state factorizes in the limit, meaning that 
\be
f^\e_{0,j} \to f_0^{\otimes j} \label{eq:I1}
\ee
as $\e\to 0$, where $f_0$ is a given one--particle probability density, then there exists a $\bar t>0$
such that, in the same limit, 
\be
f^\e_{j} (t)  \to f(t)^{\otimes j}\ \ \ \ \mbox{ for }\ t<\bar t
\label{lan}
\ee
almost everywhere in $(\RRR^3\times\RRR^3)^j$, where $f(t)$ is a solution of the Boltzmann equation 
with initial datum $f_0$. 
Eq. \eqref{eq:I1} means that $j$ particles are ``almost independent" in the low-density regime,
and \eqref{lan} shows that this property propagates at least for short times.

Note that we found convenient to recall the theorem as stated in \cite{Ki75} (or also in \cite{BLLS80,S81}),
namely without fixing the total number of particles. The advantage of this formulation in our context will be 
discussed later on. 

The time $\ol t$ is just a small fraction of the mean free flight time. Nevertheless, it is enough to show 
unambiguously that there is no contradiction between the reversible microscopic dynamics and the irreversible
behaviour described by the Boltzmann equation. 
So far, the only extension of this result to arbitrary times refers to the special 
situation of a rare cloud of gas expanding in the vacuum, \cite {IP86,IP89}.

We remind here the {\em Boltzmann equation} for the unknown  $f=f(x,v,t)$, with hard sphere kernel
and mean free path $\la$ \cite{Bo64},
\be
(\pa_t+v\cdot \nabla_x)f(x,v,t)
= \la^{-1} \int_{\RRR^3\times S^2_+} dv_1 d\o \ (v-v_1)\cdot\o \Big\{f(x,v_1',t)f(x,v',t)-
f(x,v_1,t)f(x,v,t)\Big\}
\label{BE}
\ee
\nomenclature[F0]{$f$}{Solution to the Boltzmann equation}%
where $S_+^2=\{\om \in S^2 |\ (v-v_1)\cdot\o \geq 0\},$ $S^2$ is the unit sphere in $\RRR^3$ 
(with surface measure $d\o$), $(v,v_1)$ is a pair of 
velocities in incoming collision configuration   and $(v',v_1')$ is the corresponding pair of outgoing 
velocities defined by the elastic reflection rules
\be
\begin{cases}
\displaystyle v'=v-\om [\om\cdot(v-v_1)] \\
\displaystyle  v_1'=v_1+\om[\om\cdot(v-v_1)]
\end{cases}\;.\label{eq:coll}
\ee

The proof in \cite{La75} was carried out by assuming suitable uniform estimates on the family of r.c.f.
at time zero. The available estimates deteriorate in time in such a way that, at time $\bar t$, any possibility of a uniform 
control is lost. Indeed the strategy of Lanford was based on an expansion of $ f^\e_{j} (t)$, for given $j>0$, in 
terms of a series (involving only the initial data $f^\e_{0,j}$) which is absolutely convergent, uniformly in $\e$, 
only for a short time interval. To complete the proof it was enough to exploit the term by term convergence 
holding by virtue of geometric and measure--zero arguments (see also \cite{Sp84,Uc88,Sp91,CIP94,Uk01}).

The original argument of Lanford was qualitative, in the sense that \eqref{lan} was shown 
without an explicit rate of convergence. Recently, explicit estimates on the rate of convergence have 
been obtained in \cite{GSRT12} (see also \cite{PSS13} for a different class of potentials), of the form
\be
\label{qest}
|f^\e_{j} (t)  \ -  f(t)^{\otimes j}\ | \leq C^j \e ^{\g} \ \ \ \mbox{ for }\ t<\bar t
\ee
outside a set of negligible measure in $(\RRR^3\times\RRR^3)^j$, for some $C, \g>0$.

A quantitative information is interesting in its own in spite of the time restriction.
We note also that the explicit control of the error can be used to reach hydrodynamic regimes:
see \cite{BGSR13}, where the heat equation is derived from the hard sphere dynamics in a low--density regime,
by studying one tagged particle in a gas close to equilibrium.

The purpose of the present paper is to introduce a notion of {\em correlation error} for the many--particle
system and estimate it explicitly. 

Before giving precise definitions, let us explain our main motivation.

\bigskip
\ni {\bf 1.2\ \ \ The size of chaos} 
\bigskip

\ni The result \eqref{lan} has been proved by Lanford for any fixed $j>0$. This cannot be uniform in $j$
since, for very large $j$  (for instance, if $j\simeq \e^{-2}$), $f^\e_j (t)$ will be very different from a tensor product. 
It is however natural to ask whether there exists a notion of convergence holding for $j=j(\e)$ suitably diverging with $\e$. 

A first answer is given by the quantitative analysis in \cite{GSRT12,PSS13}. In fact Equation \eqref{qest} implies
that the convergence of r.c.f. is actually true for $j\leq C_0 |\log \e|$ for a small positive $C_0$. On the other hand,
one would expect some power--law divergence, along the following heuristic argument. \label{arg:heur}

The proof of the asymptotic behavior \eqref{lan} is intimately connected with the problem of {\em propagation of chaos}, 
i.e. the conservation in time of the statistical independence of particles (provided that it holds at time zero).
Given a group of $j$ particles, consider,
for any $i = 1,2, \cdots , j$, the set $B_i$ of particles really influencing the dynamics of particle $i$ up to the time $t$.
We may assume that the cardinality of the sets $B_i$ is finite to have a correct kinetic behaviour in the limit.
For the propagation of chaos to hold, the groups $B_i$ must be disjoint.
Therefore, the probability that two given particles in the group $1,2, \cdots, j$ are dynamically correlated will be
$O(1/\langle N \rangle)$. Correspondingly, the probability that the $j$ particles do {\em not} behave as mutually 
independent will be $O(j^2 / \langle N \rangle )$, which is small for $j << \e^{-1}$
(see also \cite{APST15} for related considerations).

Our goal here is to analyze the ``size of chaos'', that is, how large can be a cluster of particles in order
to see their statistical independence. 

In the effort of going beyond the logarithmic scale we immediately realize that there is no hope to
improve estimate \eqref{qest}. Indeed even ignoring the correlations and {\em assuming} that 
$f_j^\e (t) \approx (f_1^\e)^{\otimes j}$, one cannot obtain nothing better than \eqref{qest} expanding 
$(f_1^\e)^{\otimes j}-f^{\otimes j}$. On the other hand, trivially, $(f_1^\e-f)^{\otimes j} = O(\e^{\g j})$ uniformly in $j$.
We are led then to give the following notion of error. Let a configuration of the particle system be 
$\bz_n = (z_1, \cdots, z_n)$, where $z_i=(x_i,v_i)$ are the position and the velocity of particle $i$ respectively,
and let $\bz_n(t) = (z_1(t), \cdots, z_n(t))$ be the corresponding time--evolved configuration.
Given a sequence of test functions over $\RRR^6$, denoted $\varphi_1,\varphi_2,\cdots$, we consider 
the naturally associated observables $F_1=F_1(t), F_2=F_2(t), \cdots$ defined by
\be
F_i(t)(\bz_n) = \e^2 \sum_{j=1}^n \varphi_i(z_j(t))\;.
\label{def:obs}
\ee
\nomenclature[f]{$F_i$}{Observable in the particle system, associated to the test function $\varphi_i$}%
For instance if $\varphi_i$ is the characteristic function of the set $\D$, then $F_i(t)$ is
the fraction of particles in $\D$ at time $t$.
The validity of the Boltzmann equation can be rephrased by saying that the
error $\Big(F_i(t) - \EEE^{\BB}[\varphi_i(t)]\Big)\approx 0$ 
for $\e$ small, where $\EEE^{\BB}[\varphi(t)] = \int dx dv \, \varphi(x,v)\, f(x,v,t)$. 
\nomenclature[EB]{$\EEE^{\BB}$}{Expectation with respect to the Boltzmann density}%
We look now at the product of $j$
such simultaneous deviations and compute its expected value $\EEE^\e$ in the state of the particle system. 
We will prove that, in the Boltzmann--Grad limit, there exists a constant $\a \in (0,1)$ such that, 
if $t$ is small enough, 
\be
\lim_{\e \to 0} \,\sup_{j < \e^{-\a}}\,
\Big|\EEE^\e\left[\prod_{i= 1}^j\Big(F_i(t) - \EEE^\BB[\varphi_i(t)]\Big)\right]\Big| = 0\;.
\label{eq:convMT}
\ee
Roughly speaking, with respect to \eqref{lan},
we are replacing the difference of products with the product of differences,
which is expected to be much smaller (and more difficult to control). 

Equation \eqref{eq:convMT} 
says that groups of up to $\e^{-\a}$ particles show up mutual statistical independence and
that simultaneous deviations of the particles behaviour from the Boltzmann behaviour are negligible 
in the limit.

\bigskip
\ni {\bf 1.3\ \ \ Result on correlation errors} 
\bigskip

\ni The notion of error in \eqref{eq:convMT} is closely related to what is known in statistical physics as 
{\em fluctuation} from the average value. Usually one focuses entirely on the particle system and
ignores the convergence error $\Big( \EEE^{\BB}[\varphi_i(t)] - \EEE^{\e}[F_i(t)]\Big)$. The quantity
\be
\EEE^\e\left[\prod_{i= 1}^j\Big(F_i(t) - \EEE^\e[F_i(t)]\Big)\right] \approx 0
\label{eq:FAV}
\ee
gives the $j-$th order moment of the fluctuations and is formally seen to be $O(\e^{j})$ for any fixed $j$.
For previous results on the fluctuation field in the Boltzmann--Grad limit, see \cite{BLLS80,S81,Sp83}.

To be more concrete, let us choose a collection of disjoint sets $\Delta_1, \cdots,  \Delta_j$ 
in $\RRR^{6}$ and as $\varphi_i$ the indicator function of the set $\D_i$, so that $F_i(t)=n_{\Delta_i}$
\nomenclature[nD]{$n_{\Delta}$}{Fraction of particles in the region $\Delta\subset\RRR^6$}%
is the corresponding fraction of particles in the region. Then we have
 \begin{equation}
\EEE^\e\left[ \prod_{i=1}^j \left(n_{\Delta_i} - \EEE^\e[n_{\Delta_i}] \right) \right]  = 
\int_{ \Delta_1 \times \cdots \times \Delta_j } dz_1\cdots dz_j \, E_j (z_1,\cdots, z_j ,t)\;,
\label{eq:fluctEj}
 \end{equation}
which introduces a new sequence of ($\e-$dependent) functions $E_j = E_j (\bz_j,t)$, $j=1,2,\cdots$.
Here we shall call $E_j(t)$ the {\em correlation error} of order $j$, since its size is a measure of 
the statistical dependence of $j$ distinct particles in different regions. In our work, these will be the fundamental
objects.

Technically, $E_j$ is connected to the r.c.f. by a  {\em cumulant} expansion. Explicitly, 
\bea
E_1(z_1) & =& 0\;,\nn\\
E_2(z_1,z_2) & =& f_2^\e(z_1,z_2) - f_1^\e(z_1) f_1^\e(z_2)\;, \nn\\
E_3(z_1,z_2,z_3) & =& 
f_3^\e(z_1,z_2,z_3) - f_2^\e(z_1,z_2) f_1^\e(z_3) -  f_2^\e(z_1,z_3) f_1^\e(z_2) \nn\\
&&- f_2^\e(z_2,z_3) f_1^\e(z_1) + 2  f_1^\e(z_1) f_1^\e(z_2)f_1^\e(z_3)\;, 
\label{eq:exampleDEF}
\eea
etc., or for generic $j$,
\be
E_J (t)= \sum_{K\subset J} (-1)^k  (f_1 ^{\e}(t) )^{ \otimes K}  f^\e_{J \setminus K} (t)\;,
\label{correrr1}
\ee 
\nomenclature[Ej]{$E_j$}{Correlation error of order $j$}%
where $J=\{ 1,2, \cdots , j\}$, 
\nomenclature[J]{$J$}{Set of indices of particles $\{ 1,2, \cdots , j\}$}%
$K$ is a subset of indices in $J$ ($\emptyset$ and $J$ are included in the sum,
with the convention $f^\e_{\emptyset} = 1 = (f_1^\e)^{\otimes \emptyset} $), $k=|K|$ is the cardinality of the set $K$
and, if $Q=\{i_1, \cdots, i_q\} $, one denotes
\bea
&& f^\e_{Q} (\bz_Q,t)=f^\e_{q}(z_{i_1}, \cdots, z_{i_q},t)\;,\nn\\
&& E_{Q} (\bz_Q,t)=E_{q}(z_{i_1}, \cdots, z_{i_q},t)\;,\nn\\
&& (f_1 ^{\e}(t) )^{ \otimes Q}= f_1 ^{\e}(z_{i_1},t)f_1 ^{\e}(z_{i_2},t) \cdots f_1 ^{\e}(z_{i_q},t)\;.
\label{eq:notintro}
\eea

Eq. \eqref{correrr1} can be inverted to give
\be
f^\e_J (t)= \sum_{K\subset J}(f_1 ^{\e}(t) )^{ \otimes K}  E_{J \setminus K} (t)\;,
\label{correrr}
\ee
which has a clear physical interpretation as a sum over subgroups of uncorrelated particles.
In this paper we shall adopt \eqref{correrr} as definition of correlation error. The precise connection with
fluctuations will be discussed in Section \ref{sec:proofcorconv}.

We stress again that the above definition involves the particle system only and does not refer to any 
kinetic equation. The propagation of chaos amounts to say that, for any given $j$, as $\e\to 0$
\be
E_j(0) \to 0\ \ \ \Longrightarrow \ \ \ E_j(t) \to 0 \label{eq:SI}
\ee
for $t>0$. In a sense, the correlation errors identify and strictly isolate those dynamical events which 
destroy propagation of chaos.

Actually our main result will be that, {\em in the Boltzmann--Grad limit, there exist constants $\a,\g\in (0,1)$
such that, if $t$ is small enough and provided $j < \e^{-\a}$, then
\be
\int dv_1 \cdots dv_j |E_j (z_1, \cdots, z_j, t)| \leq \e^{\gamma j}\;,
\label{bound}
\ee 
for a given configuration $x_1, \cdots, x_j$ of distinct points and $\e$ sufficiently small.}
Of course an estimate similar to \eqref{bound} has to be assumed at time zero, together with uniform estimates
on the family of r.c.f. (as in Lanford's theorem). We shall construct  explicit
examples of physically relevant initial states satisfying such hypotheses.
The convergence result \eqref{eq:convMT} will be a consequence of the main estimate \eqref{bound}.

\bigskip
\ni {\bf 1.4\ \ \ Strategy: hierarchical particle flows} 
\bigskip

\ni 
Let us now briefly comment on the difficulties. 
The only known strategy to rigorously derive estimates
on the particle system goes through a reformulation of the problem in terms of special characteristics
of a set of hierarchical equations. Such flows share all the features of the interacting dynamics of finite
groups of particles and their control is a very delicate task.

The breakdown of the statistical independence \eqref{eq:SI} is indeed due to {\em mechanical} effects. 
First of all, one should keep in mind that any given state of the real system (in particular, whatever choice of 
the time--zero state) cannot be exactly factorized, 
because of the simple hard core exclusion. This is just a static  feature. Secondly, and most importantly,
correlations between particles are generated by the dynamics itself. In the context of \cite{La75} and of the subsequent
literature, the events responsible for these dynamical correlations are called {\em recollisions}.
Their effective control is quite complicated since they generally depend on the full particle dynamics.

To be more precise, one looks at the BBGKY hierarchy, namely the set of coupled evolution equations
for $\{f^\e_j\}_{j \geq 1}$\footnote{Originally written for smooth potentials by Bogolyubov, Born, Green, 
Kirkwood and Yvon; see, e.g.,~\cite{Bog62}.}.
The iteration of the hierarchy gives an expression of $f^\e_j(t)$ in terms of a series expansion depending 
only on the initial data. Each term of this expansion is in one--to--one correspondence with a special 
trajectory of clusters of particles flowing backwards in time. Looking at such flows one 
may single out precisely the (re-)collisions that generate correlations. 

Roughly, formulas \eqref{correrr}, \eqref{bound}
will be constructed starting from the BBGKY expansion for $f^\e_j$,
 by systematically replacing such collision--events with ``free overlap--events" where the
two considered particles ignore and cross each other freely, and estimating the consequent errors
(see Section \ref{sec:plan} for a simple example). In addition, one has to extract the correlation error 
of the initial state due to the exclusion. The main technical part of the work shall consist of (i)
a suitable cluster expansion (needed to control the total number of produced terms) and (ii) geometric 
estimates for trajectories of $j$ particles showing up many recollisions.

The net result expresses $f_j^\e (t)$ as a sum of contributions \eqref{correrr}. The first, $O(1)$, is just the product 
state. Then, we sum over all possible ways of choosing two correlated particles, the remaining $j-2$ 
particles being uncorrelated. This events are $O(\e^{2\gamma })$ (actually $O(\e^2)$). 
Then we pass to the events in which three particles are correlated, which give a contribution 
$O(\e^{3\gamma })$, and so on.

Note that in this paper we derive the bound on $E_j$, as roughly explained above, exploiting  the series expansion 
for $f^\e_j$. Another possibility is to use \eqref{correrr1} and the evolution equations for $f^\e_j$ and $f_1^\e$,
but  a closed evolution equation for $E_j$ seems to be more difficult to write and to handle with. 

\bigskip
\ni {\bf 1.5\ \ \ The Enskog error} 
\bigskip

\ni Although the main technical effort in this work will concern the correlation error $E_j$,
it is important to observe that this function describes a part, but not {\em all}, of the total dynamical correlation 
between particles, the remainder being encoded in the one--point marginal $f_1^\e$. 

Working again in terms of backwards flows, one may extract from the definition of $f^\e_1$ a second 
(and last) class of recollision--events. This operation leads to define another interesting sequence of quantities, 
that is $E^{\EE}_j (\bz_j,t),$ $j=1,2,\cdots$, given by
\be
f^\e_J (t)= \sum_{K\subset J}(g^\e(t))^{ \otimes K}E^{\EE}_{J \setminus K} (t)
\label{convcorrerrE}
\ee
\nomenclature[EjE]{$E^{\EE}_j$}{Enskog error term}%
(where we extend in an obvious way the notations of \eqref{eq:notintro}) 
or by
\be
E^\EE_J (t)= \sum_{K\subset J} (-1)^k  (g^\e(t))^{ \otimes K}  f^\e_{J \setminus K} (t)\;,
\label{correrr1E}
\ee
where $g^\e(t)$ is defined by an explicit expression that does {\em not} involve any more correlations
among particles. 
Namely, $g^\e(t)$ is the series solution to the {\em Enskog equation}
(more properly, the Boltzmann--Enskog equation \cite{Bog75}), which we recall:
\bea
&& (\pa_t+v\cdot \nabla_x)g^\e(x,v,t) = \la^{-1} \int_{\RRR^3\times S^2_+} dv_1 d\o \ (v-v_1)\cdot\o  \label{EE} \\
&&\ \ \ \ \ \ \ \ \ \ \ \ \ \ \ \ 
\times \Big\{g^\e(x-\om\e,v_1',t)g^\e(x,v',t)- g^\e(x+\om\e,v_1,t)g^\e(x,v,t)\Big\}\;, \nn
\eea
\nomenclature[ge]{$g^\e$}{Solution to the Enskog equation}%
where we used the notations introduced next to \eqref{BE}. We shall therefore refer to $E^{\EE}_j (\bz_j,t)$
as the {\em Enskog error term}.

Note that if $ f^\e_j (t)$ factorizes strictly, i.e. $ f^\e_{J} (t)= (f^\e_{1}(t))^{\otimes J}$, then
$$
E^\EE_J(t)=\left((f^\e_{1} - g^\e)(t)\right)^{\otimes J}\;.
$$
More generally, the size of $E^\EE_J (t)$ is a measure of both the breakdown of propagation of chaos
and the error in the convergence of $f^\e_1$ to $g^\e$. We will show that $E^\EE_j(t)$ can be bounded 
as $E_j(t)$, i.e.
\be
\int dv_1 \cdots dv_j |E^\EE_j(t)| \leq \e^{\gamma j}
\label{basic}
\ee 
for $t$ small enough and $j \leq \e^{-\alpha}$, as soon as $f^\e_1(0)$ is assumed
to converge uniformly as a power of $\e$ to the initial datum for the Enskog equation.

In our framework, the Enskog equation appears as a {\em bridge} between the hard sphere 
dynamics and the Boltzmann equation. 
In particular, to obtain the representation \eqref{convcorrerrE}--\eqref{basic}, no
regularity property needs to be assumed for the state of the system. The Enskog picture is
what emerges from the mechanical system once we eliminate all the sources of correlation,
including both the dynamical correlation and the static correlation of the time--zero state.

\bigskip
\ni {\bf 1.6\ \ \ The Boltzmann error} 
\bigskip

\ni 
Finally, the only difference between the Enskog system described by $g^\e(t)$ and the Boltzmann system described by $f(t)$
and \eqref{BE} (with same initial data), is that interactions between particles are described as occurring at distance 
$\e$ instead that at distance zero. In other words, microscopic translations of the Enskog flow lead to the Boltzmann
flow. A simple continuity property (assumed for the initial data) implies now
\be
 \int dv_1 \cdots dv_j  |E^\BB_j(t)| \leq \e^{\gamma j}
\label{convcorrerr}
\ee
for $t$ small enough and $j \leq \e^{-\alpha}$, where the {\em Boltzmann error term} $E^{\BB}_j$ is defined by
\be
f^\e_J (t)= \sum_{K\subset J}(f(t))^{ \otimes K} E^{\BB}_{J \setminus K} (t)\;.
\label{convcorrerr1}
\ee
\nomenclature[EjB]{$E^{\BB}_j$}{Boltzmann error term}%

Note that Equations \eqref{convcorrerr}--\eqref{convcorrerr1} are a reformulation of Lanford's result together with an 
explicit representation of the error.  The restriction to short times is also the same. However if the 
Boltzmann equation is globally valid, the statistical independence 
cannot fail to hold and, in this case, we believe that our estimations are also globally valid.

The quantities $E^\BB_j (t)$, under the name ``$v$--functions'', were previously introduced in 
\cite {CDPP91,CP96,CPW98}, in dealing with kinetic limits of stochastic particle systems. 

Equation \eqref{eq:convMT} follows from a further estimate 
of contraction terms due to the fact that, for generic observables, the same particle may appear 
simultaneously in the computation of $F_i$ and $F_j$, $i \neq j$ (see Section \ref{sec:proofcorconv}).

\bigskip
\ni {\bf 1.7\ \ \ Plan of the paper} 
\bigskip

\ni The plan of the paper is the following. In Section \ref{sec:AS} we introduce the model,
formulate a precise statement of our results and add several remarks and comments.
In Section \ref{sec:hs} we introduce the preliminaries of our analysis,
i.e. the hierarchical evolution equations and the ``tree expansion''--representations of the series solutions
(our basic tool). Section \ref{proof} is devoted to the proofs. An outline of the main technical tasks 
is given in Section \ref{sec:plan}. Five appendices contain the developments of some useful, but now straightforward,
arguments.

Although rather technical, we believe that the 
proofs of this paper are interesting in itself, since they enter in the complex mechanism of the propagation 
of chaos in a constructive and quantitative manner.


\section{Assumptions and main results} \label{sec:AS}
\setcounter{equation}{0}    
\def\theequation{2.\arabic{equation}}

In this section we describe precisely our setting, fix the notation and state our main results.

\subsection{The hard sphere system} \label{sec:AS1}
\setcounter{equation}{0}    
\def\theequation{2.1.\arabic{equation}}

We consider a {\em system of hard spheres} of unit mass and of diameter $\e >0$ moving  in 
the whole space $\RRR^3$. We will denote
$$
z_i=(x_i,v_i)\in \RRR^6
$$ 
\nomenclature[zi]{$z_i$}{State (position $x_i$, velocity $v_i$) of particle $i$} %
\nomenclature[xi]{$x_i$}{Position of particle $i$} %
\nomenclature[vi]{$v_i$}{Velocity of particle $i$} %
the state of the $i$--th particle, $i=1,2, \cdots$.
For groups of particles we shall use the notation $$\bz_j = z_1,\cdots,z_j, 
\ \ \ \ \ \ \ \ \ \ \ \bz_{j,n} = z_{j+1},\cdots, z_{j+n},$$
\nomenclature[zj]{$\bz_j$}{Vector $(z_1,\cdots,z_j)$} %
\nomenclature[zjn]{$\bz_{j,n}$}{Vector $(z_{j+1},\cdots,z_{j+n})$} %
and call ``particle $i$'' a particle whose configuration is labelled by the index $i$.

We will work in the grand-canonical {\em phase space}
\be
\MM(\e) = \cup_{n\geq 0}\MM_n(\e) \;, \label{eq:pshs}
\ee
\nomenclature[MM]{$\MM$}{Grand-canonical phase space} %
where
\be
\MM_n(\e) = \Big\{\bz_n \in \RRR^{6n},\ \ |x_i-x_j| > \e,\ i\neq j\Big\}\;, \qquad \MM_0(\e)=\emptyset .
\ee
\nomenclature[MMn]{$\MM_n$}{Canonical $n-$particle phase space} %
Unless necessary we omit, for simplicity, the dependence on $\e$ of the spaces defined above.

Notice that $\MM_N,$ with $N\sim \e^{-2}$, is the canonical $N$--particle phase space used in \cite{La75} and in most
of the subsequent literature on the Boltzmann--Grad limit. In this paper we find convenient to 
consider a more general class of measures where the exact number of particles is not necessarily fixed. The advantage
of this picture will be discussed in Section \ref{subsec:CR}, Remark 6.

The equations of motion for the $n$--particle system are defined as follows.
Between collisions each particle moves on a straight line with constant velocity.
When two hard spheres collide with positions $x_i, x_j$ (at distance $\e$), normalized relative distance 
$$\o = (x_i-x_j)/|x_i-x_j|=(x_i-x_j)/\e \in S^2$$
and incoming velocities $v_i, v_j$ (that means $(v_i-v_j)\cdot\o <0$), 
these are instantaneously transformed to outgoing velocities $v'_i, v'_j$ (with $(v'_i-v'_j)\cdot\o >0$) through the relations
\bea
&&v'_i = v_i - \o[\o\cdot(v_i-v_j)]\;,\nn\\
&&v'_j = v_j + \o[\o\cdot(v_i-v_j)]\;.\label{eq:collpp}
\eea
The collision transformation is invertible and preserves the Lebesgue measure on $\RRR^{6}$.

The above prescription defines the {\em flow of the  $n$--particle dynamics}, $t \mapsto \TT_n^\e(t)\bz_n$.
\nomenclature[TTn]{$\TT_n^{\e}$}{$n-$particle hard sphere flow} %
Observe that these rules do not cover all possible situations, e.g. triple collisions are excluded. Nevertheless, 
as proved by Alexander in \cite{Ale75}, there exists a full--measure subset of $\MM_n$, over which $\TT_n^\e(t)$ 
is uniquely defined for all $t$ (see also \cite{MPPP76,CIP94}). 
Thus $\TT_n^\e(t)$ can be defined as a one--parameter group of Borel maps on $\MM_n$, 
leaving invariant the Lebesgue measure.

Notice that the flow $\TT_n^\e(t)$ is piecewise continuous in $t$ (we do not identify outgoing and incoming configurations).
If necessary, we may distinguish the limit from the future $(+)$ and the limit from the past $(-)$ by writing
$\TT_n^\e(t^\pm) \bz_n = \lim_{\e\rightarrow 0^+} \TT_n^{\e}(t\pm\e)\bz_n$. Moreover, we shall fix the convention
of right--continuity of the flow, $\TT_n^{\e}(t)\bz_n = \TT_n^{\e}(t^+)\bz_n$.

\subsection{Statistical states and kinetic limit}
\setcounter{equation}{0}    
\def\theequation{2.2.\arabic{equation}}

\ni Let us turn now to the statistical description of our system. We shall adopt a general formulation,
in the spirit of classical statistical mechanics \cite{Ru67}. 

We introduce the set of density functions over $\MM$, denoted
$\bW^{\e}_{0} =  \{W^{\e}_{0,n}\}_{n\geq 0}$, 
\nomenclature[We]{$\bW^{\e}$}{State of the hard sphere system: a collection of measures $\{W^{\e}_{0,n}\}_{n\geq 0}$} %
where $W^{\e}_{0,n}: \MM_n \rightarrow \RRR^+$ is a positive Borel function symmetric in the particle labels. 
The quantity $(1/n!)W^{\e}_{0,n}(\bz_n)$ gives the probability density of finding exactly $n$ particles in
$z_1,\cdots,z_n$. We refer to $\bW^{\e}_{0}$ as the {\em state of the particle system}.

Note that $n$, the total number of particles, is a random variable, and $(1/n!) \int  W^{\e}_{0,n}$ is its distribution. 
The normalization condition reads
\be
\sum_{n=0}^{\infty} \frac{1}{n!} \int  W^{\e}_{0,n}=1\;.
\ee

Given an initial measure over $\MM$ with density specified by $\bW^{\e}_{0}$, its evolution at time $t$ is given by the 
{\em Liouville equation}
\be
W^{\e}_{n}(\bz_n,t) = W^{\e}_{0,n}(\TT_n^{\e}(-t)\bz_n), \label{eq:liouville}
\ee
to be valid almost everywhere in $\MM_n$. This defines $\bW^{\e} (t) $, the state at time $t$.

For notational convenience, we shall sometimes extend the definition of the state to the 
whole space as 
\be
W^{\e}_{n}(\bz_n,t) = 0\mbox{\,\,\,\,\,\,\,\, if \,\,\,}|x_i-x_k|<\e
\label{eq:WenEXC}
\ee
for some $i\neq k$.

We are not interested in the positions and velocities of the entire particle system, but rather in the 
amount of single particles, couples, triples etc., in a certain configuration. Then we define next 
the vector of {\em correlation functions} over $\MM$ as $\br^\e(t)  = \{\r_j^{\e}(t)\}_{j \geq 0}$, $t\geq 0$, by
\nomenclature[rje]{$\r_j^{\e}$}{Correlation function of order $j$} %
\bea
\r_j^{\e}(\bz_j,t)=\sum_{k=0}^{\infty} \frac{1}{k!} \int_{\MM_k} dz_{j+1}\cdots dz_{j+k} 
W^{\e}_{j+k}(\bz_{j+k},t)\;.\label{eq:defcftris}
\eea

We say that {\em a state admits correlation functions} when the series in the right hand side 
of \eqref{eq:defcftris} is convergent, together with the series in the inverse formula
\bea
W^{\e}_{j}(\bz_j,t)= \sum_{k=0}^{\infty} \frac{(-1)^k}{k!} \int_{\MM_k} dz_{j+1}\cdots dz_{j+k} 
\r^{\e}_{j+k}(\bz_{j+k},t)\;.\label{eq:defcftrisinv}
\eea
In this case, the set of functions $\br^\e(t)$ describes all the properties of the system. 
Later on, we will assume explicit estimates ensuring the convergence of the series for any finite $\e$.

The normalization condition for the correlation functions is
\be
\label{norm}
\int_{\MM_j} \r_j^{\e}(\bz_j,t) d\bz_j = \E_t (n(n-1) \dots (n-j+1))=\E_0 (n(n-1) \dots (n-j+1))\;,
\ee
where $n$ is the total number of particles, and the expectation $\E_t$ is done with respect to the state $\bW^{\e} (t) $.

In this setting, the {\em Boltzmann--Grad scaling} is given by the following condition: the {\em average} number of particles
has to diverge as $\e^{-2}$, that is
\be
\lim_{\e \to 0} \e^2 \int_{\RRR^6} \r_1^{\e}(z_1,t) = \la^{-1}\;, \label{eq:scaling}
\ee
where $\la>0$ is proportional to the mean free path. From now on, we shall fix $$\la = 1$$ for notational simplicity. 

The central object of our study becomes the collection of {\em rescaled correlation functions} (r.c.f.)
defined by
\be
f_j^{\e}(\bz_j,t) = \e^{2j} \r_j^{\e}(\bz_j,t)\;. \label{eq:defrcfhs}
\ee
\nomenclature[fje]{$f_j^{\e}$}{Rescaled correlation function (r.c.f.) of order $j$} %
These are expected to be $O(1)$ as $\e\to 0$.

\subsection{Assumptions on the initial state} \label{sec:AS2}
\setcounter{equation}{0}    
\def\theequation{2.3.\arabic{equation}}

\subsubsection{Initial data for the particle system} \label{subsec: AS2'}

The state of the hard sphere system at time zero, $\bW^{\e}_{0}$, admits, as rescaled correlation functions, 
the collection $f^\e_{0,j}: \MM_j\to \RRR^+$, 
$j \geq 0$, which are by definition Borel functions, symmetric in the permutation of particles.

We assume:
\begin{hyp} \label{hyp:bound}
There exist constants $z,\b>0$ and
a function $h \in L^1(\RRR^3;\RRR^+)$ with $\esssup_x h(x)=z$,
such that the rescaled functions at time zero, $f ^{\e}_{j}(\cdot,0) \equiv f ^{\e}_{0, j}$, satisfy the bound

\be
f ^{\e}_{0, j}(\bz_j) \leq h(x_1)\cdots h(x_j)\ e^{-(\b/2)\sum_{i=1}^j v_i^2}\leq z^j\ e^{-(\b/2)\sum_{i=1}^j v_i^2}\;.
\label{eq:assbound}
\ee

\end{hyp}
Moreover,
\begin{hyp} \label{hyp:cefe0}
There exist two positive constants $\a_0,\g_0$ such that
the initial r.c.f. admit the following representation:
\be
f_{0,J}^{\e} =\sum_{H \subset J} (f_{0,1}^{\e})^{\otimes H} E^0_{J\setminus H}
\label{eq:repid1}
\ee
with $E^0_{\emptyset} = 1$, $E^0_k: \MM_k \to \RRR$ and, for $\e$ small enough,
\be
|E^{0}_K| \leq \e^{\g_0 k} z^k\ e^{-(\b/2)\sum_{i \in K} v_i^2}
\ \ \ \ \ \ \ \ \ \mbox{$\forall$ $k < \e^{-\a_0}$}\;.   \label{eq:EK}
\ee
\end{hyp}
The bound \eqref{eq:EK} holds almost everywhere in $\MM_k(\e)$. Observe that $E^0_1 = 0$.

Here we are using the same notation introduced in Section 1.3 which we recall now and that will be adopted 
throughout all the paper. We use capital latin letters ($J, H, K, \cdots$) for subsets of indices of $\{1,2,3,\cdots\}$
and corresponding small letters for the cardinality of the same sets ($j=|J|, k=|K|,\cdots$), namely, in Eq. \eqref{eq:repid1},
$J=\{1,2 \cdots j\}$ and $H=\{i_1,i_2 \cdots i_h\} \subset J$. 
In addition, $\bz_H= (z_{i_1},z_{i_2}, \cdots, z_{i_h})$ and, for given functions $f_h$, $f$, 
we abbreviate $f_H(\bz_H)=f_h (\bz_H)$ and $f^{\otimes H} (\bz_H)=\prod_{i \in H} f(z_i)$.
Finally, the conventions $f_{\emptyset} = f^{\otimes \emptyset} = 1$ are used.

Notice that, with respect to the usual hypotheses of the Lanford's theorem, we are requiring the additional 
explicit information \eqref{eq:repid1}--\eqref{eq:EK}. 
We know that the hard core exclusion, $|x_i-x_\ell|>\e$, prevents the full factorization of the state. A class of
measures which are  ``maximally factorized'', in the sense that the correlations are only those 
arising from the exclusion,  will be constructed in Appendix A.
Such a class of states fulfills the above hypotheses.

\subsubsection{Initial data for the kinetic equation} \label{subsec: AS2'''}

The initial datum for the Boltzmann and Enskog equations $f_0 = f_0(x,v)$ is a probability density 
over $\RRR^6$ ($\int_{\RRR^6} f_0 = 1$).

As regards the error bound involving the kinetic equation, we also postulate
\begin{hyp} \label{hyp:f0}
There exists a positive constant $\g_0$ such that, for $\e$ small enough,
\be
|\left(f_{0,1}^\e-f_0\right)(x,v)| \leq \e^{\g_0} \,z\, e^{-(\b/2)v^2}\;. \label{eq:conv1}
\ee
\end{hyp}
In particular, in the Boltzmann--Grad scaling, condition \eqref{eq:scaling} is satisfied.
Here the constant $\g_0$ has been chosen equal to the one in Hypothesis \ref{hyp:cefe0}
for  notational simplicity.

Putting together the hypotheses \ref{hyp:cefe0} and \ref{hyp:f0}, 
it follows that the r.c.f. of the hard sphere system admits as well the following representation
in terms of $f_0$:
\be
f^\e _{0,J} =\sum_{H \subset J} f_{0}^{\otimes H} {E}^{\BB,0}_{J\setminus H} \;,
\label{eq:sbo}
\ee
with ${E}^{\BB,0}_k: \MM_k \to \RRR$ satisfying
\be
|{E}^{\BB,0}_K| \leq \e^{\g'_0 k} z^k\ e^{-(\b/2)\sum_{i \in K} v_i^2}
\ \ \ \ \ \ \ \ \ \mbox{$\forall$ $k < \e^{-\a_0}$}\;,  \label{eq:E'K}
\ee
for some $\g'_0>0$ and $\e$ small enough.

\subsection{Results} \label{sec:AS3}
\setcounter{equation}{0}    
\def\theequation{2.4.\arabic{equation}}

We are now in a position to formulate our main results,
summarized in the following theorem. Let 
\be
\MM^x_n(\d) = \Big\{\bx_n \in \RRR^{3n},\ \ |x_i-x_j| > \d,\ i\neq j\Big\} \label{eq:defMxnd}
\ee
\nomenclature[Mxnd]{$\MM^x_n(\d)$}{Phase space of $n$ particles with mutual distance larger than $\d$}%
where
\be
\d = \e^{\theta}
\ee
and $\theta \in (0,1]$.

\begin{thm} \label{thm:MR}
Let $ \bW^{\e}_{0}$ be a state of the hard sphere system with rescaled correlation functions $f^{\e}_{0,j}$
satisfying  Hypotheses \ref{hyp:bound} and \ref{hyp:cefe0}.
Let $\bW^{\e} (t)$ be the state
evolved at time $t>0$, with r.c.f. $f^{\e}_j(t)$. There
exist positive constants $\theta,\a,\g$, a time $t^*>0$ and $\e_0 > 0$ 
such that
\be
f_{J}^{\e}(t) =\sum_{ H \subset J} (f_{1}^{\e}(t))^{\otimes H} E_{J\setminus H}(t)
 \label{eq:EKthmREP} 
 \ee
and
 \be
\int_{\RRR^{3k}} d\bv_k |E_K(t)| \leq \e^{\g k}
\ \ \ \ \ \ \ \ \ \mbox{$\forall$ $k < \e^{-\a}$}\;,  \label{eq:EKthm}
\ee
for any $t< t^*$, $\e < \e_0$ and $\bx_k \in \MM^x_k(\d)$.

Moreover, let $g^\e(t)$, $f(t)$ be the solutions to the Enskog and the Boltzmann equation respectively, 
 with $f_0$ the common  initial datum.

If $f_0$ satisfies Hypothesis \ref{hyp:f0}, then 
for any $t<t^*$, $\e < \e_0$ and $\bx_k \in \MM^x_k(\d)$,
\bea
&& f_{J}^{\e}(t) =\sum_{ H \subset J} (g^{\e}(t))^{\otimes H} E^{\EE}_{J\setminus H}(t)\;,
\label{eq:repid2thmREP}\\
&& 
\int_{\RRR^{3k}}d\bv_k|E^{\EE}_K(t)| \leq \e^{\g k}\ \ \ \ \ \ \ \ \ \mbox{$\forall$ $k < \e^{-\a}$}\;.
\label{eq:repid2thm}
\eea

If, additionally, $f_0$ is Lipschitz continuous with respect to the space variables, with 
Lipschitz constant $Le^{-(\b/2)v^2}$, $L>0$, then 
for any $t<t^*$, $\e < \e_0$ and $\bx_k \in \MM^x_k(\d)$,
\bea
&& f_{J}^{\e}(t) =\sum_{ H \subset J} (f(t))^{\otimes H} E^{\BB}_{J\setminus H}(t)\;,
\label{eq:repid3thmREP}\\
&& 
\int_{\RRR^{3k}}d\bv_k|E^{\BB}_K(t)| \leq \e^{\g k}\ \ \ \ \ \ \ \ \ \mbox{$\forall$ $k < \e^{-\a}$}\;.
\label{eq:repid3thm}
\eea
\end{thm}

As we shall see in the course of the proof, the solutions $g^\e(t)$ and $f(t)$ are local in time and 
constructed by means of a series expansion.  

Equation \eqref{eq:EKthmREP}--\eqref{eq:EKthm} is an expression for the propagation of chaos, with
an explicit representation of the error, while Equations 
\eqref{eq:repid2thmREP}--\eqref{eq:repid2thm} and 
\eqref{eq:repid3thmREP}--\eqref{eq:repid3thm} express in addition the asymptotic equivalence of the r.c.f. 
with the solution of the Enskog and the Boltzmann equation.

Moreover, the convergence to the Boltzmann equation can be also expressed in terms of deviation
from average values of observables. To this purpose, let us denote $\EEE^\e$, $\EEE^{\BB}$ the average
values with respect to the hard sphere state and the Boltzmann density respectively. 
Then the following result holds:

\begin{thm} \label{cor:MR}
Let $\varphi_i \in C_{c}(\RRR^6;\RRR)$, $i=1,2,\cdots$ be a sequence of test functions 
with $\max\left(\|\varphi_i\|_{L^\infty}, \|\varphi_i\|_{L^1_x(L^{\infty}_v)}\right) \leq G$ for some $G>0$.
Moreover, let $F_i = F_i(t): \MM \to \RRR$ be the associated sequence of observables defined by \eqref{def:obs}.
Then, if Hypotheses \ref{hyp:bound}, \ref{hyp:cefe0} and \ref{hyp:f0} hold, 
there exists  a positive constant $\a'$ such that, for any $t< t^*$,
\be
\lim_{\e \to 0} \,\sup_{j < \e^{-\a'}}\,
\Big|\EEE^\e\left[\prod_{i= 1}^j\Big(F_i(t) - \EEE^\BB[\varphi_i(t)]\Big)\right] \Big|= 0\;.
\label{eq:convMT'}
\ee
\end{thm}
The theorem will be proved in Section \ref{sec:proofcorconv}.

\subsubsection{Comments on the result} \label{subsec:CR}

\ni (1)\;
The constants $\a$ and $\g$ can be computed explicitly.
Upper bounds (certainly not optimal) will be given in Section 4.3.3.a, as a byproduct
of the proof.

In this paper we are not concerned with optimal bounds on rates of convergence
nor with the optimality of the coefficient $\a$. 
Improvements in this direction would complicate 
the proof presented in the rest of the paper.
An exception is the geometrical estimate of internal recollisions (see Lemma \ref{lem:MRresINT}),
which can be shown to be $\e^{\g_1}$ for arbitrary $\g_1<1$ by following the proof in Appendix D.

\medskip

\ni (2)\;
The limiting time $t^*$ is obtained by imposing the absolute (uniform in $\e$) convergence of the 
series expansions appearing in the proof, and is determined only by $z,\b$ (see Hypothesis~\ref{hyp:bound}).

\medskip

\ni (3)\; 
The use of the $L^1-$norm in the velocity variables
is essential in the proof of the estimate of many--recollision events (Lemmas \ref{lem:MR} and \ref{lem:MRresINT} below), 
to obtain a $k-$dependent rate of convergence such as \eqref{eq:EKthm}.
Chebyshev's inequality implies then that $|E_K(t)|\leq \e^{\bar\g k}$ for some $\bar\g>0$ (and 
similar estimates for $E^\EE$ and $E^\BB$), outside a subset of $\MM_k$ of measure smaller than 
$\e^{(\g-\bar\g)k}$. 

\medskip

\ni (4)\label{rem:DetSC}\; In particular, the comparison with the uniform estimate in Hypothesis \ref{hyp:cefe0}
shows that the set where the convergence takes place deteriorates in time. This is a feature of the Boltzmann--Grad limit.
In fact, it will be clear from the proof that, due to recollisions, the propagation of chaos
$f_{J}^{\e}(t) \to (f_{1}^{\e}(t))^{\otimes J}$ necessarily fails over the time--dependent set 
$$\Big\{\bz_J  \ \Big|\ \min_{i,k\in J }\min_{s\in (0,t)}[\left(x_i-x_k\right) - \left(v_i-v_k\right)s] = 0\Big\}.$$
Actually it can be proved that, outside this null--measure set,
$E_K(t) = O(\e^\eta)$ for some $\eta > 0$ (see \cite{PSS13}, where this is done for a system of smoothly 
interacting particles). 

\medskip

\ni (5)\; The above discussion does not influence, however, the following integrated estimate.
\begin{prop} \label{cor:EKthmOBS}
Let $\varphi_i$ be test funtions as in Theorem \ref{cor:MR}.
In the assumptions of Theorem \ref{thm:MR}, there exists $\a'>0$ such that, for $t<t^*$ and $\e$
small enough,
\be
\Big|\int_{\RRR^{6k}} d\bz_k \, \varphi(z_1)\cdots \varphi(z_k)\,E_K(\bz_k,t) \Big| \leq \e^{\g k}
\ \ \ \ \ \ \ \ \ \mbox{$\forall$ $k < \e^{-\a'}$}\;. \label{eq:EKthmOBS}
\ee
\end{prop}
Observe that, by definition (remind Eq. \eqref{eq:WenEXC}), 
since $f^{\e}_J(\bz_J,t)=0$ when two particles in $\bz_J$ are at distance smaller than $\e$, 
inside the ``excluded'' region $\RRR^{6k}\setminus\MM_k$ (of small measure) the correlation errors
will generally satisfy the bad estimate $|E_K(t)|\leq (const.)^k$ (see \eqref{correrr1}).
Equation \eqref{eq:EKthmOBS} will be derived in Section \ref{sec:proofcorconv}.

\medskip

We conclude with some comments on the choice of the setting.

\medskip

\ni (6)\; We are working with rescaled correlation functions in a grand canonical formalism 
(no fixed total number of particles $N$) in place of the more usual formulation in terms of marginals in the
canonical setting (fixed $N=\e^{-2} $).  This choice is always convenient when dealing with fluctuations and truncated 
functions (i.e. cumulant expansions), see e.g. \cite{BLLS80,S81,Sp83}. The reason is that the mere facts of
(i) fixing the number of particles $N$, and (ii) labelling the particles from $1$ to $N$ (implied in the
definition of marginal) are itself a source of correlation. Consequently, even though the r.c.f. are asymptotically 
equivalent to the marginals of the canonical setting, here additional error terms are produced
which should be expanded and estimated in order to get a quantitative result like \eqref{eq:EKthm}. We do not deal
with this problem in the present paper. 
More details on this point will be provided in Section \ref{subsec:canonical}.

\medskip

\ni (7)\;
Another simplification comes from the choice of the unbounded spatial domain $\RRR^3$.
Since we do not use dispersive properties, our analysis in the whole space can be transferred 
to the case of a bounded box
(assuming periodic or reflecting boundary conditions) with minor modifications (see Section \ref{subsec:BOX}).
One faces here two extra difficulties. The first one arises from the fact that the recollisions are 
more likely. This has been discussed in \cite{BGSR13}. The second one is that, 
as in the canonical  formalism, the total number of particles cannot exceed
a given integer, that is the close--packing number $N_{cp}$. However $N_{cp}=O(\e^{-3})$ is much larger
than the average density and the corresponding error of correlation is very small and easily tractable.


\subsubsection{Further remarks on the initial states} \label{subsec: AS2''}

In the proof of the main result we shall find more convenient to use a
representation of the initial data different from the one given in Hypothesis \ref{hyp:cefe0}.
We illustrate it in this section.

Let $S= \{1, \cdots, s \} $ be a set of indices (particles) and $\{ S_1, S_2, \cdots, S_j\}$ a partition of $S$ 
into nontrivial clusters, i.e. 
$\cup_{i=1}^j  S_i=S$, $S_i \cap S_k =\emptyset $ for $i \neq k$, $|S_i|>0$.

Denote by  $\JJ = \{ 1,\cdots, j \} $ the set of indices of the clusters $\{S_i \}$.
\nomenclature[JJ]{$\JJ$}{Set of indices of clusters $\{1,2,ááá ,j\}$}%

We introduce an expansion on products of higher order, not only $1$--point, rescaled correlation functions.
We  use a calligraphic capital letter for the subsets of $\JJ$.

\bigskip

\ni { \bf Property 1}\;\; {\it There exist two positive constants $\a_0,\g_0$ such that
the initial r.c.f. admit the following collection of representations:
\be
f^{\e}_{0,S}  =\sum_{\HH \subset \JJ}   
\left(\prod_{i \in \HH}f^{\e}_{0,S_i}\right) 
E^0_{\JJ\setminus \HH}
\label{eq:repid1int}
\ee
\nomenclature[EzKK]{$E^0_{\KK}$}{Time--zero correlation error associated to the partition in the clusters $\KK$}%
for any partition of the set $S$, where
$E^0_{\emptyset}=1$, 
$E^0_{\KK}: \MM_{k} \to \RRR$ and, for $\e$ small enough,
\be
|E^0_{\KK} | \leq \e^{\g_0 |\KK|}  z^k\ e^{-(\b/2)\sum_{i \in K} 
v_i^2}\ \ \ \ \ \ \ \ \ \mbox{$\forall$ $k < \e^{-\a_0}$}\;,
\label{eq:repid1intest}
\ee
with $|\KK| = $\ total number of elements (clusters) in $\KK$,  and $k = $\  
total number of indices (particles) in $K = \cup_{i \in \KK} S_i$. 
} 

\bigskip

Note that $E^0_{\KK}=0$ for $|\KK|=1$. We stress that $E^0_{\KK}$ and $E^0_{K}$
denote different quantities (unless all the clusters in $\KK$ are singletons).

Property 1 is actually equivalent to Hypothesis \ref{hyp:cefe0}. 
For the proof, we refer to Appendix~A.

Observe that, again, $E^0_{\KK}$ will be order 1 outside $\MM_{k}(\e)$.
As already mentioned, this is due to the hard sphere exclusion which is a first obvious source of correlation, 
namely the fact that the r.c.f. $f^\e_j (t)$ is naturally defined on $\MM_j$ and extended to zero outside.
On the other hand, we will need to compare the r.c.f. with $(f_1^\e(t) )^{\otimes j}$ which is defined in the extended 
phase space $\RRR^{6j}$. In particular it will be necessary, in the course of the proof, to embed Eq.
\eqref {eq:repid1int} in the whole space $\RRR^{6s}$ as follows:
\be
\label{expexten}
f^{\e}_{0,S}  =\bar \chi_S^0 \sum_{\HH \subset \JJ}   
\left(\prod_{i \in \HH}f^{\e}_{0,S_i}\right) 
E^0_{\JJ\setminus \HH}\;,
\ee
where
$$
\bar \chi_S^0=\prod_{\substack{i,k \in S \\ i \neq k}}\bar \chi^0_{i,k}
$$
and $ \bar \chi^0_{i,k}$ is the indicator function of the set $\{ |x_i -x_k| > \e \}$.

In Section \ref{subsec:comb} we develop a technique which allows a useful expansion of 
$\bar \chi_S^0$, for which we can prove (see Appendix A):

\bigskip

\ni { \bf Property 2}\;\; 
{\it Equation \eqref {eq:repid1int} can be extended in $\RRR^{6s}$ according to
  \be
f^{\e}_{0,S}  =\sum_{\HH \subset \JJ}   
\left(\prod_{i \in \HH}\bar \chi_{S_i}^0 f^{\e}_{0,S_i}\right) 
\bar E^0_{\JJ\setminus \HH}\;,
\label{eq:repid2intext}
\ee
\nomenclature[EzKKb]{$\bar E^0_{\KK}$}{Extension of $E^0_{\KK}$ to the whole space}%
with
\be
\label{boundEtilde}
|\bar E^0_{\KK}| \leq \sum_{\substack {\HH_1,\HH_2  \\ \HH_1 \cup \HH_2=\KK\\ \HH_1 \cap \HH_2=\emptyset}}  
\left(C^{|\HH_1|}\, |\HH_1|!\,  \chi^0_{\HH_1,\KK}\, \prod_{i \in \HH_1}\bar \chi_{S_i}^0 f^{\e}_{0,S_i}\right) \, 
\left(\bar\chi^0_{H_2} |E^0_{\HH_2}|\right)\;,
\ee
where: 

\ni (i) $\chi^0_{\HH_1,\KK}=1$ if and only if any cluster $S_i$, with $i\in\HH_1$, has, at least, one particle ``overlapping''
with another particle in $S_j$ with $j\in \KK$, $j\neq i$;

\ni (ii) $H_2 = \cup_{i \in \HH_2} S_i$.

}

\bigskip

By {\em overlap} of two particles we mean that their relative distance is smaller than $\e$.

Note that $\bar\chi^0_{H_2}$ allows to insert \eqref{eq:repid1intest} into \eqref{boundEtilde}, while the particles
contained in $\HH_1$ are constrained to lie in a very small set. Explicitly, 
\be
|\bar E^0_{\KK}| \leq z^k\, e^{-(\b/2)\sum_{i \in K} v_i^2}\, 
\sum_{\substack {\HH_1,\HH_2  \\ \HH_1 \cup \HH_2=\KK\\ \HH_1 \cap \HH_2=\emptyset}}  
C^{|\HH_1|}\, |\HH_1|!\,  \chi^0_{\HH_1,\KK}\, \e^{\g_0 |\HH_2|}\;,
\label{boundEtildeUsed}
\ee
for all $k < \e^{-\a_0}$.


\section{Hierarchies} \label{sec:hs}
\setcounter{equation}{0}    
\def\theequation{3.\arabic{equation}}

In this section we introduce the standard description of the evolution of a statistical state of particles, namely
the chain of BBGKY hierarchy equations (Sec. \ref{sec:BBGKY}). We also introduce the analogue hierarchies at the kinetic
level, which can be obtained by formally taking the limit $\e\to 0$ (Sec. \ref{sec:BHier}-\ref{sec:EHier}).
An explicit representation of the solution to the BBGKY can be given in terms of a tree expansion and of a class of 
special flows of particles evolving backwards in time. An analogous description is also possible for the Boltzmann 
(or the Enskog) evolution equation. These well known expressions, which will be our basic tool, are introduced in sections 
\ref{sec:hs2}--\ref{sec:hs2B} together with some new expansions that will have the role of an intermediate 
object in the transition towards the kinetic limit. We conclude the section with a summary of the main objects 
introduced.

\subsection{BBGKY hierarchy}  \label{sec:BBGKY}
\setcounter{equation}{0}    
\def\theequation{3.1.\arabic{equation}}

\ni We describe here the time evolution of the hard sphere system for any fixed $\e>0$.
The evolution equations for the considered quantities were first  derived formally by Cercignani in 
\cite{Ce72}. 

Assuming some explicit bound and 
sufficient smoothness, he deduced the hard sphere version of the {\em BBGKY hierarchy} of equations \cite{Bog62},
which for the rescaled correlation functions takes the form
\be
\left(\pa_t +\sum_{i=1}^j v_i \cdot \nabla_{x_i}\right) f_j^{\e}(\bz_j,t)
= \sum_{i=1}^j \int_{S^2\times\RRR^3}d\o \ dv_{j+1}\ B^\e(\o; v_{j+i}-v_i)\ 
f_{j+1}^{\e}(\bz_j,x_i+\e\o,v_{j+1},t)\;,
\ee
where
\bea
B^\e(\o; v_{j+i}-v_i)=\o \cdot (v_{j+i}-v_i)\,
\mathbbm{1}_{\{\min_{\ell=1,\cdots,j;\ell\neq i}|x_i+\o\e-x_\ell|>\e\}}(\o)\;. \label{eq:specBeBBGKY}
\eea
\nomenclature[Be]{$B^\e$}{BBGKY collision kernel}%
Here $\mathbbm{1}_A$ denotes the indicator function of the set $A$.

Notice that the difference of this formula with respect to the hierarchy written for marginals in the canonical setting 
(as for instance in \cite{Ce72}), is that the factor $\e^2(N-j)$ is absent in the right hand side ($N = $ fixed total 
number of particles). This is a small notational advantage in using correlation functions.

The {\em series solution} of the hierarchy (obtained from integration and repeated iteration of the above formula) is 
\bea
&& f_j^{\e}(t) = \sum_{n \geq 0} \int_0^t dt_1 \int_0^{t_1}dt_2\cdots\int_0^{t_{n-1}}dt_n \nn\\
&& \SS_j^\e(t-t_1)\CC^\e _{j+1}\SS_{j+1}^\e(t_1-t_2)\cdots 
\CC^\e _{j+n}\SS^\e_{j+n}(t_n) f^{\e}_{j+n}(0)\;, \label{eq:BBGexp}
\eea
where we used the definitions of {\em interacting flow operator} $\SS^\e_j(t)$ 
\nomenclature[SSej]{$\SS^\e_j$}{$j-$particle interacting flow operator}%
and {\em BBGKY collision operator} $\CC^{\e}_{j+1}$,
\nomenclature[CCej]{$\CC^{\e}_{j+1}$}{BBGKY collision operator}%
i.e. respectively 
\be
\SS^\e_j(t) f_j^{\e}(\bz_j,\cdot)=f_j^{\e}(\TT^\e_j(-t)\bz_j,\cdot)
\ee
and 
\bea
&& \CC^{\e}_{j+1} = \sum_{k=1}^j \CC^{\e}_{k,j+1}\label{eq:Cedec}\\
&& \CC^{\e}_{k,j+1}f_{j+1}^{\e}(\bz_j,\cdot) =
\int_{S^2 \times\RRR^3}d\o dv_{j+1}
B^\e(\o; v_{j+1}-v_{k}) f^{\e}_{j+1}(\bz_{j},x_k+\o\e,v_{j+1},\cdot)\;.\nn
\eea

Rigorous derivations of the hard sphere hierarchy, under rather weak assumptions on the initial measure, 
have been discussed later on, e.g. \cite{Sp85,IP87,Si13}\footnote{See also \cite{PS15}, appeared before 
revision of the present paper.}.
The latter references focus mainly on the validity of the series expansion \eqref{eq:BBGexp}.

Let us formulate the result in a form useful for our analysis. 
\begin{prop}[BBGKY series expansion] \label{prop:BBGKY}
Let $ \bW^{\e}_{0}$ be a state of the hard sphere system satisfying Hypothesis \ref{hyp:bound}.
Then the measure at any time $t>0$ has rescaled correlation functions 
$f^{\e}_j(t)$ given by Eq. \eqref{eq:BBGexp}, for almost all points in $\MM_j$.
\end{prop}
For a complete proof of the validity result as formulated above, we refer to 
\cite{Si13}\footnote{Note that the quoted result of \cite{Si13} (Corollary 2) is stated for a system of particles in 
a finite box. Given the explicit assumption on the spatial decay, Eq. \eqref{eq:assbound}, the result
can be easily established on the full space along the same lines.}.

Proposition \ref{prop:BBGKY} is the starting point of our analysis. 
All the formulas involving the r.c.f. at positive times will be valid only almost everywhere.

\subsection{Boltzmann hierarchy} \label{sec:BHier}
\setcounter{equation}{0}    
\def\theequation{3.2.\arabic{equation}}

Now we want to give a picture of the Boltzmann equation which can be conveniently compared to \eqref{eq:BBGexp}.

Suppose that $f$ is a solution to the Boltzmann equation \eqref{BE} (with $\la=1$).
Consider the products
\be
f_j(\bz_j,t)=f(t)^{\otimes j}(\bz_j) = f(z_1,t)f(z_2,t)\cdots f(z_j,t)\;. \label{eq:fjtdef}
\ee
\nomenclature[fj]{$f_j$}{$j-$particle function solving the Boltzmann hierarchy}%
The family of $f_j$ solves then the hierarchy of equations ({\em Boltzmann hierarchy}):
$$
\left(\pa_t+\sum_{i=1}^j v_i \cdot \nabla_{x_i}\right)f_j = \CC_{j+1}f_{j+1}\;,
$$
where
\bea
&& \CC_{j+1} =\sum_{k=1}^j \CC_{k,j+1} \label{eq:defBco} \\
&& \CC_{k,j+1} = \CC^{+}_{k,j+1} - \CC^{-}_{k,j+1} \nn\\
&& \CC^{+}_{k,j+1}f_{j+1}(\bz_j,\cdot) = \int_{S^2_+\times\RRR^3}d\o dv_{j+1}
(v_k-v_{j+1})\cdot\o f_{j+1}(z_1,\cdots,x_k,v'_k,\cdots,z_j,x_k,v'_{j+1},\cdot) \nn\\
&& \CC^{-}_{k,j+1}f_{j+1}(\bz_j,\cdot) = \int_{S^2_+\times\RRR^3}d\o dv_{j+1}
(v_k-v_{j+1})\cdot\o f_{j+1}(z_1,\cdots,x_k,v_k,\cdots,z_j,x_k,v_{j+1},\cdot)\;, \nn
\eea
\nomenclature[CCj]{$\CC_{j+1}$}{Boltzmann hierarchical collision operator}%
with
\be
\begin{cases}
\displaystyle v'_k=v_{k}-\om [\om\cdot(v_k-v_{j+1})] \\
\displaystyle  v'_{j+1}=v_{j+1}+\om[\om\cdot(v_k-v_{j+1})]
\end{cases}
\ee
and
\be
S^2_+ = \{\o\ |\ (v_k - v_{j+1})\cdot\o \geq 0\}\;.
\ee

The corresponding series solution reads
\bea
&& f_j(t)= \sum_{n\geq 0}\int_0^t dt_1 \int_0^{t_1} dt_2 \cdots \int_0^{t_{n-1}}dt_n \nn\\
&& \ \ \ \ \ \ \ \ \ \ \ \ \ \ \ \ \cdot\SS_j(t-t_1)\CC _{j+1}\SS_{j+1}(t_1-t_2)\cdots \CC _{j+n}\SS_{j+n}(t_n) f_{0,j+n}\;,
\label{eq:fjexp}
\eea
where now $\SS_j(t)$ is the {\em free flow operator}, defined as
\be
\SS_j(t)f_j(\bz_j,\cdot) = f_j(x_1-v_1t,v_1,\cdots,x_j-v_jt,v_j,\cdot)\;, \label{eq:ffodef}
\ee
\nomenclature[SSj]{$\SS_j$}{$j-$particle free flow operator}%
and 
\be
f_{0,j} = f_0^{\otimes j} \label{eq:indB}
\ee
are the initial data.

The absolute convergence of this formula has been discussed in \cite{La75} and holds (over all $\RRR^6$)
only for a sufficiently small time. We shall give a proof, for completeness, in Section \ref{sec:proofpre} (Proposition
\ref{prop:STE}).
This implies, in particular, local existence and uniqueness 
of the solution to the time--integrated version of the Boltzmann 
hierarchy in the class of continuous functions such that $|f_j(t)|\leq c^j e^{-c'\sum_{i=1}^jv_i^2}$ for some $c,c'>0$. Moreover,
in the case of initial product states, factorization is propagated in time, each factor being the local solution to the time--integrated 
Boltzmann equation (see formula \eqref{eq:fBfact} below). 

The similarity of \eqref{eq:fjexp} and \eqref{eq:BBGexp} follows from the well known decomposition of the collision 
operator into its positive and negative part.

\subsection{Enskog hierarchy} \label{sec:EHier}
\setcounter{equation}{0}    
\def\theequation{3.3.\arabic{equation}}

We provide here an intermediate item between the BBGKY hierarchy and the Boltzmann hierarchy, 
that is the so called {\em Enskog hierarchy}. As mentioned in the introduction, this will turn to be useful in the sequel. 

Let $g^{\e}$ be a solution to the {\em Enskog Equation} \eqref{EE} (with $\la=1$).
Proceeding as above, the products
\be
g^{\e}_j(\bz_j,t)=g^{\e}(t)^{\otimes j}(\bz_j) \label{eq:tfjtdef}
\ee
\nomenclature[gej]{$g^{\e}_j$}{$j-$particle function solving the Enskog hierarchy}%
satisfy 
\be
\left(\pa_t+\sum_{i=1}^j v_i \cdot \nabla_{x_i}\right)g^{\e}_j = \CC^{\EE}_{j+1}g^{\e}_{j+1}\;,
\ee
where the definition of $\CC^{\EE}_{j+1}$ is induced by that of the collision operator on the right hand side of \eqref{EE}
(the symbol $\EE$ stands for ``Enskog'', while we drop the dependence on $\e$).
\nomenclature[CCj]{$\CC^{\EE}_{j+1}$}{Enskog hierarchical collision operator}%

Deriving the corresponding series solution and performing a change of variables $\o \to - \o$ inside the positive part
of $\CC^{\EE}_{j+1}$ (see the next section for details), one obtains easily
\bea
&& g^{\e}_j(t)= \sum_{n\geq 0}\int_0^t dt_1 \int_0^{t_1} dt_2 \cdots \int_0^{t_{n-1}}dt_n \nn\\
&& \ \ \ \ \ \ \ \ \ \ \ \ \ \ \ \ \cdot\SS_j(t-t_1)\CC^{\EE}_{j+1}\SS_{j+1}(t_1-t_2)\cdots\CC^{\EE}_{j+n}\SS_{j+n}(t_n) 
g^{\e}_{0,j+n}\;,
\label{eq:fjexpE}
\eea
where the collision operator can be written as 
\be
\CC^{\EE}_{j+1}g^{\e}_{j+1}(\bz_j,\cdot) =\sum_{k=1}^j \int_{S^2 \times\RRR^3}d\o dv_{j+1}
(v_{j+1}-v_{k})\cdot\o\, g^{\e}_{j+1}(\bz_{j},x_k+\o\e,v_{j+1},\cdot)\;,
\label{eq:Ecollop}
\ee
and 
\be
g^{\e}_{0,j} =f_0^{\otimes j} \label{eq:indE}
\ee
are the initial data (which in this paper will be assumed, for simplicity, equal to the initial data 
of the Boltzmann hierarchy).

Notice that the operator $\CC^{\EE}_{j+1}$ is identical to $\CC^{\e}_{j+1}$ introduced in \eqref{eq:Cedec}, 
except for the fact that the former allows particles to overlap.

Local existence, uniqueness and propagation of chaos are discussed exactly as for the Boltzmann 
hierarchy (see the comment after \eqref{eq:indB}, and formula \eqref{eq:fEfact} below).


\subsection{The tree expansion} \label{sec:hs2}
\setcounter{equation}{0}    
\def\theequation{3.4.\arabic{equation}}

In this section we shall follow mainly \cite{PSS13} Sec. 6, adapting discussions and notation therein 
to the simpler case of hard spheres. Our purpose is to rewrite formulas \eqref{eq:BBGexp} and 
(in the next section) \eqref{eq:fjexp} in a convenient and more explicit way.

We start from \eqref{eq:BBGexp}, which we write as
\bea
&& f_j^{\e}(t) = \sum_{n \geq 0}{\sum_{\bk_n}}^* 
\int_0^t dt_1 \int_0^{t_1}dt_2\cdots\int_0^{t_{n-1}}dt_n \nn\\
&&\ \ \ \ \ \ \ \ \ \cdot\SS_j^\e(t-t_1)\CC^{\e}_{k_1,j+1}\SS_{j+1}^\e(t_1-t_2)\cdots 
\CC^{\e} _{k_n,j+n}\SS^\e_{j+n}(t_n) f^{\e}_{0,j+n}\;, \label{eq:tfnjexp}
\eea
where
\bea
&& {\sum_{\bk_n}}^* = \sum_{k_1=1}^j \sum_{k_2=1}^{j+1}\cdots \sum_{k_n=1}^{j+n-1}\;. \label{eq:specialsumk}
\eea

We introduce the {\bf{\em $n-$collision, $j-$particle tree}}, denoted $\G(j,n)$, 
\nomenclature[Gjn]{$\G(j,n)$}{$n-$collision, $j-$particle tree}%
as the collection of integers
$k_1,\cdots,k_n$ that are present in the sum \eqref{eq:specialsumk}, i.e.
\be
k_1\in I_j, k_2 \in I_{j+1}, \cdots, k_n\in I_{j+n-1}\;,\ \ \ \ \ \ \mbox{with\ \ \ \ \ \ $I_s=\{1,2,\cdots,s\},$}
\label{eq:defsumtreesb}
\ee
so that
\be
{\sum_{\bk_n}}^* = \sum_{\G(j,n)}\;. \label{eq:defsumtrees}
\ee

\begin{figure}[htbp] 
\centering
\includegraphics[width=5in]{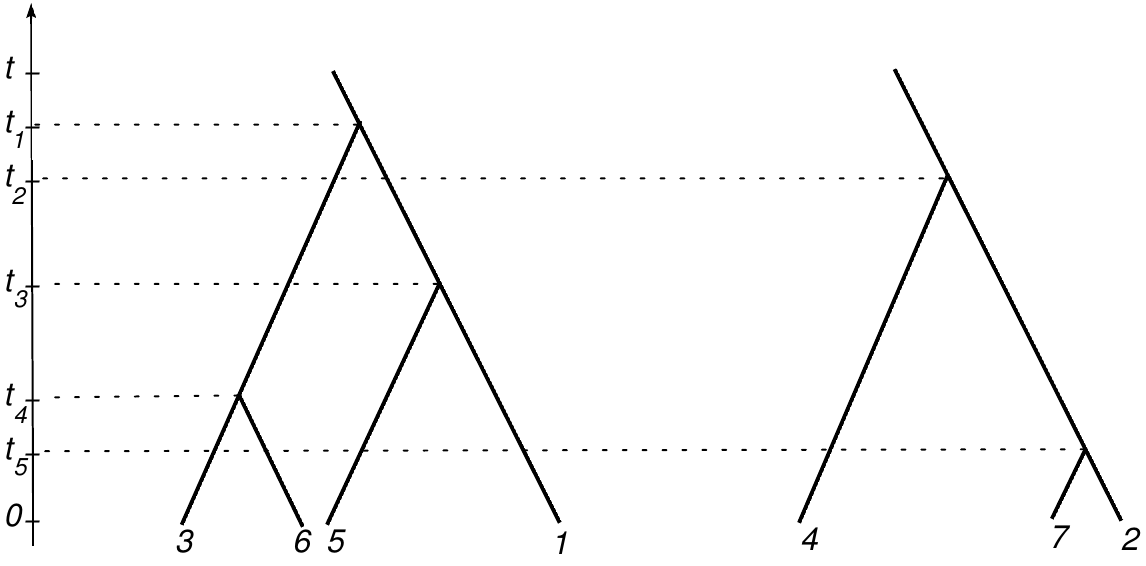} 
\caption{The two--particle tree $\G(2,5)=(1, 2, 1, 3, 2).$ The tree associated to $1$ is $\G_1 = (1,1,3)$,
while $\G_2 = (2,2)$.}
\label{fig:treedef}
\end{figure}

The name ``tree'' is justified by its natural graphical representation,
which we explain by means of an example: see Figure \ref{fig:treedef} corresponding to $\G(2,5)$ 
given by $1, 2, 1, 3, 2.$

\ni In the figure, we have also drawn a time arrow in order to associate times to the nodes of the trees:
at time $t_i$ the line $j+i$ is ``created''. 
Lines $1$ and $2$ of the example, existing for all times, are called
``root lines''.

\subsubsection{The interacting backwards flow (IBF)} 

Given a $j$--particle tree  $\G(j,n)$ and fixed a value of all the integration variables in the expansion 
\eqref{eq:tfnjexp} (times, unit vectors, velocities), we associate to them a special ($\e-$dependent)
trajectory of particles, which we call {\em interacting backwards flow} (IBF in the following), since it will be 
naturally defined by going backwards in time. The rules for the construction of this evolution are explained in what 
follows.

\medskip
First, we introduce a notation for the configuration of particles in the IBF, by making use of Greek alphabet, i.e.
$\bze^\e(s)$, where $s \in [0,t]$ is the time. 
\nomenclature[zee]{$\bze^\e(\cdot)$}{Interacting backwards flow}%
Note that there is no label specifying the number of particles. This 
number depends indeed on the time. If $s \in (t_{r+1},t_r)$ (with the convention $t_0=t, t_{n+1}=0$), there are 
exactly $j+r$ particles:
\be
\bze^\e(s) = (\z_1^\e(s),\cdots,\z_{j+r}^\e(s)) \in \MM_{j+r}\ \ \ \ \ \mbox{for }s \in (t_{r+1},t_r)\;, \label{eq:IBFnot}
\ee
with
\be
\z_i^\e(s)=(\xi^\e_i(s),\eta^\e_i(s))\;,
\ee
the positions and velocities of the particles being respectively
\bea
&& \bxi^\e(s) =(\xi_1^\e(s), \cdots, \xi_{j+r}^\e(s))\;, \nn\\
&& \bet^\e(s)=(\eta_1^\e(s), \dots, \eta_{j+r}^\e(s))\;. \label{eq:IBFnot'}
\eea
\nomenclature[xe]{$\bxi^\e(\cdot)$}{Positions in the IBF}%
\nomenclature[ee]{$\bet^\e(\cdot)$}{Velocities in the IBF}%

Our final goal is to write Eq. \eqref{eq:tfnjexp} in terms of the IBF (to be defined below), i.e.:
\be\label{fjN-grad}
f_j^{\e}(\bz_j, t)= \sum_{n \geq 0} \sum _{\G (j,n)} \T^\e (\bz_j,t)
\ee
where $\T^\e(\bz_j,t)$ is the {\em value of the tree} $\G(j,n)$ {\em with configuration } $\bz_j$ {\em at time} $t$, {\em for
the interacting flow},
\be
\T^\e (\bz_j,t)= \int d \L (\bt_n , \bo_n , \bv_{j,n}) \prod_{i=1}^n 
B^\e(\o_i; v_{j+i} - \eta^\e_{k_i} (t_i)) f^{\e}_{0,j+n}(\bze^\e(0))\;, \label{eq:Teszt}
\ee
\bea
&& \bt_n = t_1,\cdots,t_n \;,\nn\\
&& \bo_n = \o_1,\cdots,\o_n \;, \nn\\
&& \bv_{j,n} = v_{j+1},\cdots,v_{j+n} \;,
\eea
\nomenclature[tn]{$\bt_n$}{Times of scattering (creation) in backwards flow}%
$d\L$ is the measure on $\RRR^n\times S^{2n}\times\RRR^{3n}$
\be
d\Lam ({\bf  t}_n , \bo_n ,\bv_{j,n})= \mathbbm{1}_{\{t>t_1>t_2 \dots >t_n>0\}} dt_1\dots dt_n
d\o_1\dots d\o_n dv_{j+1}\dots dv_{j+n}\;,
\ee
\nomenclature[dL]{$d\L$}{Integration measure in the tree expansion}%
and we use the shorthand notation
\bea
B^\e(\o_i; v_{j+i}-\eta^\e_{k_i} (t_i))=\o_i \cdot (v_{j+i}-  \eta^\e_{k_i} (t_i))
\mathbbm{1}_{\{|\xi^\e_{j+i}(t_i)-\xi^\e_{k}(t_i)| > \e\  \forall k\neq k_i\}}\;.\label{eq:defBe}
\eea
In other words, in the generic term $\T^\e (\bz_j,t),$ the initial datum $ f^{\e}_{0,j+n}$ is integrated,
with the suitable weight, over all the possible time--zero states of the IBF associated to $\G(j,n) $.

In formula \eqref{eq:Teszt}, the triple $(t_i,\o_i,v_{j+i})$ may be thought as associated 
to the node of $\G(j,n)$ where line $j+i$ is created (see Figure \ref{fig:treedef}). 
\nomenclature[tov]{$(t_i,\o_i,v_{j+i})$}{Triple describing a scattering (creation) in backwards flow}%
In the rest of the paper, we shall abbreviate further 
\be
\int d \L (\bt_n , \bo_n , \bv_{j,n}) \prod B^\e = 
\int d \L (\bt_n , \bo_n , \bv_{j,n}) \prod_{i=1}^n B^\e(\o_i; v_{j+i} - \eta^\e_{k_i} (t_i)) \;, \label{eq:Tesztbis}
\ee
where the $ \eta^\e_{k_i} (t_i)$ in the factors $B^{\e}$ have to be computed through 
the rules specified below, 
starting from the set of variables $(\bt_n , \bo_n , \bv_{j,n})$, the corresponding $j$--particle tree (whose nodes
are labeled by $(\bt_n , \bo_n , \bv_{j,n})$), together with the associated value of $\bz_j, t$.

\medskip
Let us construct $\bze^\e(s)$ for a fixed collection of variables $\G(j,n), \bz_j,\bt_n,\bo_n,\bv_{j,n}$, with
\be
t\equiv t_0 > t_1 > t_2 > \cdots > t_n > t_{n+1}\equiv 0\;, \label{eq:ordertimes}
\ee
and $\bo_n$ satisfying a further constraint that will be specified soon. The root lines of the $j$--particle tree 
are associated to the first $j$ particles, with configuration $\z_1^\e,\cdots,\z_j^\e.$ 
Each branch $j+\ell$ ($\ell=1,\cdots,n$) represents a new 
particle with the same label, and state $\z_{j+\ell}^\e.$ This new particle appears, going backwards in time, at 
time $t_\ell$ in a collision configuration with a previous particle (branch) $k_\ell\in\{1, \cdots, j+\ell -1\},$ with either
incoming or outgoing velocity.
 
More precisely, in the time interval $(t_r,t_{r-1})$ particles $1,\cdots,j+r-1$ flow according to the usual dynamics
$\TT^\e_{j+r-1}.$ This defines $\bze^\e_{j+r-1}(s)$ starting from $\bze^\e_{j+r-1}(t_{r-1}).$ At time $t_r$ the 
particle $j+r$ is ``{\em created} ''   by particle $k_r$ in the position
\be
\xi_{j+r}^\e(t_r)= \xi_{k_r}^\e(t_r)+ \o_r \,\e \label{eq:addedpos}
\ee
and with velocity $v_{j+r}$. This defines $\bze^\e(t_r) = (\z_1^\e(t_r),\cdots,\z^\e_{j+r}(t_r)).$
 
The characteristic function in the collision operator \eqref{eq:Cedec}--\eqref{eq:specBeBBGKY} 
(or the characteristic function in \eqref{eq:defBe}),
is a constraint on $\o_r$ ensuring that two hard spheres cannot be at distance smaller than $\e$.

\begin{figure}[htbp] 
\centering
\includegraphics[width=6in]{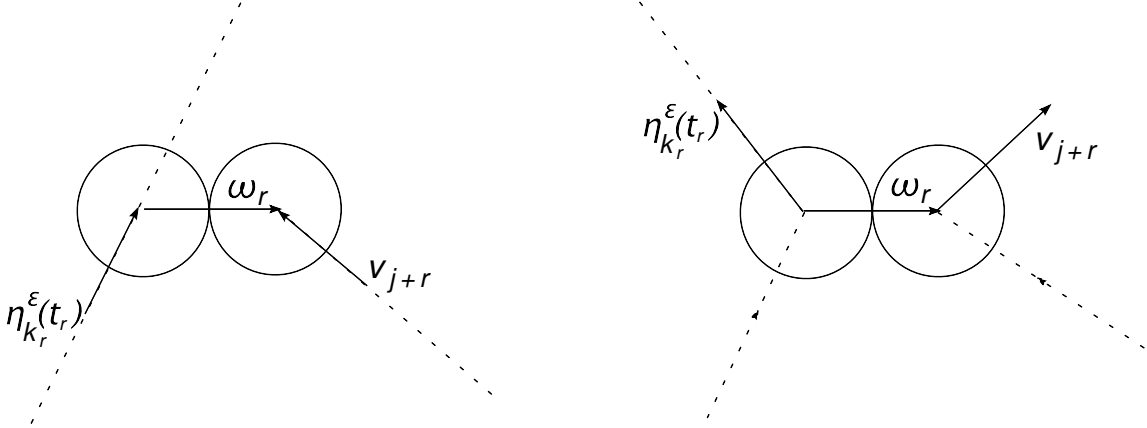} 
\caption{At time $t_r,$ particle $j+r$ is {\em created} by particle $k_r,$ either in incoming 
($\o_r \cdot (v_{j+r}-\eta^\e_{k_r} (t_r)) \leq 0$) or in outgoing
($\o_r \cdot (v_{j+r}-\eta^\e_{k_r} (t_r)) \geq 0$) collision configuration. 
Particle $k_r$ is called the {\em progenitor} of particle $j+r.$}
\label{fig:creations}
\end{figure}

Next, the evolution
in $(t_{r+1},t_r)$ is contructed applying to this configuration the dynamics $\TT^\e_{j+r}$ (with negative times).
We have two cases. If $\o_r \cdot (v_{j+r}-\eta^\e_{k_r} (t_r)) \leq 0$, then the velocities 
are incoming and no scattering occurs, namely for times  $s< t_r$ the pair of particles moves backwards freely with velocities 
$\eta^\e_{k_r}(t_r)$ and $v_{j+r}$. If  instead $\o_r \cdot (v_{j+r}-\eta^\e_{k_r}(t_r)) \geq 0$, the 
pair is post--collisional. Then the presence of the interaction in the flow $\TT^\e_{j+r}$ forces the pair to perform a 
(backwards) instantaneous collision. The two situations are depicted in Fig. \ref{fig:creations}.

Proceeding inductively, the IBF is constructed for all times $s\in [0,t]$.

\subsubsection{Recollisions and factorization}

Observe that between two creation times $t_r, t_{r+1}$ any pair of particles among the existing $j+r$ 
can possibly interact. These interactions are called {\em recollisions}, because 
they may involve particles that have already interacted at some creation time (in the future) with another 
particle of the IBF. In our language, recollisions are the ``interactions different from creations''.

\medskip
Let us focus now in more detail on the structure of the backwards flow and on the mechanisms of correlation.

\medskip
We observe preliminarily that the graphical representation of a $n-$collision, $j-$particle
tree $\G(j,n)= (k_1,\cdots,k_n)$ consists of $j$ connected components. Each of these components 
is associated to a root line $i \in \{1,2,\cdots,j\}$ and collects
$n_i$ nodes $i_1,i_2,\cdots,i_{n_i}$. In particular, we have the following map:
\bea
&& \G(j,n) \longrightarrow \bG_j = \G_1,\cdots,\G_j\;, \label{eq:mapTsT}\\
&& \G_i = (k^i_1,\cdots,k^i_{n_i})\;,\ \ \ \ \ k^i_h = k_{i_h}\;.\nn
\eea
\nomenclature[Gi]{$\G_i$}{Tree generated by particle $i$}%
{\em In the sequel we will call simply {\bf tree} (generated by $i$) the collection of integers $\G_i$.}
In the example of Figure \ref{fig:treedef} one has $\G_1 = (1,1,3), \G_2 =(2,2)$.

Note that the map \eqref{eq:mapTsT} is not invertible, since the collection $\bG_j$ does {\em not}
specify the ordering of particles belonging to different trees. A one--to--one correspondence is instead
the following:
\be 
n, \ \G(j,n),\ \bt_n \longleftrightarrow n_1,\G_1,\bt^{1}_{n_1},\cdots,n_j,\G_j,\bt^{j}_{n_j}\;,
\label{eq:corrTsT}
\ee
where $$\bt^{i}_{n_i} = t^{i}_1,\cdots, t^{i}_{n_i}\;,\ \ \ \ \ t^i_h = t_{i_h}\;.$$
Clearly $n=\sum_i  n_i$.

For a given sequence of trees $\bG_j$, there are several $j-$particle trees $\G(j,n)$ having 
$\bG_j$ as image of the map \eqref{eq:mapTsT}. However summing the time--ordered product over such trees
$\G(j,n)$ is equivalent to a free time integration leaving only the partial ordering dictated by the sequence 
$\bG_j$. Namely it holds:
\bea
\sum_{\G(j,n)}\int  \mathbbm{1}_{\{t>t_1>t_2 \dots >t_n>0\}} dt_1\dots dt_n \, F
= \prod_{i=1}^j \sum_{\G_i} \int \mathbbm{1}_{\{t>t^i_1>t^i_2 \dots >t^i_{n_i}>0\}} dt^i_1\dots dt^i_{n_i} \, F
\nn\\ \label{eq:sfpoftr}
\eea
where $F=F(\bG_j,\bt_n)$.

Applying this property to the expansion \eqref{fjN-grad}, we obtain the following factorization result:
\bea
&& f_j^{\e}(\bz_j, t)=\sum_{n \geq 0} \sum_{\G(j,n)} \int d \L (\bt_n , \bo_n , \bv_{j,n}) \prod B^\e f^{\e}_{0,j+n}(\bze^\e(0)) \nn\\
&& = \prod_{i=1}^j\left( \sum_{n_i, \G_i} \int d\L (\bt^i_{n_i} , \bo^i_{n_i} , \bv^i_{1,n_i}) \right)
\prod B^\e  f^{\e}_{0,j+n}(\bze^\e(0))\;.
\label{eq:reds}
\eea
In \eqref {eq:reds}, the triples in $(\bt^i_{n_i} , \bo^i_{n_i} , \bv^i_{1,n_i})$ are associated to the nodes 
of the tree $\G_i$, while the IBF (hence the integrand $\prod B^\e  f^{\e}_{0,j+n}$) 
has to be computed with the rules specified in the previous subsection.

\medskip
With the notations introduced above (see in particular Figure \ref{fig:treedef}), 
it should be clear that each particle of the IBF ``belongs'' to exactly one tree $\G_i$.
Therefore we may distinguish two types of recollisions. The {\em internal recollisions}, occurring among particles 
of the same tree and the {\em external recollisions}, occurring between particles  belonging to different trees. 
Because of the external recollisions, we say that different trees are {\em correlated}, in the sense that their interacting 
backwards flows are not pairwise independent. 

\medskip
\ni {\bf Remark.} \label{rem:PF}
Formula \eqref {eq:reds} shows a {\em partial factorization}: a full factorization is prevented by  the correlations 
of the initial datum $f^{\e}_{0,j+n}$, the forbidden (external) overlaps of created particles at the creation times 
(written in $B^\e$) and, more importantly, the external recollisions in the IBF. 
If we simply ignore these effects and replace $f^{\e}_{0,j+n}$ with a tensor product, then \eqref{eq:reds}
becomes a completely factorized expression. 

\medskip
From now on, in handling formula \eqref {eq:reds} and similar ones established in the sequel, we 
will use intensively the notations
\be
\G_i = \mbox{tree generated by particle $i \in \{1,\cdots,j\}$} \;, \label{eq:notGi}
\ee
which is a ($n_i$--collision) tree with associated configuration $z_i$ at time $t$,
\be
(\bt^i_{n_i} , \bo^i_{n_i} , \bv^i_{1,n_i}) = \mbox{collection of triples associated to the nodes of $\G_i$}\;,
\ee
and
\be
S(i) = \mbox{set of particles associated to $\G_i$}\;. \label{eq:notationS(i)}
\ee
\nomenclature[Si]{$S(i)$}{Set of particles belonging to the tree $\G_i$}%
Moreover,
\be
S(K) = \cup_{i}S(i)\;, \label{eq:notSK}
\ee
where $K$ is any subset of $\{1,\cdots,j\}$.

\subsection{Factorized expansions} \label{sec:hs2B}
\setcounter{equation}{0}    
\def\theequation{3.5.\arabic{equation}}

\subsubsection{The {\em uncorrelated} IBF} \label{sec:tbzeS}

Using the symmetry of the state, we could change notation in the integrals \eqref{eq:reds}, by substituting
$ \bze^\e(0) $ with $\left( \bze^\e_{S(1)}(0), \cdots, \bze^\e_{S(j)}(0) \right)$, where $\bze^{\e}_{S(i)} =
\{\z^{\e}_k\;;\ k \in S(i)\}$. As already pointed out, however, configurations $\bze^\e_{S(i)}$ with 
different values of $i$ are correlated through the external recollisions.

Let us introduce a different notion of backwards flow, in which the correlations among
different groups $\bze^\e_{S(i)}$ are ignored. Suppose that we want the tree $\G_i$ to 
be ``uncorrelated''. Then, for all $k \in S(i)$, we substitute
the IBF $\z^{\e}_k(s)$ with the evolution
\be
\tilde \z^\e_{k}(s)\;, \label{eq:tbzeS}
\ee
to be constructed as $\z^{\e}_k(s)$ with the additional prescription that {\em its external recollisions 
are ignored} (see Figure \ref{fig:PSflows} at page \pageref{fig:PSflows}).
\nomenclature[zeet]{$\tilde\bze^\e(\cdot)$}{Uncorrelated interacting backwards flow}%
The constraint excluding overlaps of created particles in $\G_i$ with particles of different trees
at the moment of creation, has to be also ignored. Notice that \eqref{eq:tbzeS} is a function of the only 
$z_i, \G_i,\bt^i_{n_i}, \bo^i_{n_i} , \bv^i_{1,n_i}$.

For instance if we want {\em all} the trees in the expansions \eqref{eq:reds} to be uncorrelated in the above
sense, we shall replace $ \bze^\e(s) \to \tilde \bze^\e(s)$ inside the formula and require that:

\noindent -- factors $B^\e$ associated to different trees become completely 
independent;

\noindent -- the initial data are evaluated in the time--zero configuration $\tilde\bze^\e(0)=
(\tilde\bze^\e_{S(1)}(0), \cdots, \tilde\bze^\e_{S(j)}(0))$ 
(with $\tilde \bze^{\e}_{S(i)} = \{\tilde \z^{\e}_k\;;\ k \in S(i)\}$), that is a collection of $j$ independent objects.
The resulting quantity
would differ from the tensorized product 
$f_1^{\e}(t)^{\otimes j}(\bz_j)$ {\em only} because of the correlations assumed for the initial r.c.f. $f^{\e}_{0,j+n}$.

\subsubsection{The Enskog backwards flow (EBF)} \label{sec:EBF}

Even after replacing the IBF with the uncorrelated flow in \eqref{eq:reds},
there is still a nontrivial correlation among particles of the {\em same} tree. This is due to the 
internal recollisions in $\tilde\bze^\e$, among particles of each set $S(i)$.
To get rid of them, one has to introduce the completely uncorrelated backwards flow
\be
\z^{\EE}_{k}(s) \label{eq:tbzeS0}
\ee
(where $\EE$ stands for ``Enskog'')
for all $k \in S(i)$, to be constructed as $\tilde \z^{\e}_{k}(s)$ with the additional prescription that 
{\em its internal recollisions are ignored}, together with the constraint excluding overlaps of created 
particles at the moment of creation. 
\nomenclature[zeeE]{$\bze^\EE(\cdot)$}{Enskog backwards flow}%

The evolution $\bze^{\EE}$ will be called {\em Enskog backwards flow} (EBF). In this flow, particles
are created at distance $\e$ (from their progenitor), 
but they may reach a distance smaller than $\e$ during the evolution (in particular,
its time--zero state $\bze^{\EE}(0)$ varies in $\RRR^{6(j+\sum_i n_i)}$).

Alternatively, we may say that the EBF is constructed exactly as the IBF, except for the following differences:

\ni - the interacting dynamics $\TT^\e$ is replaced by the simple free dynamics;

\ni - there is no constraint on $\o_r$.

The name ``Enskog'' is due to the obvious connection with the Enskog equation. 
Indeed, Eq. \eqref{eq:fjexpE}--\eqref{eq:Ecollop}
can be written explicitly
\be
g^{\e}_j (\bz_j,t)= \sum_{n=0}^\infty\sum _{\G (j,n)} \T^\EE (\bz_j,t)
\label{eq:treeexpErev}
\ee
where $\T^\EE(\bz_j,t)$ is the {\em value of the tree} $\G(j,n)$ {\em with configuration } $\bz_j$
{\em at time} $t$, {\em for the Enskog flow}
\be
\T^\EE (\bz_j,t)= 
\int d \L (\bt_n , \bo_n , \bv_{j,n}) \prod B^{\EE} g^{\e}_{0,j+n}(\bze^{\EE}(0))\;,
\label{eq:treeexpE}
\ee
with $\prod B^{\EE}  = \prod_{i=1}^n B(\o_i; v_{j+i}-\eta^{\EE}_{k_i} (t_i))$,
\be
B(\o_i; v_{j+i}-\eta^{\EE}_{k_i} (t_i))=\o_i \cdot (v_{j+i}-  \eta^{\EE}_{k_i} (t_i))\;.
\label{eq:defBboltz}
\ee
\nomenclature[B]{$B$}{Boltzmann collision operator}%

Note that  the EBF allows a complete factorization, whenever the initial datum does. Namely
if $g^{\e}_{0,j}=(f_0)^{\otimes j}$ for all $j$,  
the expansion above gives immediately
\be
g^{\e}_j (\bz_j,t) = \prod_{i=1}^j\left( \sum_{n_i, \G_i} \int d\L (\bt^i_{n_i} , \bo^i_{n_i} , \bv^i_{1,n_i}) 
\prod B^{\EE} g^{\e}_{0,1+n_i}(\bze^{\EE}_{S(i)}(0)) \right) \;,
\label{eq:fEfact}
\ee
where $\bze^{\EE}_{S(i)} = \{\z^{\EE}_k\;;\ k \in S(i)\}$.

\subsubsection{The Boltzmann backwards flow (BBF)}

The previous discussion can be repeated, with minor changes, for the case of the Boltzmann series 
\eqref{eq:fjexp}. The interacting backwards flow is now substituted by the {\em Boltzmann backwards flow} (BBF)
$\bze(s)\;.$
\nomenclature[zeeB]{$\bze^\BB(\cdot)$}{Boltzmann backwards flow}%
For it, we use the same notations of  \eqref{eq:IBFnot}--\eqref{eq:IBFnot'} with the superscript $\e$ 
omitted. 

Since the collision operator \eqref{eq:defBco} is splitted into a gain and a loss term,
then, together with the sum over $\G(j,n)$, we have an additional $\sum_{\bs_n}$ with $\bs_n=(\s_1, \cdots, \s_n)$, $\s_i=\pm$.
To have a compact expression, we change variables $\o \to - \o$ inside the positive part of the collision operators. 
As a result, in each term of the expansion, $\s_i$ fixes the sign of the product $\o_i \cdot (v_{j+i}-\eta_{k_i} (t_i))$
(where the relative velocity at the moment of creation appears).
Note that the same procedure has to be followed when deriving \eqref{eq:fjexpE}--\eqref{eq:Ecollop} from \eqref{eq:tfjtdef}.

The BBF turns out to be defined exactly as the IBF, 
except for the following differences:

\ni - the interacting dynamics $\TT^\e$ is replaced by the simple free dynamics;

\ni - in the right hand side of \eqref{eq:addedpos} the second term is missing, i.e. the created particle appears at the same
position of its progenitor: $\xi_{j+r}(t_r)= \xi_{k_r}(t_r)$;

\ni - there is no constraint on $\o_r$ other than the one implied by the value of $\s_r;$

\ni - if $\s_r=+,$ to determine the configuration of particles in $(t_{r+1},t_r),$ {\em before} applying free evolution we have to
change velocities according to $(\eta_{k_r}(t_r^+),v_{j+r}) \to (\eta_{k_r}(t_r^-),\eta_{j+r}(t_r^-)),$ where $\to$ denotes the
elastic scattering rule with scattering vector $\o_r$. 
We recall that, in our conventions, $\eta_{k_r}(t_r) \equiv \eta_{k_r}(t_r^+)$ (which indicates the
limit from the future, while $\eta_{k_r}(t_r^-)$ indicates the limit from the past).

Eq. \eqref{eq:fjexp} can then be rewritten:
\be\label{fj}
f_j (\bz_j,t)= \sum_{n=0}^\infty\sum _{\G (j,n)} \T (\bz_j,t)\;,
\ee
where $\T(\bz_j,t)$ is the {\em value of the tree} $\G(j,n)$ {\em with configuration } $\bz_j$ {\em at time} $t$, {\em for
the Boltzmann flow}, 
\be
\T (\bz_j,t)= \int d \L (\bt_n , \bo_n , \bv_{j,n})\prod B\, f_{0,j+n}(\bze(0))\;, \label{eq:TBf}
\ee
with $\prod B = \prod_{i=1}^n B(\o_i; v_{j+i} - \eta_{k_i} (t_i))$ and
\bea
&& B(\o_i; v_{j+i}-\eta_{k_i} (t_i))=\sum_{\s_i} \s_i |\o_i \cdot (v_{j+i}-  \eta_{k_i} (t_i))|
\mathbbm{1}_{\{\s_i\o_i \cdot (v_{j+i}-\eta_{k_i} (t_i)) \geq 0\}} \nn\\
&& \ \ \ \ \ \ \ \ \ \ \ \ \ \ \ \ \ \ \ \ \ \ \ \ \ \ = \o_i \cdot (v_{j+i}-  \eta_{k_i} (t_i))\;.
\eea

Note that, in the final formula, the difference between gain and loss collision operators is hidden inside
the {\em rule} for the construction of the BBF, which depends, as explained above, on the sign of each 
product $\o_i \cdot (v_{j+i}-  \eta_{k_i} (t_i))$. (A similar consideration can be made for the case of Enskog.)

Note also that 
\be
\bet = \bet^\EE \label{eq:betBbetEE}
\ee
and 
\be
B = B^\EE\;, \label{eq:equalityB}
\ee
the only difference between the BBF and the EBF being due to the position in space of created particles.

As before, \eqref{fj} can be immediately written in the form
\be
f_j (\bz_j,t) =  \prod_{i=1}^j \left(\sum_{n_i, \G_i}\int d\L (\bt^i_{n_i} , \bo^i_{n_i} , \bv^i_{1,n_i}) \prod B \right)
f_{0,j+n}(\bze (0)) \;, \label{eq:fBfact}
\ee
which shows a complete factorization in the case of factorized initial data.

\subsection{Summary} \label{sec:hs2Bsum}
\setcounter{equation}{0}    
\def\theequation{3.6.\arabic{equation}}

We have introduced: 

\medskip

(i) the tree expansion for the evolution of the hard sphere system, solution to the BBGKY hierarchy of equations:
see Eq.s \eqref{fjN-grad}--\eqref{eq:Teszt} (equivalently, \eqref{eq:reds});

(ii) an ``uncorrelated'' tree expansion described in Section \ref{sec:tbzeS};

(iii) the tree expansion for the Enskog equation, solution to the Enskog hierarchy: Eq.~\eqref{eq:treeexpErev}--\eqref{eq:treeexpE};

(iv) the tree expansion for the Boltzman equation, solution to the Boltzmann hierarchy: Eq.s \eqref{fj}--\eqref{eq:TBf};

\medskip

\ni and, correspondingly:

\medskip

(i') the interacting backwards flow, $\bze^\e(s)$, expressing the evolution of the rescaled correlation functions
of the hard sphere system; 

(ii') the partially uncorrelated flow $\tilde\bze^\e(s)$, obtained from the IBF by ignoring the external recollisions;

(iii') the Enskog backwards flow $\bze^{\EE}(s)$, obtained from the IBF by ignoring all the recollisions;

(iv') the Boltzmann backwards flow, $\bze(s)$, describing the evolution of functions obeying the Boltzmann hierarchy,
and obtained from the EBF by making the particles interact at distance zero instead than $\e$.

\medskip

 \ni See Figure \ref{fig:PSflows} below.

The flows in (iii') and (iv') will be used to prove convergence of the hard sphere system to the Enskog and the Boltzmann
equation, while (ii') will be enough for the proof of propagation of chaos.

\begin{figure}[htbp] 
\centering
\includegraphics[width=4in]{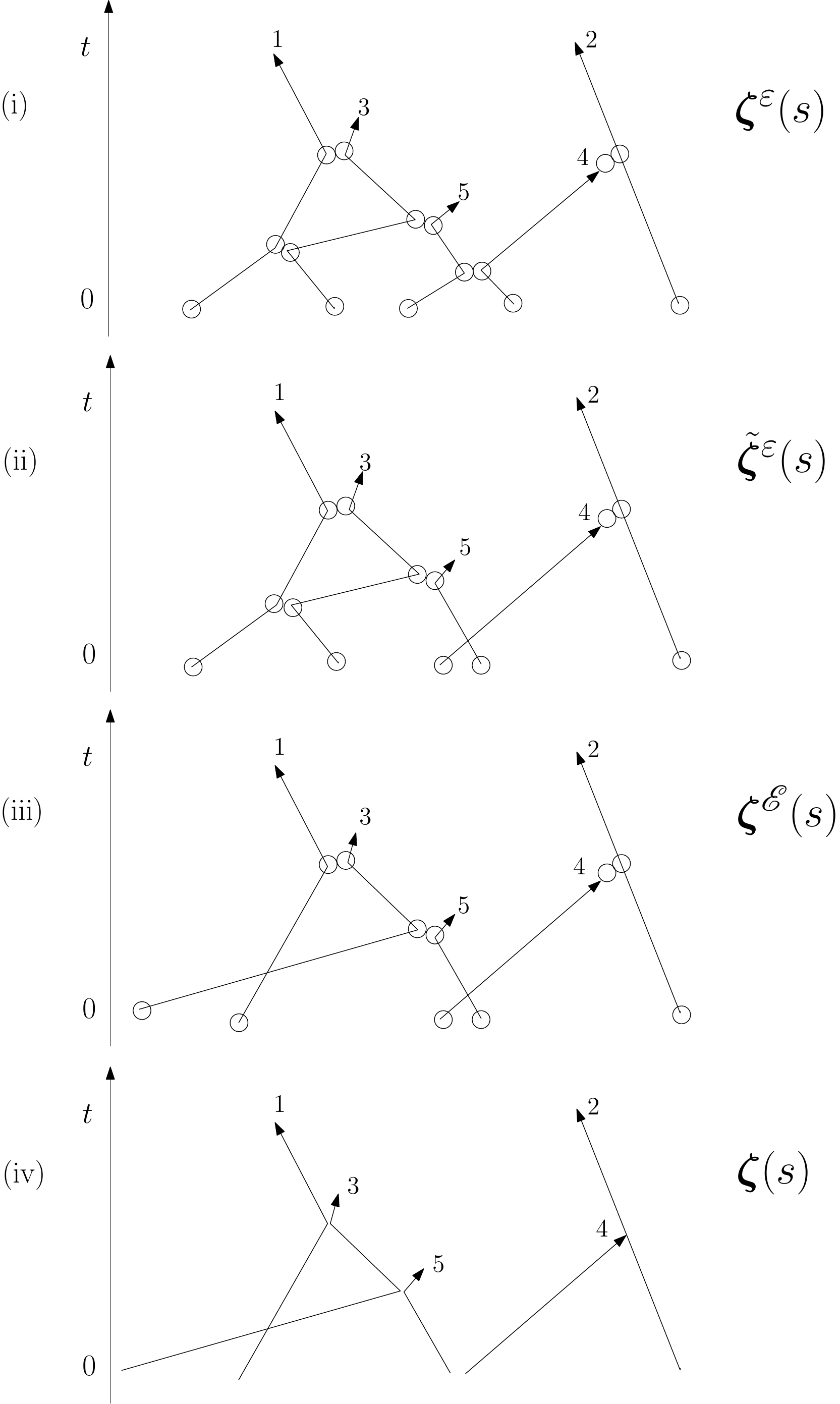} 
\caption{An example of trajectories drawn by the four types of backwards flow introduced, 
in the case of the two--particle tree $\G(2,3)=(1, 2, 3)$, for fixed values of the
variables (here $\bz_2, \bt_3, \bo_3,\bv_{2,3}$). The hard spheres of diameter $\e$ are pictured
at the creation times and at the recollision times.}
\label{fig:PSflows}
\end{figure}


\section {Proof} \label{proof}

\subsection{Basic estimates} \label{sec:proofpre}
\setcounter{equation}{0}    
\def\theequation{4.1.\arabic{equation}}

In this section we briefly remind the short time basic estimate given by Lanford, in a form well suited for our purposes. 

The following well known property of the BBGKY, Enskog and Boltzmann
series expansions introduced in Section \ref{sec:hs}, is preliminary to our work.

\begin{prop}[short time estimates] \label{prop:STE}
If the initial data $f^{\e}_{0,j}$ and $f_{0,j}$ are bounded as in \eqref{eq:assbound}, 
then the absolute convergence of the expansions \eqref{fjN-grad}, \eqref{eq:treeexpErev}
(uniformly in $\e$) and \eqref{fj} holds, for any $t < \bar t = \bar t(z,\b)$.
\end{prop}

The proof,
reported here for completeness, reduces immediately to the bound given by the following lemma,
which is stated in a somewhat general form.

\begin{lem} \label{lem:STE}
Let $a=1,2$. There exist constants $\bar t,\bar C>0$ (depending on $z,\b$)
such that, for any $t<\bar t$, the following estimate holds:
\bea
\sum_{n \geq 0} z^{j+n}\sum _{\G (j,n)} \int d \L (\bt_n , \bo_n , \bv_{j,n}) \left(\prod
|B^\e|\right)^a \ e^{-(\b/2)\sum_{i\in S(J)}(\eta^{\e}_i(0))^2}
\leq \bar C^j e^{-(\b/4)\sum_{i\in J}v_i^2}\;. \nn\\
\label{eq:STE} 
\eea
The same result holds when $B^\e, \bze^\e$ are replaced by $B^{\EE},\bze^{\EE}$ 
(Enskog flow) or $B,\bze$ (Boltzmann flow).
\end{lem}
We remind that $J = \{1,\cdots,j\}$ and, by the notation \eqref{eq:notSK}, $S(J) = \{1,2,\cdots,j+n\}$.

In the case $a=1$, this shows that the expansions of Proposition \ref{prop:STE} are also 
absolutely convergent in the norm $\esssup_{\bx_j}\int d\bv_j$.

\medskip
The case $a=2$ in the above lemma implies the following result, which will be used  in Appendix C
to simplify the expression of formulas in the recollision estimates. 
This procedure was already applied in \cite{PSS13}.
\begin{cor}  \label{cor:Btrick}
Let $F \leq 1$ be any positive measurable function of the variables $\bz_j$, $\bt_n,$ 
$\bo_n,$ $\bv_{j,n}$. Let $N>0$ and $\theta_1>0$.
There exists $\bar C'>0$ such that, for any $t<\bar t$, 
\bea
&& \int d\bv_j\sum_{n = 0}^Nz^{j+n} \sum _{\G (j,n)} \int d \L (\bt_n , \bo_n , \bv_{j,n}) \left(\prod
|B^\e|\right) \   e^{-(\b/2)\sum_{i\in S(J)}(\eta^{\e}_i(0))^2}\  F  \nn\\
&& \leq (\bar C')^j \e^{\theta_1 j} + \e^{-\theta_1 j}\sum_{n = 0}^N z^{j+n}
\sum _{\G (j,n)} \int d\bv_j d \L (\bt_n , \bo_n , \bv_{j,n})\nn\\
&& \ \ \ \ \ \ \ \ \ \ \ \cdot  e^{-(\b/2)\sum_{i\in S(J)}(\eta^{\e}_i(0))^2}\ \prod_{i=1}^n
\mathbbm{1}_{\{|\xi^\e_{j+i}(t_i)-\xi^\e_{k}(t_i)| > \e\ 
\forall k\neq k_i\}} \ F\;. \label{eq:Btrick}
\eea
The result holds also when $B^\e, \bze^\e$ are replaced by $B^{\EE},\bze^{\EE}$ 
(Enskog flow) or $B,\bze$ (Boltzmann flow).
\end{cor}
To deduce the corollary, it is enough to observe that the integral on the l.h.s., when restricted to 
the set such that $\prod |B^\e| > \e^{-\theta_1 j}$, is bounded by $\e^{\theta_1 j}$ times the integral
with respect to $d\bv_j$ of the left hand side in \eqref{eq:STE} with $a=2$. Applying the lemma, 
we obtain the result by taking $\bar C' = \bar C (4\p/\b)^{3/2}$.

\medskip
\ni {\bf Proof of Lemma \ref{lem:STE}.} 
The conservation of energy at collisions implies 
\be
\sum_{i\in S(J)}(\eta^{\e}_i(0))^2 = \sum_{i=1}^{j+n}v_i^2\;. \label{eq:consentr}
\ee
In particular $\sum_{k_i=1}^{j+i-1}\left(\eta_{k_i}^\e(t_i)\right)^2\leq\sum_{i=1}^{j+n}v_i^2$.
Then, using the expression of $B^\e$ (see \eqref{eq:defBe}) and \eqref{eq:specialsumk},
\be
\sum_{\G(j,n)}\left(\prod |B^\e|\right)^a \leq a^n \prod_{i=1}^n \Big[(j+n)|v_{j+i}|^a+
(j+n)^{\frac{2-a}{2}}\Big(\sum_{l=1}^{j+n} v_l^2\Big)^{\frac{a}{2}}\Big]\;.
\ee
Moreover,
\bea
&& \Big(\sum_{l=1}^{j+n} v_l^2\Big)^{\frac{1}{2}} e^{-\frac {\beta }{4n}\sum_{i=1}^{j+n} v_i^2  } \leq \frac{1}{\sqrt 2}\;,\nn\\
&& \Big(\sum_{l=1}^{j+n} v_l^2\Big) e^{-\frac {\beta }{4n}\sum_{i=1}^{j+n} v_i^2  } \leq \frac{4n}{e\b}\;.
\eea

Replacing these estimates in the l.h.s. of \eqref{eq:STE}, it follows that we can bound it by
\be
 e^{-(\b/4)\sum_{i\in J}v_i^2}\sum_{n\geq0}2^nz^{n+j} \int d\L \ \prod_{i=1}^n\left((j+n)|v_{j+i}|^a e^{-\frac{\b}{4}v_{j+i}^2}
+ \left( \sqrt{\frac{j+n}{2}} + \frac{4n}{e\b}\right)e^{-\frac{\b}{4}v_{j+i}^2}\right)\;. \label{eq:proofSTE}
\ee
The integral on the velocities factorizes so that
\be
\mbox{\eqref{eq:proofSTE}} \leq e^{-(\b/4)\sum_{i\in J}v_i^2} 
\sum_{n\geq0} C(z,\b)^{j+n}\frac{t^n}{n!} (j+n)^n  \label{eq:estBproof1}
\ee
for a suitable constant $C(z,\b)>0$
(which can be explicitly computed in terms of gaussian integrals). Since 
\be
\frac{(j+n)^n}{n!} \leq \frac{(j+n)^{j+n}}{(j+n)!} \leq e^{j+n}\;, \label{eq:nnnfact}
\ee
we have that \eqref{eq:estBproof1} is bounded by a geometric
series. Hence choosing 
\be
\label {shtime}
\bar t < \frac{1}{C(z,\b)e}\;,
\ee
we obtain \eqref{eq:STE}.

The cases of the Enskog and of the Boltzmann flow are treated in the same way. \qed

\medskip
\ni {\bf Remark.} Lemma \ref{lem:STE} and Proposition \ref{prop:STE} imply immediately that 
the correlation errors $E_K,E^{\EE}_K,E^{\BB}_K$ introduced in Theorem \ref{thm:MR} are bounded 
by $(const.)^k$, uniformly in $\e$, for all $t<\bar t$. Note that this is also true in the regions
$\RRR^{6k} \setminus \MM_k$, as soon as the definition of the r.c.f. is extended there as $f^{\e}_J(\bz_J,t)=0$.

\subsection{Plan of the proof} \label{sec:plan}
\setcounter{equation}{0}    
\def\theequation{4.2.\arabic{equation}}

In this section we outline the main technical difficulties in proving our result and give some intuitive explanation of the 
strategies developed to overcome them. We concentrate on estimate \eqref{eq:EKthm} which is the 
main result of the paper. There are three main issues:

Step 1: Combinatorics;

Step 2: Ordering of recollisions;

Step 3: Estimate of the single recollision event.

\ni In Step 1 we construct a perturbative expression for the correlation error $E_K$
and control the number of terms. Steps 2, 3 deal with the estimate of such an expression. 
All these steps present peculiar difficulties and we shall discuss them separately.

\medskip
\ni {\bf General strategy.} 
We start from the explicit formula for the evolution of rescaled correlation functions $f^\e_j$, that is the tree expansion
described in Section \ref{sec:hs2}, see Eq. \eqref{eq:reds}. It is important to keep in mind
the structure of this formula. We have: (a) a sum over $j$ binary tree graphs; (b) an integral over characteristic 
flows of type $\bze^\e$ associated to the trees (see Figure \ref{fig:PSflows}-i above).

Our first purpose is to manipulate directly the formula and {\em reorganize} it into the cumulant type 
expansion \eqref{correrr}. 

Reconstructing formula \eqref{correrr} from the tree expansion means to {\em extract}, among the 
$j$ trees, different {\em subsets} of independent one--particle trees. 
``Independent'' here has a precise meaning, namely the 
{\em value} of the tree does {\em not} depend on particles external to the tree. 
Conversely, ``correlated'' trees are not independent trees. Now remind that (see Section \ref{sec:tbzeS}),
by working with formula \eqref{eq:reds} we are in a favourable position, since the sources of correlation become 
totally explicit. 

There are two different effects: the propagation in time of the initial 
correlations due to the hard sphere exclusion, and the dynamical correlations induced by the 
external recollisions. 
As an example, consider the two--particle function $f^\e_2$. An associated trajectory is Figure 
\ref{fig:PSflows}-i where the trees $\G_1, \G_2$ are correlated through the external recollision. 
To get independent trees we would need: (1) to replace $\bze^\e$ by $\tilde\bze^\e$, 
where the particles of different trees ignore and cross each other freely ({\em overlap});
(2) to replace the time--zero distribution $f_{0,5}^\e(z_1,\cdots,z_5)$ with $f^\e_{0,3}(z_1,z_3,z_5)
f^\e_{0,2}(z_2,z_4)$. Point (1) can be readily achieved by the following elementary addition/subtraction
procedure: for any function of the flow $F$ (here $F= F(\bz_2,\bt_3,\bo_3,\bv_{2,3})$), it holds that
(here $\int = \int d\L(\bt_3,\bo_3,\bv_{2,3})$)
\be
\int F(\bze^\e) = \int F\chi^{rec}_{1,2}(\bze^\e) + \int F(1-\chi^{rec}_{1,2})(\bze^\e)
=  \int F(\tilde\bze^\e) + \int [F\chi^{rec}_{1,2}(\bze^\e) - F\chi^{ov}_{1,2}(\tilde\bze^\e)]\;,
\nn
\ee
where $\chi^{rec}_{1,2}\, (\chi^{ov}_{1,2})$ is the indicator function of the recollision (overlap) condition
between the two trees.
The last integral, which will be $O(\e^{2\g})$ for some $\g>0$, is part of the correlation error $E_2$, and to
have the complete expression it suffices now to add the error term in point~(2).

\medskip
\ni {\bf Step 1.} 
The extension of this procedure to the case of $j>2$ particles leads to a combinatorial problem.
One could write $1 = \prod_{i<\ell}[\chi^{rec}_{i,\ell} + (1-\chi^{rec}_{i,\ell})]$ and proceed as above.
For any pair of recolliding trees we hope to gain a factor $O(\e^\gamma )$. But for $k$ recolliding 
trees one would get $\sim 2^{\frac {k(k-1)}{2}}$ terms, and therefore nothing better than 
$E_K \sim 2^{\frac {k(k-1)}{2}} \e^{\g k}$, which is a good estimate only for $k\leq O(-\log \e)$.
To reach $k\sim O(\e^{-\alpha})$, we use a simple graph expansion procedure (Lemma \ref{lem:CEoG}
below) allowing to replace the bad counting factor by $k!$.

In Section \ref{subsec:comb} we introduce this technique (in a sense reminiscent of the cluster 
expansion in equilibrium statistical mechanics) in an abstract form, and then apply it to both
the dynamical and the initial correlation.

Let us explain here the method in a few words. Suppose to have a set $J$ of trees. 
Some of them recollide externally, say those in $L_0\subset J$, while the other trees
$L = J \setminus L_0$ do not. We indicate these conditions with the indicator 
functions of the flow: $\chi^{rec}_{L_0}$, $\bar\chi^{rec}_{L,J}$. The latter function
makes obviously the trees in $L$ {\em not} independent. To make them independent we 
need to substitute $\bze^\e \to \tilde\bze^\e$: we produce an error
\be
F \bar\chi^{rec}_{L,J} = F - \sum_{\emptyset\neq L_1\subset L}
F \chi^{ov}_{L_1,L_0\cup L_1}\, \bar\chi^{rec}_{L \setminus L_1,J}
\nn
\ee
where everything is evaluated in $\tilde\bze^\e$,
and $\chi^{ov}_{L_1,L_0\cup L_1}$ means that {\em all}
the trees in $L_1$ are constrained to overlap with some other 
tree in $L_0 \cup L_1$. 
From the above errors, we hope to gain $O(\e^{\g |L_1|})$. The trees $L \setminus L_1$
are still correlated, but now we just iterate the formula. Then, it is easy to see that the total number of
terms with $q$ overlapping conditions grows as $q!$ (see Appendix B), i.e.
we are essentially writing
\be
1 = \sum_{L_0 \subset J} \chi^{rec}_{L_0}\,\bar\chi^{rec}_{L,J} \approx
\sum_{\substack{L_0 \subset J \\ÊQ\subset J\setminus L_0}} C^q q! \,\chi^{rec}_{L_0} \,\chi^{ov}_{Q,Q\cup L_0}
\label{eq:plan3}
\ee
for some constant $C>0$. The result is summarized in the picture. 

\begin{figure}[htbp] 
\centering
\includegraphics[width=3in]{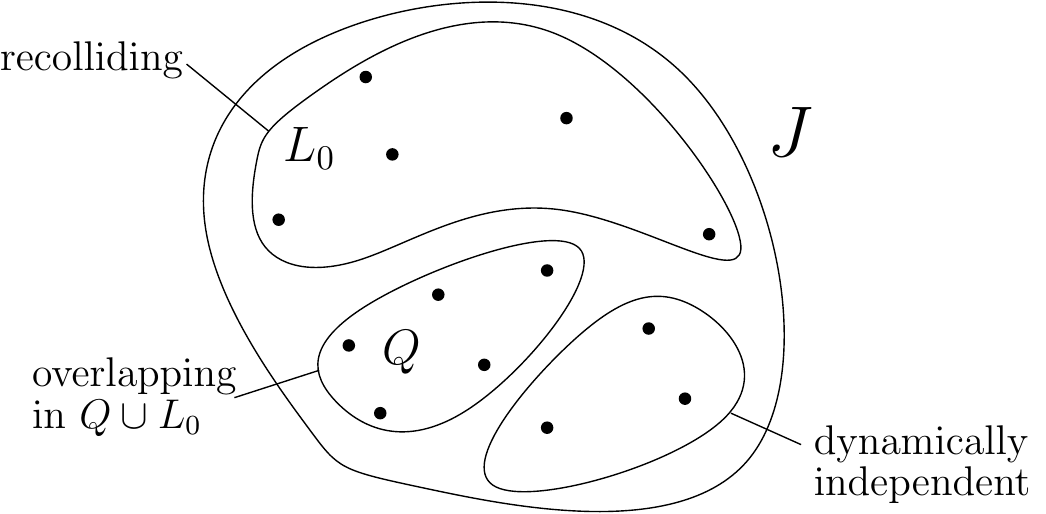} 
\label{fig:GEpic}
\end{figure}

The trees in $J \setminus (L_0\cup Q)  $ are {\em free} of dynamical conditions and evaluated in the uncorrelated 
flow $\tilde\bze^\e$, therefore they are correlated only through the initial condition. Such a correlation, which is due to the hard 
sphere exclusion at time zero, can be treated by the same method (with the ``recollision condition'' replaced 
by the ``overlap of spheres at time zero''). This makes the final expression for the correlation error $E_K$ slighty
more complicated than what can be guessed from \eqref{eq:plan3} (see \eqref{eq:errorEexp} below).

\medskip
\ni {\bf Step 1.1} \label{sec:Step11}
Once derived the expression for $E_K$ it becomes clear that we need to face an estimate
of events with many external recollisions and overlaps, like $\int \chi^{rec}_{L_0}\chi^{ov}_{Q,Q\cup L_0}(\cdots)$.
This will be the object of steps 2 and 3. A preliminary, crucial operation is the substitution of the integrand
$(\cdots)$ with a simplified expression. 
This is based on estimates known from previous papers. In Section \ref{sec:REF} we will show that
the Hypotheses on the initial data and the introduction of several cutoffs allow to replace the integrands in the
expansion for $E_K$ with a positive, bounded, compactly supported function of the {\em energy of trees}.

\medskip
\ni {\bf Step 2.} Let us call now $F = F(K)$ a nice function of the energy of the trees in $K$.
Consider, for simplicity ($Q = \emptyset$ above), the estimate of $K$ recolliding trees with $n$ created particles:
$\int \chi^{rec}_{K}\,F(K)$, from which we want to gain a factor $\e^{\g k}$.
This is a delicate point because, while it is understood how to estimate a single (internal or external) recollision 
(see \cite{GSRT12, PSS13} and Appendix D of the present paper) it is not obvious at all that, in case of $k$ 
recolliding trees -- implying at least $k/2$ external recollisions --, one can gain the $\e^{\g k} $ by means 
of a sufficiently large number of integrations. 

The $k/2$ recollision conditions are, of course, {\em not independent}. Therefore, we need to proceed carefully.
First, we {\em order} the recollisions in time. Secondly, for any possible sequence, we estimate the recollisions
one by one {\em iteratively}, following the time order.

To clarify this better, let us consider the following possible ordering. Going backwards in time,
the first two trees to collide are $\G_2$ and $\G_1$. Going further backwards, the first collision involving a {\em new}
tree is between $\G_3$ and $\G_2$, the second is between $\G_4$ and $\G_3$ and so on up to the
last collision of the chain, between $\G_k$ and $\G_{k-1}$. It is natural to say that the trees $\a$ and $\b$ are in 
a relation ``bullet-target'' when the {\em first} external recollision of $\a$ going backwards in time occurs with $\b$.
In the case considered, we have a sequence $(\a_1,\b_1), (\a_2,\b_2), \cdots,(\a_{k-1},\b_{k-1})$, where 
$(\a_i,\b_i) =(\G_{i+1},\G_i)$ and the first external recollision of $\a_i$ occurs in the future with respect to the first 
external recollision of $\a_{i+1}$.
For instance for $k=4$:

\begin{figure}[htbp] 
\centering
\includegraphics[width=4.5in]{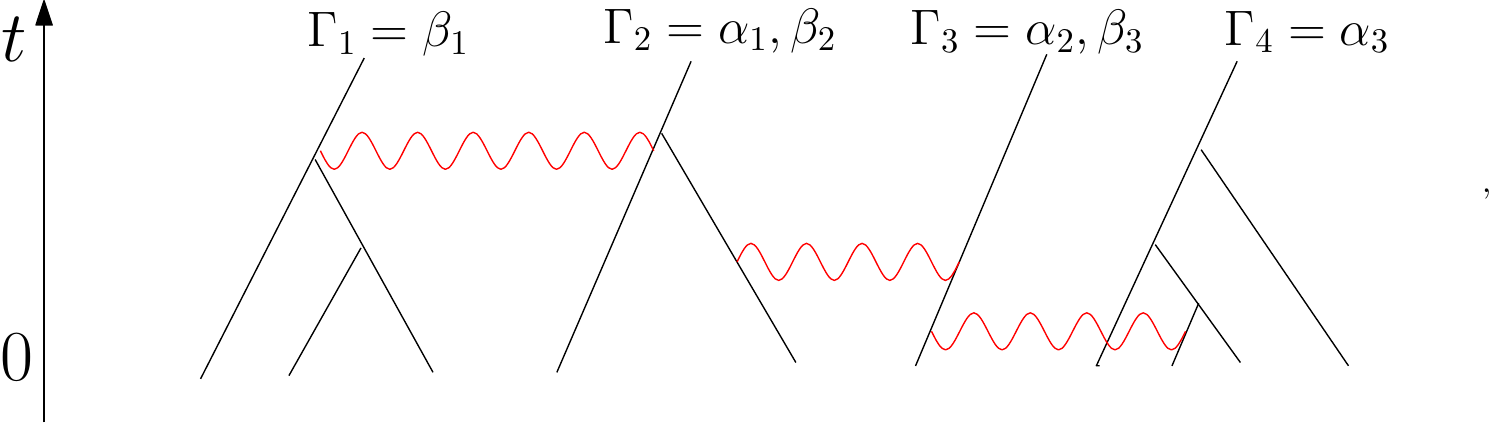} 
\end{figure}

\ni where we represented with red wavy lines the recollisions characterizing the bullets.

The velocities at time $t$ of the particles generating the trees are denoted by $v_1,v_2, \cdots, v_{k}$ respectively. 
Now {\em fix} all the integration variables but those relative to the last tree $\G_{k}$. Then we can integrate with respect 
to the latter variables (including $v_{k}$) with {\em no interference} with the other constraint, thanks to the
fact that the recollision $\G_{k} - \G_{k-1}$ is the last one in backward order. We gain a small factor $\e^{\g_1}$
by this integration (see Lemma \ref{lem:MR}) and we can iterate the procedure to obtain the desired result.

More generally, to handle with $\int \chi^{rec}_{K}\,F(K)$ we shall introduce, in Section \ref{sec:order}, a ``table of recollisions''
$\{(\a_1,\b_1), (\a_2,\b_2), \cdots,(\a_{\ell},\b_{\ell})\}$, characterizing one of all possible choices of the couples bullet--target 
and of the orderings of bullets in time. Whatever is the sequence, we know that $\ell \geq k/2$, and
we reduce to an ordered integral (see Eq. \eqref{eq:orderMRst} below)
which allows to estimate the constraints one by one (integrate out the $\a_{\ell}-$variables first,
then the $\a_{\ell-1}-$variables, and so on up to $\a_1$).

\medskip
\ni {\bf Step 3.} 
The price we pay for the approach in Step 2 is that
we need now to control the single overlap of a given bullet against one target tree whose particles 
perform a very complicated trajectory, possibly due to several recollisions.  
In fact, even if we ignore the internal recollisions of the bullet tree (as can be actually done producing a small
error, see appendix D), we are still left with the following kind of challenge
\begin{figure}[htbp] 
\centering
\includegraphics[width=5in]{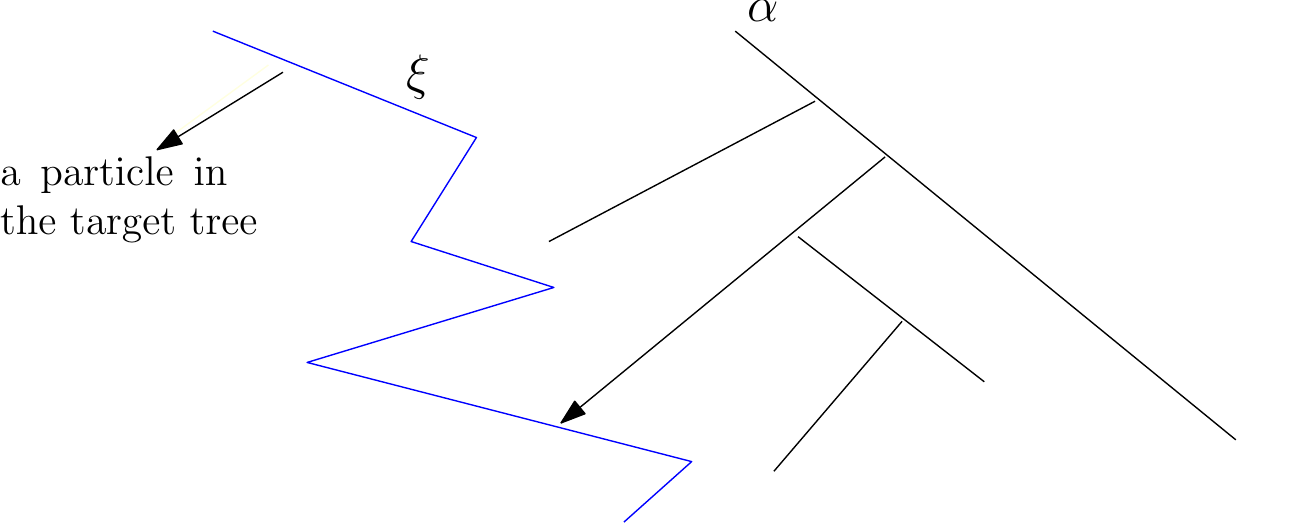} 
\end{figure}

\ni namely, the {\em geometry} of the constraint is more complex with respect
to the one in \cite{GSRT12, PSS13}. In the latter references, targets move freely, while here they may 
have an uncontrolled number of recollisions.

The key ingredients to deal with such an estimate are: (i) a parametrization of the constraint
in terms of the {\em relative velocities} of the bullet tree at the creation times; (ii) to exploit the variable 
$v_\a$, namely the velocity of the {\em root} of the bullet tree. Indeed, using (i), 
the constraint may be rewritten as $\mathbbm{1}(v_{\a}\in \T^{\e}_\xi)$, where $\T^{\e}_\xi$
is a {\em thin tube} around a curve of parametric equation given by $\xi$ 
(which is frozen) and by all the variables (but $v_{\a}$) spanning the flow of the bullet
(see Eq. \eqref{eq:eqTUBE} below).

Observe that no scattering transform is required in this procedure. On the other hand, 
it is crucial to integrate in the velocities $\int d\bv_k$ as stated in the main theorem.
For this reason, we cannot treat the internal recollisions with the same method.

\subsection{Step 1: Combinatorics } \label{subsec:comb}
\setcounter{equation}{0}    
\def\theequation{4.3.\arabic{equation}}

In this section we develop the graph expansion technique, Lemma \ref{lem:CEoG} below, which will serve
to reconstruct the representation \eqref{eq:EKthmREP} with an explicit expression for $E_K(t)$.
We find convenient to discuss this method of expansion in an abstract formulation first, since it 
will be used twice in the sequel, that is, it will be applied to the BBGKY series
\eqref{eq:reds} (Section \ref{sec:proof1}) and to the initial data (proof of Property 2 in Appendix A,
applied in Section \ref{sec:proof1'}). Next, the discussion in Sections \ref{sec:REF}--\ref{sec:proof5}
will reduce the proof of the theorem to an estimate of many--recollision events.

\medskip
Let us start with some classical definitions.

\begin{defi}\ 

(i) A graph over a set ${\cal I} =\{a_1 \dots a_n\}$ of vertices, is a collection of edges (links)
$\{ \ell_{i,j}\}_{i \neq j}$, where $\ell_{i,j}$ takes the values $1,0$ if  the vertices $a_i$ and $a_j$ are 
connected or not respectively. 

(ii) ${\cal G}$ is the family of all graphs over ${\cal I}$.

(iii) We introduce the following characteristic functions on ${\cal G}$: 
$$
\chi_{i,K}=1
$$
if and only if the vertex $a_i$ is connected with some vertex in $K\subset {\cal I}$;
$$
\bar \chi_{i,K}=1-\chi_{i,K}\;;
$$
and, for $H\subset {\cal I}$,
$$
\chi_{H,K}=\prod_{i \in H} \chi_{i,K}\;,
$$
$$
\bar \chi_{H,K}=\prod_{i \in H} \bar \chi_{i,K}\;.
$$
\end{defi}
\nomenclature[cHK]{$\chi_{H,K}, \bar \chi_{H,K}$}{Generic constraints on graphs}%
Observe that $\chi_{H,K}=1$ if and only if any vertex of $H$ is connected with some vertex
in $K$, and $\bar\chi _{H,K}=1$ if and only if any vertex in $H$ is not connected with any vertex in $K$.
Note also that a vertex cannot be self connected, i.e. $\chi_{i,i} = 0$ and $\bar\chi_{i,i} = 1$.

\begin{figure}[htbp] 
\centering
\includegraphics[width=1.5in]{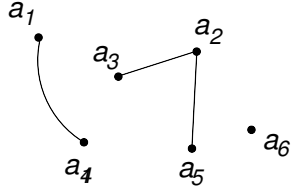} 
\caption{Graph $\ell_{1,4}=\ell_{2,3}=\ell_{2,5}=1$, and different $\ell_{i,j}=0$.}
\label{fig:graph}
\end{figure}

\begin{lem} \label{lem:CEoG}
Let $L\subset {\cal I}$ and $L_0={\cal I}\setminus L$. Then,
\be
\bar \chi _{L,L\cup L_0} =\sum_{Q\subset L} R(Q,L_0)
\label{Rabs}
\ee
where, for some pure constant $C > 0$,
\be
|R(Q, L_0)| \leq  C^q\, q!\, \chi_{Q, Q\cup L_0 }\;.
\ee
\end{lem}
We remind the notation $q = |Q|$ and the convention
$\bar\chi_{\emptyset,\cdot} = \chi_{\emptyset,\cdot} = 1$.

Note that each term of the expansion \eqref{Rabs} does not depend on $L \setminus Q$,
i.e. there is no condition on the vertices of this set (they are ``free'' vertices).

The proof of the Lemma is a simple algebraic computation and is deserved to Appendix~B.

\subsubsection{Expanding the dynamical constraints} \label{sec:proof1}

We start by rewriting the formula, introduced in Section \ref{sec:hs2}, yielding the reduced correlation 
functions at time $t$, namely
\be
f^\e_J(\bz_J,t )=\sum_{n \geq 0} \sum_{\G(j,n)} \int d \L \prod B^\e f^{\e}_{0,S(J)}\;,
\label{eq:usualexp}
\ee
where we abbreviate
\be
\int d\Lam = \int d\L (\bt_{n} , \bo_{n} , \bv_{j,n})\;,
\label{eq:factdL}
\ee
\be
f^\e_{0, S(J)} = f^\e_{0, |S(J)|}(\bze^\e _{S(J)}(0))\;,
\ee
$\G(j,n)$ denotes the set of $j-$particle trees with $n$ created particles,
$(\bt_{n} , \bo_{n} , \bv_{j,n})$  are the collections of node variables in the tree
and $S(J)$ denotes the set of indices of the particles created in the backwards 
flow $\bze^\e$ at time $0$. 
Clearly, $|S(J)|=j+n$.

\newpage
\ni {\bf 4.3.1.a\ \ \ Selection of the recolliding set} 
\bigskip

\ni Let us focus on the external recollisions. 

Consider the map \eqref{eq:mapTsT}.
We say that two trees, say $\Gamma_i$  and  $\Gamma_k$ (or, briefly, $i$ and $k$) 
recollide if there is a particle in $S(i)$ which collides with a particle in $S(k)$.

We introduce the characteristic function $\chi^{rec}_{i,K}$ defined by:
$$
\chi^{rec}_{i,K}=1
$$
if and only if the tree $i$ recollides with some tree in $K\subset J$. This depends of course
on the IBF  $\bze^\e$. Also, we introduce
$$
\chi^{rec}_{K}=\prod_{i \in K} \chi^{rec}_{i,K}\;,
$$
so that $\chi^{rec}_{K}=1$ if and only if all the trees in $K$ recollide with some other tree in $K$. 
\nomenclature[crK]{$\chi^{rec}_{L_0}$}{Recollision constraint}%
Finally, 
$$
\bar \chi^{rec}_{i,K}=1-\chi^{rec}_{i,K}
$$
and, for $H \subset J$,
$$
\bar \chi^{rec}_{H,K}=\prod_{i \in H} \bar \chi^{rec}_{i,K}\;.
$$
That is, $\bar \chi _{H,K} ^{rec}=1$ if and only if the trees in $H$  do not recollide with any tree in $K$.

With these definitions, one has
\be
1=\sum_{L_0\subset J}  \chi_{L_0}^{rec}\bar \chi _{J\setminus L_0,J} ^{rec}\;.
\label{part}
\ee
Observe that, if $L_0 \neq \emptyset$, $l_0 = |L_0|\geq 2$.

Inserting this partition of unity into \eqref{eq:usualexp}, we find
\be
f^\e_J(\bz_J,t )=\sum_{n \geq 0} \sum_{\G(j,n)}
\sum_{L_0 \subset J} \int d\Lam \ \prod B^\e\   \chi_{L_0}^{rec}\bar \chi _{J\setminus L_0,J} ^{rec} 
\, f^\e_{0, S(J)}\;.
\label{eq:usualexp1}
\ee

\bigskip
\ni {\bf 4.3.1.b\ \ \ Mixed backwards flow} 
\bigskip

\ni From the discussion in Section \ref{sec:plan} it should be clear that, to treat \eqref{eq:usualexp1}, we need
to introduce a {\em mixed} backwards flow, in which the particles of the trees in $L_0$
are evolved by taking into account all the recollisions among themselves, while the particles belonging to the trees
in $J \setminus L_0$ are evolved through the flow $\tilde \bze^\e$, i.e. by ignoring their external recollisions
(see Figure \ref{fig:PSflows} above).
We shall indicate such a flow
\be
(\bze^{ (L_0)}, \tilde \bze^{ (J \setminus L_0)})\;, \label{eq:mixIBF}
\ee
where $\bze^{(L_0)}$ is the flow of particles of the trees in $L_0$
and $\tilde\bze^{(J\setminus L_0)}$ is the flow of particles of the trees in $J \setminus L_0$.
Note that we are ignoring the dependence on $\e$ (now clear from the context) to unburden the
notation.

Let $i\in H \subset J \setminus L_0$ and $K \subset J$. We introduce the following characteristic functions:
$$
\chi^{ov}_{i,K}=1
$$
if and only if the tree $i$ {\em overlaps} with some tree in $K \subset J$ in the dynamics \eqref{eq:mixIBF}
(in the sense that some particle in $S(i)$ reaches a distance smaller than $\e$ from some other particle
in $S(K)$); moreover we set
$$
\chi^{ov}_{H,K}=\prod_{i \in H}\chi^{ov}_{i,K},
$$
$$
\bar \chi^{ov}_{i,K}=1-\chi^{ov}_{i,K},
$$
$$
\bar \chi^{ov}_{H,K}=\prod_{i \in H} \bar \chi^{ov}_{i,K}.
$$
\nomenclature[coHK]{$\chi^{ov}_{H,K}$}{Overlap constraint}%
That is, $\chi_{H,K}^{ov} =1$ if and only if all the trees in $H$ overlap with some tree in $K$ 
while  $\bar\chi_{H,K}^{ov}=1$ if and only if all the trees in $H$ do not overlap with any tree in $K$. 

Finally, we write (with a small abuse w.r.t. the notation \eqref{eq:defBe})
\be
\prod B^\e(\bze^{ (L_0)}, \tilde \bze^{ (J \setminus L_0)})
=\mathbbm{1}_{L_0}\,\tilde{\mathbbm{1}}_{J \setminus L_0}\,
\prod_{\substack{i=1 \\ (k_i\in L_0)}}^n \o_i \cdot (v_{j+i}-  \eta^\e_{k_i} (t_i))\,
\prod_{\substack{i=1 \\ (k_i\in J \setminus L_0)}}^n \o_i \cdot (v_{j+i}-  \tilde\eta^\e_{k_i} (t_i))\;,
\label{eq:defBemix}
\ee
where: 

-- $\mathbbm{1}_{L_0}$ is the characteristic function ensuring that the particles created in the trees 
$\bG_{L_0}$ do not overlap among each other at the moments of creation; 

-- $\tilde{\mathbbm{1}}_{K \setminus L_0}$ 
is the characteristic function ensuring that the particles created in the trees $\bG_{K \setminus L_0}$ do not overlap ``internally'' 
(i.e. with particles of the same tree) at the moments of creation. 

With these definitions, the following trivial identity holds:
\bea
\Big(\prod B^\e\,\chi_{L_0}^{rec}\bar \chi _{J\setminus L_0,J} ^{rec} 
\, f^\e_{0, S(J)}\Big) (\bze^\e )=
\Big(\prod B^\e\,\chi_{L_0}^{rec}\bar \chi _{J\setminus L_0,J} ^{ov} 
\, f^\e_{0, S(J)}\Big) (\bze^{ (L_0)}, \tilde \bze^{ (J \setminus L_0)}) \;,\nn\\
\label{iden}
\eea
which inserted into \eqref{eq:usualexp1} leads to
\be
f^\e_J(\bz_J,t )=\sum_{n \geq 0} \sum_{\G(j,n)}
\sum_{L_0 \subset J} \int d\Lam \ \prod B^\e\  
\chi_{L_0}^{rec}\bar \chi _{J\setminus L_0,J} ^{ov} \, f^\e_{0, S(J)}\;, \label{eq:papertp}
\ee
with the integrand function calculated via the flow \eqref{eq:mixIBF}.

\bigskip
\ni {\bf 4.3.1.c\ \ \ Application of Lemma \ref{lem:CEoG}} 
\bigskip

\ni Up to now, we just changed notation in \eqref{eq:usualexp1}.

Next we  apply Lemma \ref{lem:CEoG} to the case: $\II = J$, $\bar\chi= \bar\chi^{ov}$ and $L = J \setminus L_0$.
We obtain
\be
\bar \chi _{J\setminus L_0,J} ^{ov}= \bar \chi^{ov} _{L,L\cup L_0} =\sum_{Q\subset L} R^{ov}(Q,L_0),
\label{eq:AL45}
\ee
where
\be
|R^{ov}(Q, L_0)| \leq  C^q q! \chi^{ov}_{Q, Q\cup L_0 }
\label{eq:boundRov}
\ee
for some $C>0$. 

Inserting the above expansion in \eqref {eq:papertp}, we find
\be
f^\e_J(\bz_J,t )=\sum_{L_0 \subset J} \sum_{Q\subset J\setminus L_0} \sum_{\substack{n \geq 0 \\ \G(j,n)}}
\int d\Lam\ \prod B^\e \chi_{L_0 }^{rec} R^{ov}(Q,L_0) \, f^\e_{0, S(J)}\;.
\label{eq:intexpprid}
\ee
Each tree in $Q$ must obey an overlap--condition in order that $R^{ov}\neq 0$,
while the trees in $L_0$ must recollide among themselves.
In contrast, the trees $J\setminus (L_0\cup Q)$ are {\em free}, in the sense that there is 
no condition over them, so that they are not dynamically correlated (see the figure at page \pageref{fig:GEpic}).

Of course, the latter are still correlated through the initial data $f^\e_{0, S(J)}$.
Actually if the initial data were factorizing, then the algebraic part of our proof would finish here by extracting 
the leading term (namely, the trees which are free) which would reconstruct exactly the factorized part in \eqref{eq:EKthmREP}.

\subsubsection{Expanding the initial correlation: final expression for $E_K(t)$} \label{sec:proof1'}

To eliminate the additional correlation due to the initial datum, we expand it according to Property 2--Eq.
\eqref {eq:repid2intext}\footnote{It is now clear that we need an expansion in the {\em extended} 
phase space because the mixed backwards flow \eqref{eq:mixIBF} allows overlapping particles.}
with respect to the following tree--dependent partition of $S(J)$:
$$
\{ S_i \} _{ i \in J \setminus (L_0\cup Q) }\,,\, S( L_0\cup Q)
$$
where $S_i=S(i)$ is the set of indices of the particles in the tree $\G_i$. 
For this particular partition, Eq. \eqref {eq:repid2intext} yields
\bea
\label{indatum}
f^\e_{0,S(J)} =\sum_ {H \subset J \setminus (Q\cup L_0)}  \left(\prod_{i \in  H} f^\e_{0,S(i)} \right) 
\left(\bar E^0_{\{ S_i \} _{ i \in J \setminus H }, S( Q\cup L_0)}
+ \bar E^0_{\{ S_i \} _{ i \in J \setminus H}}
f^\e_{0, S(Q \cup L_0)}\right)\;,  \nn\\
\eea
which holds in the extended phase space $\RRR^{6|S(J)|}$. Notice that we are assuming now 
the convention \eqref{eq:WenEXC} for the initial data $f^\e_{0,S(J)}$, therefore 
we omit the characteristic functions $\bar \chi_{S(i)}^0$, $\bar\chi_{S(Q \cup L_0)}^0$.

Inserting this equation into \eqref {eq:intexpprid} we readily find the final result
\be
\label{expt}
 f_{J}^{\e}(t) =\sum_{ H \subset J} (f_{1}^{\e}(t))^{\otimes H} E_{J\setminus H}(t)\;,
 \ee
where the correlation error has the expression:
\bea
E_{K}(t) && =\sum_{\substack{L_0,Q\\Ê\subset \ K \\ \mbox{{\scriptsize disjoint}}}}
\ \sum_{n \geq 0} \sum_{\G(k,n)} \int d\L \prod B^\e \chi_{L_0 }^{rec}  \, R^{ov} (Q,L_0)
\nn\\&&\,\,\,\,\,\,\cdot\Big(\bar E^0_{\{ S(i) \} _{ i \in K \setminus (Q\cup L_0) }, S( Q\cup L_0)}
+ \bar E^0_{\{ S(i) \} _{ i \in K\setminus (Q\cup L_0)}}  f^\e_{0, S(Q \cup L_0)}\Big)\;,\nn\\
\label{eq:errorEexp}
\eea
with $\int d\Lam = \int d\L (\bt_{n} , \bo_{n} , \bv_{k,n})$ and the integrand calculated through 
the mixed flow $(\bze^{ (L_0)}, \tilde \bze^{ (K \setminus L_0)})$.

\subsubsection {Step 1.1: Reduction to energy functionals} \label{sec:REF}

Let us provide a first preliminary estimate of the above formula for $E_K$, by using
the available informations on $R^{ov}$ and the initial data.
As announced on page \pageref{sec:Step11}, our purpose here is to replace the integrand
in \eqref{eq:errorEexp} with a simplified expression depending on the energy of flows only.
To do this we will need to introduce some cutoff parameters.

We recall first the estimates at disposal: 

\ni -- from Lemma \ref{lem:CEoG}, one has the combinatorial
bound \eqref{eq:boundRov} for $R^{ov} (Q,L_0)$; 

\ni -- to control the initial data, we use Hypothesis
\ref{hyp:bound} and \eqref{boundEtildeUsed} to obtain, for $k+n< \e^{-\a_0}$,
\be
\Big|\bar E^0_{\{ S(i) \} _{ i \in K\setminus (Q\cup L_0)}} f^\e_{0, S(Q \cup L_0)}\Big|
\leq z^{k+n}\ e^{-(\b/2)\sum_{i \in S(K)} v_i^2}  \sum_{ B \subset K\setminus(Q \cup L_0)}
C^b b!  \chi^0 _{B,K\setminus (Q \cup L_0)}\e^{\g_0 (k-q-l_0-b)}\\ \label{eq:insBisfce}
\ee

\ni and similar estimate for the other term in \eqref{eq:errorEexp}. Again we use the conventions
$b=|B|$, $q=|Q|$, $l_0=|L_0|$ and remind that $\chi^0 _{B,K\setminus (Q \cup L_0)}=1$
if and only if all the trees in $B$ have a particle overlapping, at time zero, with some particle in a different
tree belonging to $K\setminus (Q \cup L_0)$.
        
Inserting this information into \eqref{eq:errorEexp} we can establish the following result. 
Set $$\HH_K := \sum_{i\in S(K)}v_i^2$$ (twice) the energy of the trees in $K$
\nomenclature[HK]{$\HH_K$}{Energy of the trees in $K$}%
and 
\be
F_{\theta_3} = F_{\theta_3}(K) := e^{-(\b/2) \HH_K} \,\mathbbm{1}_{\HH_K \leq \e^{-\theta_3}}\;.
\label{eq:Fth2}
\ee
\nomenclature[Ft2]{$F_{\theta_3}$}{A cutoffed function of the energy}%
\begin{lem} \label{prop3}
Let $\theta_1, \theta_2, \theta_3 >0$. There exist $\a, \g$ such that, 
for $k < \e^{-\a}$, $t<\bar t$ and $\e$ sufficiently small,
 \bea
\int d\bv_K |E_K(t)|  
&& \leq \frac{3\e^{\g k}}{4}
+ (zC)^k \,k! \, \e^{-\theta_1 k} \sum_{\substack{L_0,Q\\Ê\subset \ K \\ \mbox{{\scriptsize disjoint}}}}
\ \e^{\g_0(k-q-l_0)}\nn\\
&& \cdot
\sum_{n=0}^{\log\e^{-\theta_2 k}}  z^n
\sum_{\G(k,n)} \int d \L \,
d\bv_k \,\mathbbm{1}_{L_0}\,\tilde{\mathbbm{1}}_{K \setminus L_0}\,
\chi_{L_0 }^{rec} \chi^{ov}_{Q, K } \, 
F_{\theta_3}(K)\;.\nn\\
\label{eq:reordMCI}
\eea
\end{lem}
Formula \eqref{eq:reordMCI} summarizes the combinatorial part of our proof. 
For each pair $L_0, Q$, we gained a factor $\e^{\g_0 (k-l_0-q)} < \e^{\g (k-l_0-q)}$ if $\g < \g_0$. 
The remaining $\e^{\g (l_0+q)}$ necessary to achieve the main theorem must be obtained from the 
constraints $ \chi_{L_0 }^{rec}  \, \chi^{ov}_{Q,K}$ and this requires to control the ``many--recollision integral''
on which we focus in Section \ref{subsub:MRE}.

In \eqref{eq:reordMCI}, $C>0$ is a pure constant as in Lemma \ref{lem:CEoG} (but larger) not depending on any parameter
introduced. The characteristic functions $\mathbbm{1}_{L_0}, \tilde{\mathbbm{1}}_{K \setminus L_0}$ 
(ensuring well posedness of the mixed flow $(\bze^{ (L_0)}, \tilde \bze^{ (K \setminus L_0)})$ on which
$\chi_{L_0 }^{rec} \chi^{ov}_{Q, K }$ is evaluated) have been defined after \eqref{eq:defBemix}. 

Note that we have introduced several cutoffs: see the list below. 

The proof of Lemma \ref{prop3} is now rather straightforward and is postponed to Appendix~C.

\bigskip
\ni {\bf 4.3.3.a\ \ \ List of parameters} 
\bigskip

\ni 
We collect here, for the reader's convenience, a list of positive parameters entering in the proof of the
main theorem and the conditions they have to obey according to our estimates.

\ni -- $z$ and $\b$ fix the norm of the initial data (see Hypothesis \ref{hyp:bound}).

\ni -- $\a_0$ and $\g_0$ describe the correlation error estimates satisfied by the initial state
(Hypothesis~\ref{hyp:bound}).

\ni -- The truncation on the physical space appearing in Theorem \ref{thm:MR} is given by 
$\d=\e^{\theta}$ with 
\be
\theta < 1/4\;.
\label{eq:thetas14}
\ee
The reason of this restriction is technical and related to Step 3 of the proof (see Section \ref{sec:Tube}).

\ni -- $t^*$ is the limiting time of absolute convergence of the expansions involved in the proofs
and it will be determined by $z, \b$. We shall not optimize the value of $t^*$ (for more details, see Section 
\ref{subsec:time}).

\ni -- As already mentioned, the parameters $\a$ and $\g$ entering in the main theorem
and describing the correlation error at time $t$, will have to satisfy the following conditions:
\begin{equation}
\label{par1}
\left\{
\begin{aligned}  
&\a < \min \left[\,\a_0\,,\, (1/3) a(\g_0)\,,\, 1/4-(1/3) a(\g_0)-(2/3) \theta\,\right] \\
&\g < \min \left[\,a(\g_0)\,,\, 3/4 - a(\g_0)-2\theta\,\right] - 3\a
 \end{aligned}\right. \;,
\end{equation}
where 
\be
a(\g_0) = \min[\g_0/2,1/4]\;.
\label{eq:ag0}
\ee
These conditions are used in the proof of Lemma \ref{prop3} and in Section \ref{sec:proof5}, 
and are associated to the choice \eqref{par2} below.

\ni -- $\theta_1$ is the cutoff parameter entering in Corollary \ref{cor:Btrick} and related
to the truncation of cross--section factors. 

\ni -- $\theta_2$ is the cutoff parameter bounding the number of creations in a collection of trees
(see Eq. \eqref{eq:reordMCI}).

\ni -- $\theta_3$ is the cutoff parameter controlling high energies (see Eq. \eqref{eq:Fth2}).

For the sake of concreteness the new parameters are fixed according to
\begin{equation}
\label{par2}
\left\{
\begin{aligned}  
&\theta_1  =   a(\g_0) \\
& \theta_2 = 1/ (2 \log(\bar t C(z,\b) e)^{-1})   \\
& \theta_3 = 1/5  \\
 \end{aligned}\right. \;,
\end{equation} 
where the constant $C( z, \b )$ and the time $\bar t$ appear in Lanford's estimate, see \eqref{shtime}.

\subsubsection{Many--recollision estimate} \label{subsub:MRE}

\begin{prop}\label{basicprop}
There exist constants $C_1 = C_1(z,\b) >0$ and $\g_1>0$ 
such that, for all $n \leq \e^{-3/4} \log\e^{-\theta_2}$, 
$Q, L_0 \subset K$, $Q\cap L_0=\emptyset$ and $\e$ sufficiently small,
\be
z^n
\sum_{\G(k,n)} \,\int d \L\, d\bv_k \, 
\mathbbm{1}_{L_0}\,\tilde{\mathbbm{1}}_{K \setminus L_0}\,
\chi_{L_0 }^{rec}\, \chi^{ov}_{Q, K } \, F_{\theta_3}(K)\\
\leq C_1^k\,k^{k}\,(n+k)^k\,(C_1 t)^n\, \e^{\g_1 \frac{q + l_0}{2}}
\label{eq:integralMR}
\ee
for $\bx_k \in \MM^x_k(\d)$.  
\end{prop}
Sections \ref{sec:order} and \ref{sec:prooflemMR} are devoted to the proof of Proposition \ref{basicprop}.

\medskip
\ni {\bf Remark on $\g_1$.}
The coefficient $\g_1$ comes from the geometrical estimates of recollisions (Lemmas \ref{lem:MR} and \ref{lem:MRresINT} below) which imply an arbitrary choice in the interval $\g_1 \in (0, \min [1,2-4\theta-(5/2)\theta_3])$ 
(see \eqref{eq:g1boundGL}).
\medskip

\subsubsection{Proof of \eqref{eq:EKthm}} \label{sec:proof5}

We show here how to conclude the proof of \eqref{eq:EKthm}.

Let us fix the truncation parameters as in \eqref{par2} 
and assume \eqref{eq:thetas14}--\eqref{eq:ag0}
(see also the Remark on page \pageref{rem:choicepar}).

Notice that \eqref{par1} implies $\a < 3/4$ and hence $n  \leq \e^{-3/4} \log\e^{-\theta_2}$
in \eqref{eq:reordMCI}.
By Lemma \ref{prop3} and Proposition \ref{basicprop}, choosing $t^* < (eC_1)^{-1}$ we deduce that, for $t<t^*$,
\bea
&& \int_{\RRR^{3k}} d\bv_k |E_K(t)|
\leq \frac{3\,\e^{\g k}}{4}+ 
(zC)^k \,k^{k} \,\e^{-\theta_1 k}\,3^k\,C_1^k \, k^{k}\,
\sum_{n\geq 0}(n+k)^k\,(C_1 t^*)^n \, \e^{\min[\g_0,\g_1/2] k}\nn\\
&&\ \ \ \ \ \ \ \ \ \ \ \ \ \ \ \ \ \ \ \ \ \leq \frac{3\,\e^{\g k}}{4}+ 
\left(\sum_{n\geq 0}\frac{(n+k)^k}{k!}(C_1 t^*)^n\right)\,(3zCC_1)^k \,k^{3k}\, \e^{\left(\min[\g_0,\g_1/2] - \theta_1\right)k}\nn\\
&&\ \ \ \ \ \ \ \ \ \ \ \ \ \ \ \ \ \ \ \ \ \leq \frac{3\,\e^{\g k}}{4}+ 
\left(\sum_{n\geq 0}(e C_1 t^*)^n\right)\,(3zCC_1e)^k \,k^{3k}\, \e^{\left(\min[\g_0,\g_1/2] - \theta_1\right)k}\nn\\
&&\ \ \ \ \ \ \ \ \ \ \ \ \ \ \ \ \ \ \ \ \ \leq \frac{3\,\e^{\g k}}{4}+ 
\left(\sum_{n\geq 0}(e C_1 t^*)^n\right)\,(3zCC_1e)^k \,\e^{\left(\min[\g_0,\g_1/2] - \theta_1-3\a\right)k}\;.\nn\\
\label{eq:almconc}
\eea
In the third step we applied \eqref{eq:nnnfact} while in the fourth step we used $k<\e^{-\alpha}$.
We conclude that 
\eqref{eq:EKthm} holds for $\e$ small enough if 
\be
\g < \min[\g_0,\g_1/2] - \theta_1-3\a \;.
\label{eq:intrg1}
\ee
By the above Remark on $\g_1$ and \eqref{par2}, this is ensured by
\be
\g <  \min[2a(\g_0), 3/4 - 2\theta] - a(\g_0)-3\a\;.
\ee
\qed

\subsection{Step 2: Ordering of multiple recollisions}\label{sec:order}
\setcounter{equation}{0}    
\def\theequation{4.4.\arabic{equation}} 

This section is devoted to the proof of Proposition \ref{basicprop}. After suitable ordering in time
of the recollision/overlap constraints, we will show in Section \ref{sec:RSRE} that the estimate of a single recollision event 
(Lemma \ref{lem:MR}, proved in Section 4.5) allows to conclude the proof.

Let us focus on the constraint $\chi_{L_0 }^{rec}\, \chi^{ov}_{Q, K }(\bze^{ (L_0)}, \tilde \bze^{ (K \setminus L_0)})$,
where the flow in the argument is the mixed flow entering in the definition of correlation error, Section 4.3.1.b.
To simplify the proof, we will treat simultaneously recollisions and overlaps with a unique method.

\begin{defi}[table of recollisions] \label{def:TR}
Let $L_0,Q \subset K$, $L_0 \cap Q = \emptyset$.
A ``table of recollisions'' associated to $(K,L_0,Q)$
is a set of couples 
$$
(\bm \a, \bm \b):= \{ (\a_1, \b_1), \cdots  , (\a_\ell, \b_\ell)\}\;,
$$
\nomenclature[ab]{$(\bm \a, \bm \b)$}{Table of recollisions}%
with $\a_i\in  L_0 \cup Q$ and $\b_i \in K$, such that:

\ni -- \,\,\,$(\cup_{i=1}^\ell \a_i) \cup (\cup_{i=1}^\ell \b_i) \supset L_0 \cup Q\;;$

\ni -- \,\,\,$\a_i\neq\a_1,\cdots,\a_{i-1}, \b_1,\cdots,\b_{i-1}$ \text { for all } $i=1, \cdots, \ell$\;.

\ni We call ``bullet'' a particle of type $\a$ and ``target'' a particle of type $\b$.
\end{defi}
According to this definition, the bullets are always new with respect to the previous array.

We shall apply the above abstract definition to the case when the bullets $\a$ and the targets $\b$ are indices of the particles generating 
the trees $\Gamma_{\alpha}$, $\Gamma_{\beta}$ (see Figure \ref{fig:TR} below). Remind that an external recollision/overlap 
between $\Gamma_{\alpha}$ and $\Gamma_{\beta}$ indicates a recollision/overlap 
between a pair of particles of the two trees.

\begin{figure}[htbp] 
\centering
\includegraphics[width=3in]{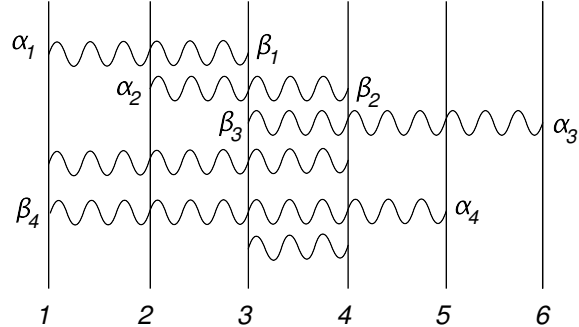} 
\caption{A scheme for a table of recollisions associated to $K = \{1,2,\cdots,6\}$, $Q \cup L_0 = \{1,2,3,5,6\}$.
Here $\ell = 4$. The vertical lines can be associated to particles (trees) in a backwards flow (time flowing upwards)
and the wavy lines
to their external recollisions / overlaps. In this case, the fourth and the last wavy lines represent recollisions
(or overlaps) that do not appear in the table.}
\label{fig:TR}
\end{figure}

\medskip
\ni {\bf Remarks}

1) All the particles in $L_0 \cup Q$ are either bullets or targets (or both)
and each particle can be the target for several bullets.
Conversely, each particle can be the bullet for at most one target, namely the $\a_i$ 
are all distinct and 
\be
| (\bm \a, \bm \b)| = \ell \geq (q+l_0)/2\;. \label{eq:boundCC}
\ee

2) A rough bound on the total number of tables of recollision associated to $(K,L_0,Q)$ is
the following:
\be
\sum_{ (\bm \a, \bm \b)}\, \leq (q+l_0)! \, k^{q+l_0}\leq k! \, k^k \;.
\label{eq:bNCC}
\ee

3) We can construct one particular table with the following explicit procedure.
Fix $(\bze^{ (L_0)}, \tilde \bze^{ (K \setminus L_0)})$ such that $\chi_{L_0 }^{rec}\, \chi^{ov}_{Q, K }=1$.
The first backwards external recollision or overlap identifies the couple $(\a_1,\b_1)$
(up to the exchange $\a_1 \leftrightarrow \b_1$, if both particles belong to $L_0 \cup Q$). 
Going further backwards in time, we consider the first external 
recollision / overlap involving at least one tree in $L_0 \cup Q$ and different from $\a_1,\b_1$.
This identifies the couple $(\a_2,\b_2)$, with the following constraint.
If one (and only one) of the two trees involved is $\a_1$ or $\b_1$, we set such tree $=\ \b_2$, and its partner 
$=\ \a_2$. We iterate this procedure until all the particles in $L_0 \cup Q$ have received a name. 
See Figure \ref{fig:TR} for an example.
\medskip

We shall say that a table of recollisions $(\bm \a, \bm \b)$ is ``realized'' if and only if the mixed dynamics 
$(\bze^{ (L_0)}, \tilde \bze^{ (K \setminus L_0)})$ is well defined (see the Remark on the existence of flows below) and, for all $i=1,\cdots,\ell$: 

\ni (a) the {\em first} (backwards) recollision/overlap of the tree $\G_{\a_i}$ 
occurs in $(0,t)$ with the tree $\G_{\b_i}$; 

\ni (b) the {\em first} (backwards) recollision/overlap of the tree $\G_{\a_i}$ occurs {\em in the past} 
with respect to the first (backwards) recollisions/overlaps of $\G_{\a_{i'}}$, $i' < i$.

\begin{lem} \label{lem:RIpre}
The following 
\be
\mathbbm{1}_{L_0}\,\tilde{\mathbbm{1}}_{K \setminus L_0}\,
\chi_{L_0 }^{rec}\, \chi^{ov}_{Q, K }
\leq \sum_{(\bm \a, \bm \b)}  \chi^{(\bm \a, \bm \b)}
\label{eq:TR}
\ee
holds true, where $\chi^{(\bm \a, \bm \b)}$ denotes the indicator function of the event
for which the table of recollisions $(\bm \a, \bm \b)$ is realized.
\end{lem} 

\medskip
\ni {\bf Proof of Lemma \ref{lem:RIpre}.}
It follows immediately (by subadditivity) from Definition \ref{def:TR} and the Remark 3 above. \qed

\medskip
The following inequality allows to estimate the integrations in the left hand side 
of \eqref {eq:integralMR} iteratively.
\begin{lem} \label{lem:RI}
The following
\be
\chi^{(\bm \a, \bm \b)}\left(\bze^{ (L_0)}, \tilde \bze^{ (K \setminus L_0)}\right) \leq  \prod_{i=1}^\ell \chi^{(\a_i,\b_i)}
\left(\bze^{(L_0 \setminus \{ \a_\ell, \a_{\ell-1},\cdots, \a_i \})}, \tilde\bze^{(K \setminus (L_0\cup\{ \a_\ell, \a_{\ell-1},\cdots, \a_i \}))}, \tilde\bze^{(\a_i)}\right) \label{eq:ri}
\ee
holds true, where $\chi^{(\a_i,\b_i)} = 1$ if and only if the first backwards overlap
of the tree $\G_{\a_i}$ occurs in $(0,t)$ with the tree $\G_{\b_i}$.
\end{lem}
Notice that the mixed flow in the argument of $\chi^{(\a_i,\b_i)}$ is {\em not} the same as in the left hand side,
namely the value of $\chi^{(\a_i,\b_i)}$ is computed as if the trees $\{ \a_\ell, \a_{\ell-1},\cdots, \a_{i+1} \}$
were absent.

\medskip
\ni {\bf Remark (existence of flows).} The requirement of well posedness of the flows involved in the above expressions
(forbidden overlaps at creations) is implicitly absorbed in the definition of $\chi^{(\a,\b)}, \chi^{(\bm\a,\bm\b)}$. 
Note that $\chi^{(\a,\b)}$ is a function of the indicated flow
{\em only} through its history in the time interval $(s,t)$ where $s$ is the overlap time of $\a$
with $\b$ (if any, and zero otherwise). 
Therefore, existence of the flows in $(0,s)$ is not required. The same is true for 
$\chi^{(\bm\a,\bm\b)}$, being in this case $s$ the overlap time of $\a_\ell$ with $\b_\ell$.

\medskip
\ni {\bf Proof of Lemma \ref{lem:RI}.} We observe that
\be
\chi^{(\bm \a, \bm \b)}\left(\bze^{ (L_0)}, \tilde \bze^{ (K \setminus L_0)}\right) =  
\chi^{(\bm \a, \bm \b)}\left( \bze^{ (L_0 \setminus \a_\ell)}, \tilde \bze^{ (K \setminus (L_0\cup\a_\ell))},\tilde\bze^{(\a_\ell)}\right)\;,
\label{eq:subUBF}
\ee
that is, we can always use 
the uncorrelated dynamics notation for the flow of the bullet. It follows that
\bea
\chi^{(\bm \a, \bm \b)}(\bze^{ (L_0)}, \tilde \bze^{ (K \setminus L_0)})
\leq \chi^{(\bm \a, \bm \b)_{\ell-1}}( \bze^{ (L_0 \setminus \a_\ell)}, \tilde \bze^{ (K \setminus (L_0\cup\a_\ell))})\,
\chi^{(\a_\ell,\b_\ell)}( \bze^{ (L_0 \setminus \a_\ell)}, \tilde \bze^{ (K \setminus (L_0\cup\a_\ell))}, \tilde\bze^{(\a_\ell)})\nn\\
\label{eq:scpi}
\eea
where $(\bm \a, \bm \b)_{\ell-1}= \{ (\a_1, \b_1), \cdots  , (\a_{\ell-1}, \b_{\ell-1})\}$
(and of course $(\bm \a, \bm \b) = (\bm \a, \bm \b)_{\ell}$).
Note that in the r.h.s. the overlap of $\G_{\a_\ell}$ with $\G_{\b_\ell}$
can occur at any time in $(0,t)$ (we forget item (b) above).
The constraint $\chi^{(\bm \a, \bm \b)_{\ell-1}}$ is now computed
ignoring the history of the bullet tree~$\G_{\a_\ell}$.

Recursive application of 
\eqref{eq:scpi} leads to the claim. \qed

\subsubsection {Reordering of the integrations in \eqref{eq:integralMR}} 

Applying Lemmas \ref{lem:RIpre} and \ref{lem:RI} to the left hand side of \eqref{eq:integralMR}, one finds
\bea
&& z^n
\sum_{\G(k,n)} \,\int d \L\, d\bv_k \, 
\mathbbm{1}_{L_0}\,\tilde{\mathbbm{1}}_{K \setminus L_0}\,
\chi_{L_0 }^{rec}\, \chi^{ov}_{Q, K } \, F_{\theta_3}(K)\\
&& \ \ \ \ \ \ \ \ \ \ 
\leq  z^n \sum_{(\bm \a, \bm \b)}\, \sum_{\G(k,n)} \,\int d \L\, d\bv_k \, \prod_{i=1}^\ell \chi^{(\a_i,\b_i)}\,F_{\theta_3}(K)\nn\\
&& \ \ \ \ \ \ \ \ \ \ 
= z^n \sum_{(\bm \a, \bm \b)}\,\sum_{\substack{n_1,\cdots,n_k, \\ n = \sum_i n_i}}\,
\sum_{\G_1,\cdots,\G_k}  \,\int d \L_1 \cdots d \L_k\,d\bv_k \, \prod_{i=1}^\ell \chi^{(\a_i,\b_i)}\,F_{\theta_3}(K)\nn
\eea
where the $\chi^{(\a_i,\b_i)}$ is evaluated via the flow
\be
(\bze^{(L_0 \setminus \{ \a_\ell, \a_{\ell-1},\cdots, \a_i \})},
\tilde\bze^{(K \setminus (L_0\cup\{ \a_\ell, \a_{\ell-1},\cdots, \a_i \}))}, \tilde\bze^{(\a_i)})\;.
\nn\label{eq:mfs2}
\ee
In the last equality, which follows from the factorization of trees \eqref{eq:sfpoftr},
we introduced $d\L_i =  \int d\L (\bt^i_{n_i} , \bo^i_{n_i} , \bv^i_{1,n_i})$.

The function $F_{\theta_3}$, defined by \eqref{eq:Fth2}, satisfies $F(K_1\cup K_2) \leq F(K_1)F(K_2)$.
Setting $A = K \setminus \{\a_1,\a_2,\cdots,\a_{\ell }\}$, $\bG_A = \{ \G_i \}_{i\in A}$ and 
$d\L_A =  \prod_{i\in A} \int d\L (\bt^i_{n_i} , \bo^i_{n_i} , \bv^i_{1,n_i})$, it follows that 
\bea
&& z^n
\sum_{\G(k,n)} \,\int d \L\, d\bv_k \, 
\mathbbm{1}_{L_0}\,\tilde{\mathbbm{1}}_{K \setminus L_0}\,
\chi_{L_0 }^{rec}\, \chi^{ov}_{Q, K } \, F_{\theta_3}(K) \label{eq:orderMRst} \\ 
&& \leq z^n \sum_{(\bm \a, \bm \b)}\,\sum_{\substack{n_1,\cdots,n_k, \\ n = \sum_i n_i}}\,
\sum_{\bG_A}  \,\int d \L_A \, d\bv_A \, F_{\theta_3}(A)\nn\\
&&\cdot\sum_{\G_{\a_1}}  \,\int d \L_{\a_1} \,dv_{\a_1}\, \chi^{(\a_1,\b_1)}\,F_{\theta_3}(\a_1)
\sum_{\G_{\a_2}}  \,\int d \L_{\a_2} \,dv_{\a_2}\,\cdots
\sum_{\G_{\a_\ell}}  \,\int d \L_{\a_\ell} \,dv_{\a_\ell}\, \chi^{(\a_\ell,\b_\ell)}\,F_{\theta_3}(\a_\ell)\;.\nn
\eea

\subsubsection {Proof of Proposition \ref{basicprop}} \label{sec:RSRE}

\ni {\bf 4.4.2.a\ \ \ Single--recollision estimate} 
\bigskip

\ni Our purpose is to estimate iteratively the integrals in \eqref{eq:orderMRst}. The result we need for the 
single step is the following.

\begin{lem}[estimate of one external recollision] \label{lem:MR}
Let $s \to \eta(s)$ be a piecewise constant function from $(0,t)$ to $\RRR^3$
such that $|\eta| \leq \e^{-\theta_3/2}$, with discontinuity points in the finite set $T = \{\t_1,\t_2,\cdots\}$.
Let $s \to \xi(s)$ be a piecewise free trajectory in $\RRR^3$ with velocity
\be
\frac{d\xi}{ds}= \eta\label{eq:trajFIXmr}
\ee
except on $T$. In a subset of $T$, jumps of entity $\e$ may occur 
($\,|\xi(\t_i^+)-\xi(\t_i^-)|=\e\,$). Let $n \leq \e^{-3/4} \log\e^{-\theta_2}$ be the maximum number of such jumps. 
Fix $x_1 \in \RRR^3$ and assume $|x_1-\xi(t)| > \d = \e^{\theta}$. 
Then, there exist $D>0$ and $\g_1>0$ such that, for all
$n_1 \leq n$ and $\e$ small enough,
\be
\sum_{\Gamma(1,n_1)} \int d\L \, dv_{1} 
\,\chi^{ov}_{\xi}(\bze^\e) \, F_{\theta_3}(1)
\leq  (Dt)^{n_1}\, \e^{\g_1}\;, \label{eq:MRres}
\ee
where $\, \bze^\e = (\z^\e_1,\cdots,\z^\e_{1+n_1})$ is the IBF associated to the $1$--particle, $n_1-$collision
tree $\Gamma(1,n_1)$ with $\z^\e_1(t) = x_1$, and $\chi^{ov}_{\xi}$ is the indicator function of the event 
$$\Big\{\exists
\,s\in (0,t)\ |\ \DD(t-s) < \e\Big\}$$ where, for $s\in(t_i,t_{i-1}]$,
$\DD(s) = \DD[\bze^\e](s):= \min_{k=1,\cdots,i} \,  |\xi(s) - \xi^{\e}_k(s)|$\;.
\end{lem}
Note that it is implicit in the definition of $\chi^{ov}_{\xi}$ that the IBF is well defined 
(no internal overlaps at the creation times) up to the first $s$ verifying the condition.
The constant $D$ depends only on the parameter $\b$ appearing in $F_{\theta_3}$.

The Lemma is proved in Section \ref{sec:prooflemMR}.

\bigskip
\ni {\bf 4.4.2.b\ \ \ Virtual trajectories} 
\bigskip

\ni Following \cite{PSS13}, we introduce a global notion of trajectory which will be convenient in the next subsection
to apply iteratively the previous lemma. This will be also crucially used in the geometrical estimate
of Step 3.

Loosely speaking, a virtual trajectory is a trajectory of a given particle in the IBF (or other flow)
{\em extended} up to time $t$. We shall use for it an upper--index notation. For instance, $\z^{\e,i}(s)$ 
coincides with $\z_i^\e(s)$ for $s>0$ and up to the time of creation of $i$; thereafter
is extended by the trajectory of its progenitor up to its creation time, and so on. 

\begin{figure}[htbp] 
   \centering
   \includegraphics[width=4.5in]{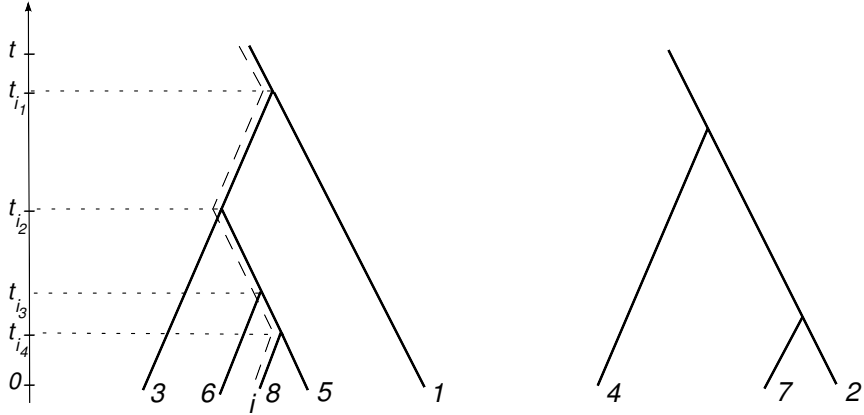} 
   \caption{The line closest to the dashed line is the path $p_i$ in the tree $\G(2,6),$ with $i=8.$ 
  The states of the particle associated to it 
via the flow $\bar\bze$ form the``virtual trajectory of $i$''.}
   \label{fig:ivirtual}
\end{figure}

\begin{defi}[virtual trajectory] \label{def:VT}
Consider particle $i$ in the graph of a tree $\G(k,n) = (k_1,\cdots,k_n)$. 
Let $\bt_n = t_1,\cdots,t_n$ be the sequence of times associated to the nodes of the tree.

\ni (i) A polygonal path $p_i$ is uniquely defined if we walk on the tree by going forward in time, starting 
from the time--zero endpoint of line $i$ and going up to the root--point at time $t$ (e.g. Figure \ref{fig:ivirtual}). 

\ni (ii) Let $t_{i_1},\cdots,t_{i_{n^i}}$ be the decreasing subsequence of $t_1,\cdots,t_n$, made of the times 
corresponding to the nodes met by following the path $p_i$ ($n^i$ being the number of 
such nodes, with the convention $i_0=0, t_{i_0}=t$). Then, for any backwards flow $\bar\bze$ which 
can be constructed from $\G(k,n),\bt_n$
\footnote{In this definition, $\bar\bze$ can be either the IBF $\bze^\e$, the uncorrelated flow
$\tilde\bze^\e$, the EBF $\bze^\EE$ or a mixed flow. 
We shall use it in different contexts.}, 
we call
{\bf virtual trajectory associated to particle $i$ in the flow}, and indicate it with upper indices
$\bar\z^i(s)=(\bar\xi^i(s),\bar\eta^i(s))\in\RRR^{6}$, $s\in[0,t]$, the trajectory given by:
\be
\bar\z^i(s) =
\begin{cases}
\displaystyle \bar\z_i(s)\mbox{\ \ \ \ \ \ for\ }s\in[0,t_{i_{n^i}}) \\
\displaystyle \bar\z_{k_{i_{r}}}(s)\mbox{\ \ \ \ for\ }s\in[t_{i_{r}},t_{i_{r-1}}),\ \ \ 0<r \leq n^i
\end{cases}\;.\label{eq:ivirtualdef}
\ee
\nomenclature[zeeta]{$\bar\z^i(\cdot)$}{Virtual trajectory of particle $i$ in the flow $\bar\z$}%
\end{defi}
Note that the virtual trajectory is piecewise--free, and built up with pieces of trajectories of (different) 
particles of $\bar\bze$. Instantaneous jumps of entity $\e$ occur at creation times, when 
the name of the particle in the flow $\bar\bze$ changes (e.g. $t_{i_1}$, $t_{i_2}$ and $t_{i_4}$ in 
Figure \ref{fig:ivirtual}).
Only during the time of existence of particle $i$ in the flow, $\bar\z^i(s)=\bar\z_i(s)$ holds.

\bigskip
\ni {\bf 4.4.2.c\ \ \ Iterative estimate of multiple recollisions} 
\bigskip

\ni Let us focus on $\chi^{(\a_\ell,\b_\ell)}$ (defined in Lemma \ref{lem:MR}). 
Remind the notation \eqref{eq:notationS(i)}.
If $\chi^{(\a_\ell,\b_\ell)} =~1$,
there exists $i \in S(\b_\ell)$ such that $\tilde\bze^{(\a_\ell)}$ overlaps with the trajectory of particle $i$
in the mixed flow $( \bze^{ (L_0 \setminus \a_\ell)}, \tilde \bze^{ (K \setminus (L_0\cup\a_\ell))})$.
In particular, $\a_\ell$ overlaps with the virtual trajectory of $i$ in the flow (Definition \ref{def:VT}
applied to the mixed flow). 

We denote such virtual trajectory by $\hat\z^{i}=(\hat\xi^{i},\hat\eta^{i})$. Then,
\be
\sum_{\G_{\a_\ell}}  \,\int d \L_{\a_\ell} \,dv_{\a_\ell}\, \chi^{(\a_\ell,\b_\ell)}\,F_{\theta_3}(\a_\ell)
\leq \sum_{i \in S(\b_\ell)} \sum_{\G_{\a_\ell}}  \,\int d \L_{\a_\ell} \,dv_{\a_\ell}\,
\chi^{ov}_{\hat\xi^{i}}(\bze^\e) \,F_{\theta_3}(\a_\ell)\;,
\ee
where the function $\chi^{ov}$ is defined in Lemma \ref{lem:MR} and $\bze^\e$ is now the
IBF associated to the $1$--particle, $n_{\a_\ell}-$collision tree $\Gamma_{\a_\ell}$.

The presence of the functions $F_{\theta_3}$ in \eqref{eq:orderMRst} ensures that $|\hat\eta^{i}| \leq \e^{-\theta_3/2}$.
Furthermore, $\bx_k \in \MM^x_k(\d)$ ensures $|x_{\a_\ell}-\hat\xi^{i}(t)| > \d = \e^{\theta}$.
Therefore, we are in the position to apply Lemma \ref{lem:MR} and we deduce
\be
\sum_{\G_{\a_\ell}}  \,\int d \L_{\a_\ell} \,dv_{\a_\ell}\, \chi^{(\a_\ell,\b_\ell)}\,F_{\theta_3}(\a_\ell)
\leq (n_{\b_\ell}+1)\, (Dt)^{n_{\a_\ell}} \e^{\g_1}\;.
\label{eq:qwertz}
\ee
Inserting into \eqref{eq:orderMRst} we obtain
\bea
&& z^n
\sum_{\G(k,n)} \,\int d \L\, d\bv_k \, 
\mathbbm{1}_{L_0}\,\tilde{\mathbbm{1}}_{K \setminus L_0}\,
\chi_{L_0 }^{rec}\, \chi^{ov}_{Q, K } \, F_{\theta_3}(K) \\ 
&& \leq \e^{\g_1}\,z^n \sum_{(\bm \a, \bm \b)}\,\sum_{\substack{n_1,\cdots,n_k, \\ n = \sum_i n_i}}\,
(n_{\b_\ell}+1)\, (Dt)^{n_{\a_\ell}} \,\sum_{\bG_A}  \,\int d \L_A \, d\bv_A \, F_{\theta_3}(A)\nn\\
&&\cdot\sum_{\G_{\a_1}}  \,\int d \L_{\a_1} \,dv_{\a_1}\, \chi^{(\a_1,\b_1)}\,F_{\theta_3}(\a_1)\,\cdots
\sum_{\G_{\a_{\ell-1}}}  \,\int d \L_{\a_{\ell-1}} \,dv_{\a_{\ell-1}}\, \chi^{(\a_{\ell-1},\b_{\ell-1})}\,F_{\theta_3}(\a_{\ell-1})\;.\nn
\eea

We repeat the above discussion {\em ad litteram} for $\chi^{(\a_{\ell-1},\b_{\ell-1})}$, and so on up to $ \chi^{(\a_1,\b_1)}$.
Since $\ell \geq (q+l_0)/2$,
the result is:
\bea
&& z^n
\sum_{\G(k,n)} \,\int d \L\, d\bv_k \, 
\mathbbm{1}_{L_0}\,\tilde{\mathbbm{1}}_{K \setminus L_0}\,
\chi_{L_0 }^{rec}\, \chi^{ov}_{Q, K } \, F_{\theta_3}(K) \\ 
&& \leq  \e^{\g_1 \frac{q + l_0}{2}}\,z^n \sum_{(\bm \a, \bm \b)}\,\sum_{\substack{n_1,\cdots,n_k, \\ n = \sum_i n_i}}\,
\,\prod_{i=1}^{\ell}(n_{\b_i}+1)\,\, (Dt)^{n_{\a_i}} \,\sum_{\bG_A}  \,\int d \L_A \, d\bv_A \, F_{\theta_3}(A)\;.
\nn
\eea

Using \eqref{eq:nnnfact}, the last sum over trees is bounded by 
$e^{|A|}\,(e\,4\pi\, (2\pi/\b)^{3/2} \,t)^{\sum_{i \in A} n_i}\;.$
Since $A = K \setminus \{\a_1,\a_2,\cdots,\a_{\ell }\}$ and $\sum_{i=1}^\ell n_{\a_i} + \sum_{i\in A}n_i = n$,
there holds
\bea
&& z^n
\sum_{\G(k,n)} \,\int d \L\, d\bv_k \, 
\mathbbm{1}_{L_0}\,\tilde{\mathbbm{1}}_{K \setminus L_0}\,
\chi_{L_0 }^{rec}\, \chi^{ov}_{Q, K } \, F_{\theta_3}(K) \nn\\
&&\leq  \e^{\g_1 \frac{q + l_0}{2}}\,(C'_1)^k\,(C'_1 t)^n\, \sum_{(\bm \a, \bm \b)}\,
\sum_{\substack{n_1,\cdots,n_k, \\ n = \sum_i n_i}}\,
\prod_{i=1}^{\ell}(n_{\b_i}+1)
\eea
for suitable $C'_1 = C'_1(z,\b)$, that is
\bea
&& z^n
\sum_{\G(k,n)} \,\int d \L\, d\bv_k \, 
\mathbbm{1}_{L_0}\,\tilde{\mathbbm{1}}_{K \setminus L_0}\,
\chi_{L_0 }^{rec}\, \chi^{ov}_{Q, K } \, F_{\theta_3}(K) \nn\\
&&\leq  \e^{\g_1 \frac{q + l_0}{2}}\,(C'_1)^k\,(C'_1 t)^n\, \sum_{\substack{\a_1,\cdots,\a_\ell \\ \a_i \neq \a_j}}\,
\sum_{\substack{n_1,\cdots,n_k, \\ n = \sum_i n_i}}\,
\prod_{i=1}^{\ell}\sum_{\b_i\in K}(n_{\b_i}+1)\nn\\
&& = \e^{\g_1 \frac{q + l_0}{2}}\,(C'_1)^k\,(C'_1 t)^n\, \sum_{\substack{\a_1,\cdots,\a_\ell \\ \a_i \neq \a_j}}\,
\sum_{\substack{n_1,\cdots,n_k, \\ n = \sum_i n_i}}\, (n+k)^\ell\nn\\
&& \leq \e^{\g_1 \frac{q + l_0}{2}}\,(C'_1)^k\,(C'_1 t)^n\, k! \, e^{k+n}\,
(n+k)^k\;,
\eea
having used $$\sum_{\substack{n_1,\cdots,n_k, \\ n = \sum_i n_i}}1\leq e^{k+n}\;.$$
Eq. \eqref{eq:integralMR} follows
by taking $C_1 = e\, C'_1$. \qed

\subsection{Step 3: Estimate of an external recollision}\label{sec:prooflemMR}
\setcounter{equation}{0}    
\def\theequation{4.5.\arabic{equation}} 

In this section, we prove Lemma \ref{lem:MR}. The proof is organized in seven steps which will
be discussed in separate subsections.

\subsubsection {Substitution of the IBF with the EBF} 

The flow $\bze^\e$ in the left hand side of \eqref{eq:MRres} involves internal recollisions, 
which is convenient to eliminate first.
Let 
\be
\chi^{int} = \chi^{int}(\bt_{n_1} , \bo_{n_1} , \bv_{1+n_1}) = 1 \label{eq:defChiir}
\ee
if and only if:

-- either an overlap at a creation time occurs (ill--defined $\bze^\e$), or

-- the IBF $\bze^\e$ delivers an internal recollision.
\begin{lem}[estimate of the internal recollision]  \label{lem:MRresINT} 
There exists a constant $D>0$ such that, for any 
$\g_1 < 1$ and $\e$ small enough,
\be
\sum_{\Gamma(1,n_1)} \int d\L \, dv_{1} 
\,\chi^{int}
\, e^{-(\b/2)\sum_{i\in S(1)}v_i^2} 
\leq  (Dt)^{n_1} \frac{\e^{\g_1}}{2}\;. \label{eq:MRresINT}
\ee
\end{lem}
The control of the internal recollisions is well known (see \cite{GSRT12, PSS13}) so that
we deserve the proof to Appendix D. We note, incidentally, that the present estimate is optimal ($\g_1 \lesssim 1$).

Lemma \ref{lem:MRresINT} allows to bound the l.h.s. in \eqref{eq:MRres} with a much simpler expression, i.e.
\bea
&& \sum_{\Gamma(1,n_1)} \int d\L \, dv_{1} 
\,\chi^{ov}_{\xi}(\bze^\e) \, F_{\theta_3}(1)
\leq  (Dt)^{n_1} \frac{\e^{\g_1}}{2} 
+ \sum_{\Gamma(1,n_1)} \int d\L \, dv_{1} 
\,\chi^{ov}_{\xi}(\bze^\e) \,(1-\chi^{int})\,F_{\theta_3}(1)\nn\\
&&\ \ \ \ \ \ \ \ \ \ \ \ \ \ \ \ \ \ \ \ \ \ \ \ \ \ \ \ \ \ \ \ \ \ \ \ \ \ \ \equiv (Dt)^{n_1} \frac{\e^{\g_1}}{2} 
+ \sum_{\Gamma(1,n_1)} \int d\L \, dv_{1} 
\,\chi^{ov}_{\xi}(\bze^\EE) \,(1-\chi^{int})\,F_{\theta_3}(1)\nn\\
&&\ \ \ \ \ \ \ \ \ \ \ \ \ \ \ \ \ \ \ \ \ \ \ \ \ \ \ \ \ \ \ \ \ \ \ \ \ \ \ \leq (Dt)^{n_1} \frac{\e^{\g_1}}{2} 
+ \sum_{\Gamma(1,n_1)} \int d\L \, dv_{1} 
\,\chi^{ov}_{\xi}(\bze^\EE) \,F_{\theta_3}(1)\nn\\
\label{eq:AMmia}
\eea
where $\, \bze^\EE = (\z^\EE_1,\cdots,\z^\EE_{1+n_1})$ is the EBF associated to the $1$--particle, 
$n_1-$collision tree $\Gamma(1,n_1)$, introduced in Section \ref{sec:EBF} (figure at page \pageref{fig:PSflows}, (iii)).

\subsubsection {Integration over virtual trajectories} 

We shall reduce the problem to the estimate
of an integral spanning a single virtual trajectory; see Section 4.4.2.b.

Condition $\chi^{ov}_{\xi}(\bze^\EE)=1$ indicates the event ``$\DD[\bze^\EE](t-s)<\e$ for some $s\in (0,t)$'',
which in turn implies 
\be
\mbox{``$|\xi(s)-\xi^\EE_{i}(s)|<\e$ for some $i\in \{1,\cdots,n_1+1\}$ and some $s\in (0,t_{i-1})$''}\;.
\label{eq:condovEx}
\ee
Note now that such event depends actually not on the full EBF, but {\em just}
on the virtual trajectory $\z^{\EE,i}$.
Consequently, we may integrate out all the variables
which are {\em not} entering in the construction of $\z^{\EE,i}(s)$. 

According to
Definition \ref{def:VT}, for any given $\G(1,n_1), i$,
calling $n^i$ the number of nodes encountered by $\z^{\EE,i}$ and
$i_1,i_2,\cdots,i_{n^i}$ their names (ordered as increasing sequence), the integration variables
describing completely the virtual trajectory are:
$$v_1,t_{i_1},\cdots,t_{i_{n^i}},\o_{i_1},\cdots,\o_{i_{n^i}},v_{i_1},\cdots,v_{i_{n^i}} \longrightarrow \z^{\EE,i}\;,$$
which we rename here for convenience:
$$v_1,t^{1},\cdots,t^{n^i},\o^{1},\cdots,\o^{n^i},v^{1},\cdots,v^{n^i} \longrightarrow \z^{\EE,i}\;.$$

With this notation and 
\be
\HH^{i}_1 := v_1^2+ \sum_{k=1}^{n^{i}}(v^{k})^2\;,
\label{eq:eni1OEX}
\ee
we get
\bea
&& \sum_{\Gamma(1,n_1)} \int d\L \, dv_{1} 
\,\chi^{ov}_{\xi}(\bze^\EE) \,F_{\theta_3}(1) \label{eq:MRxiE2}\\
&& \leq \sum_{\Gamma(1,n_1)}\sum_{i=1}^{n_1+1} \frac{(D't)^{n_1-n^{i}}}{(n_1-n^{i})!}
\int_0^t dt^1\int_0^{t^1} dt^2\cdots \int_0^{t^{n^i-1}} dt^{n^i}
\int d\o^1\cdots d\o^{n^{i}} \nn\\
&&\ \ \ \cdot\int dv_1 dv^1\cdots dv^{n^{i}}
\mathbbm{1}_{\mbox{$\{\inf_{s \in (0,t)} |\xi(s) - \xi^{\EE,i}(s)| < \e\}$}}
\,e^{-(\b/2) \HH^i_1} 
\, \mathbbm{1}_{\HH^{i}_{1} \leq \e^{-\theta_3}}\;, \nn
\eea
where $D' = 4\pi\, (2\pi/\b)^{3/2}$.

\subsubsection {A change of variables: relative velocities}

Let us focus on the last line in Eq. \eqref{eq:MRxiE2}.
To integrate over the characteristic function, it is convenient to use
the variable $v_1$ together with the relative velocities at the creation times,
which we introduce in what follows. 

The virtual trajectory $\z^{\EE,i}$ has piecewise constant velocity, with $n^{i}$ jumps 
at the creation times. We call $$\eta^1,\eta^2,\cdots,\eta^{n^i+1}$$ the values 
assumed by the velocity, 
namely $$\eta^1=v_1\;,$$ $$\eta^k \equiv \eta^{\EE,i}((t^{k-1})^-) \equiv \eta^{\EE,i}(s) 
\mbox{\ \ \ \ \ \ for\ \ \ \ \ \ } s\in (t^k,t^{k-1})\;.$$
The relative velocities at creations are then:
\bea
&& V_1 = v^1-\eta^1\;, \nn\\
&& V_2 = v^2 -\eta^2\;, \nn\\
&& \cdots \nn\\
&& V_{n^i} = v^{n^i} - \eta^{n^i}\;.\nn
\eea

Note that $v^k$ are velocities of {\em added} particles at the moment of their creation.
In particular, $\eta^k$ is {\em independent} of $v^k$, so that the previous relations can be regarded as simple translations
and
\bea
\int dv_1 dv^1\cdots dv^{n^{i}}
\mathbbm{1}_{\mbox{$\{\inf_{s \in (0,t)} |\xi(s) - \xi^{\EE,i}(s)| < \e\}$}}
\,e^{-(\b/2) \HH^i_1} 
\, \mathbbm{1}_{\HH^{i}_{1} \leq \e^{-\theta_3}} \nn\\
= \int dv_1 \, dV_1\cdots dV_{n^{i}}\,
\mathbbm{1}_{\mbox{$\{\inf_{s \in (0,t)} |\xi(s) - \xi^{\EE,i}(s)| < \e\}$}}
\,e^{-(\b/2)\HH^i_1} 
\, \mathbbm{1}_{\HH^{i}_{1} \leq \e^{-\theta_3}}\;,
\label{eq:poiutzre}
\eea
where now $\z^{\EE,i}(s)$ and $\HH^{i}_{1}$ have to be computed by using $V_1,\cdots,V_{n^i}$.

The energy function reads:
\bea
\HH^{i}_{1} = 
v_1^2 + \sum_{k=1}^{n^{i}}(V_k + \eta^k)^2\;,
\label{eq:energymaled}
\eea
which we want to express completely in terms of the new integration variables.
To do this, observe that each jump of velocity in the virtual trajectory $\z^{\EE,i}$, i.e. 
$$\eta^k = \eta^{\EE,i}((t^{k})^+)\to \eta^{k+1}= \eta^{\EE,i}((t^{k})^-)\;,$$ can be
of two types, determined uniquely by the structure of the tree $\G(1,n_1)$
(and corresponding for instance to nodes $i_2$ (type 1) and $i_3$ (type 2) in Figure \ref{fig:ivirtual} at page
\pageref{fig:ivirtual}).
That is:
\begin{figure}[htbp] 
\centering
\includegraphics[width=6in]{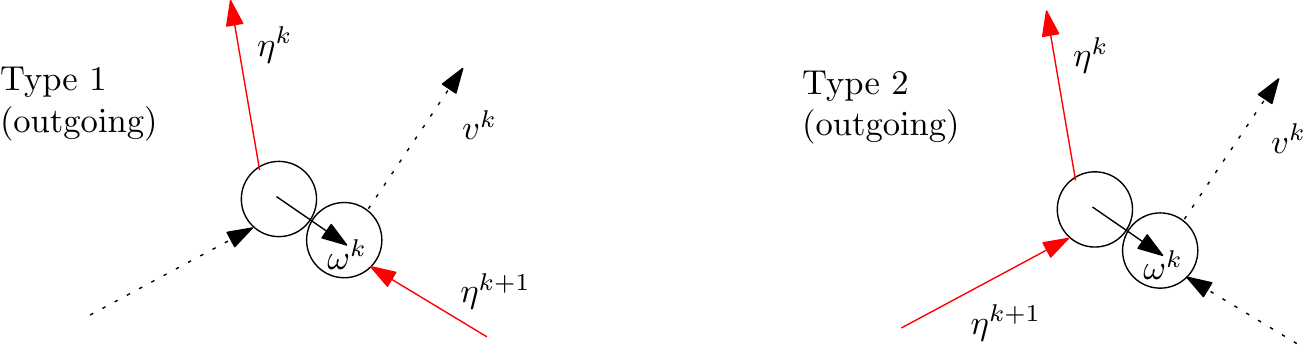} 
\label{fig:Etypes}
\end{figure}
\begin{itemize}
\item {\bf Type $1$}. 
The position jumps according to $\xi^{\EE,i}((t^{k})^+) \to \xi^{\EE,i}((t^{k})^-)=
\xi^{\EE,i}((t^{k})^+)+\e\o^k$; 
the velocity jumps according
to 
\be
 \eta^{k+1} - \eta^{k} = 
\begin{cases}
\displaystyle  P_{\o^k}^{\perp} V_k = V_k - \o^k (\o^k\cdot V_k)\ \ \ \ \ \ \ \ \ 
(\o^k\cdot V_k)\geq 0\ \ \ \mbox{(outgoing collision)}\\
\displaystyle  V_k \ \ \ \ \ \ \ \ \ \ \ \ \ \ \ \ \ \ \ \ \ \ \ \ \ \ \ \ \ \ \ \ \ \ \ \ \ \ \ 
(\o^k\cdot V_k) < 0\ \ \ \mbox{(incoming collision)}
\end{cases}\;;\nn
\ee
\item {\bf Type $2$}. The position does not jump: $\xi^{\EE,i}((t^{k})^+) = \xi^{\EE,i}((t^{k})^-)$; the velocity jumps according
to 
\be
 \eta^{k+1} - \eta^{k} = 
\begin{cases}
\displaystyle  P_{\o^k}^{\parallel} V_k = \o^k (\o^k\cdot V_k)\ \ \ \ \ \ \ \ \ \ \ \ \ \ 
(\o^k\cdot V_k)\geq 0\ \ \ \mbox{(outgoing collision)}\\
\displaystyle  0 \ \ \ \ \ \ \ \ \ \ \ \ \ \ \ \ \ \ \ \ \ \ \ \ \ \ \ \ \ \ \ \ \ \ \ \ \ \ \ 
(\o^k\cdot V_k) < 0\ \ \ \mbox{(incoming collision)}
\end{cases}\;.\nn
\ee
\end{itemize}
To have a compact notation, we write the above transformation as
\be
\eta^{k+1} - \eta^{k} = P^k V_k\;,
\label{eq:basCEFv}
\ee
which implies
\be
\HH^{i}_{1} = \sum_{k=0}^{n^{i}}\left(V_k + \sum_{h=1}^{k-1}P^hV_h + v_1\right)^2\;,
\ee
with the convention $V_0 = 0$. 

Equation \eqref{eq:basCEFv} allows to construct iteratively the whole trajectory of 
$\z^{\EE,i}$, starting from $\z^{\EE,i}(t) = (x_1, \eta^1) = (x_1,v_1)$.

\subsubsection {Energy bounds} \label{sec:LBE}

We collect here some energy estimates that will be used later on.

First, observe that
\be
a_k := V_k + \sum_{h=1}^{k-1}P^hV_h  \label{eq:transak}
\ee
is a $v_1-$independent quantity and therefore
\bea
\inf_{v_1}\HH^{i}_{1} &&= \inf_{v_1} \sum_{k=0}^{n^{i}}\left(a_k + v_1\right)^2\nn\\
&& =  \inf_{v_1} \left(\sum_{k=0}^{n^{i}}a_k^2 + (n^i + 1) v_1^2 + 2 v_1 \cdot \sum_{k=0}^{n^{i}}a_k\right)\nn\\
&& \geq \sum_{k=0}^{n^{i}}a_k^2 - \frac{\left(\sum_{k=0}^{n^{i}}a_k\right)^2}{n^i+1}\;.
\label{eq:SNCDJSCH}
\eea

Moreover, the conservation of energy at collisions implies $|\eta^k|^2 + |v^k|^2 \geq |\eta^{k+1}|^2$. In particular,
by Eq. \eqref{eq:eni1OEX}, for any $k = 1,\cdots, n^i+1$ one has
\be
\HH^{i}_1 \geq v_1^2+ \sum_{q=1}^{k-1}(v^{q})^2 =  |\eta^1|^2+ \sum_{q=1}^{k-1}(v^{q})^2
\geq |\eta^2|^2 + \sum_{q=2}^{k-1}(v^{q})^2\geq \cdots \geq |\eta^{k}|^2\;,
\ee
so that $\HH^{i}_{1} \leq \e^{-\theta_3}$ leads to
\be
|\eta^{k}|\leq \e^{-\theta_3/2}
\label{eq:boundeta^k}
\ee
for all $k$ and
\be
\Big|\sum_{k=1}^r P^kV_k\Big| \leq 2\e^{-\theta_3/2}
\label{eq:conPkVk}
\ee
for all $r \in \{0,1,\cdots,n^i\}$.

Finally, for $s \in (t^{r+1},t^{r})$, the quantity
\bea
&& \sum_{k=1}^r P^kV_k(t^k-s) = (\eta^2-\eta^1)(t^1-s)+\cdots + (\eta^{r+1}-\eta^{r})(t^r-s)\nn\\
&&\ \ \ \ \ \ \ \ \ \ \ \ \ \ \ \ \ \ \ \ \ \  = -\eta^1 (t-s) + \eta^1(t-t^1)+\eta^2(t^1-t^2)+\cdots + \eta^{r+1}(t^r-s)\nn\\
\eea
is bounded uniformly in $s, r$ by
\bea
&& \sup_{s\in (0,t)}\Big|\sum_{k=1}^r P^kV_k(t^k-s)\Big| \leq 2 \max_k|\eta^k| t \leq 2t \e^{-\theta_3/2}\;.
\label{eq:conPkVkbis}
\eea

Let 
\be
\AA := \Big\{ \mbox{\eqref{eq:conPkVk} and \eqref{eq:conPkVkbis} are satisfied}\Big\}
\label{eq:defAA}
\ee
and notice that this is $v_1-$independent.
Using \eqref{eq:SNCDJSCH}, \eqref{eq:poiutzre} and \eqref{eq:MRxiE2}, we arrive to
\bea
&& \sum_{\Gamma(1,n_1)} \int d\L \, dv_{1} 
\,\chi^{ov}_{\xi}(\bze^\EE) \,F_{\theta_3}(1) \nn\\
&& \leq \sum_{\Gamma(1,n_1)}\sum_{i=1}^{n_1+1} \frac{(D't)^{n_1-n^{i}}}{(n_1-n^{i})!}
\int_0^t dt^1\int_0^{t^1} dt^2\cdots \int_0^{t^{n^i-1}} dt^{n^i}
\int d\o^1\cdots d\o^{n^{i}} \nn\\
&&\ \ \ \cdot \int\, dV_1\cdots dV_{n^{i}}\,e^{-(\b/2)\sum_{k}a_k^2 + 
\frac{(\b/2)}{n^i+1}(\sum_{k}a_k)^2}
\,\mathbbm{1}_{\AA} \nn\\
&&\ \ \ \cdot \int dv_1 
\mathbbm{1}_{\mbox{$\{\inf_{s \in (0,t)} |\xi(s) - \xi^{\EE,i}(s)| < \e\}$}}
\, \mathbbm{1}_{\HH^{i}_{1} \leq \e^{-\theta_3}}\;.
\label{eq:nspccm}
\eea
We shall study the latter integral in the next subsection.

\subsubsection {The overlap constraint as an integral over ``tubes''} \label{sec:Tube}

We denote by $|\cdot|$ the volume of the set $\cdot$ in $\RRR^3$. We 
also introduce a small $\e-$dependent quantity 
\be
R^\e= \e^{1-\theta - \theta_3/2} \,4t
\label{eq:defReps}
\ee
and require $\theta + \theta_3/2 < 1$.
Next we prove the following estimate.
\begin{lem} \label{lem:SPAN} In the assumptions of Lemma \ref{lem:MR}, one has
\be
\int dv_1 \mathbbm{1}_{\mbox{$\{\inf_{s \in (0,t)} |\xi(s) - \xi^{\EE,i}(s)| < \e\}$}}
\, \mathbbm{1}_{\HH^{i}_{1} \leq \e^{-\theta_3}} \leq \frac{|\T^{\e}_\xi|}{t^3}
\, \mathbbm{1}_{|\xi(t)-x_1| \leq  3t \e^{-\theta_3/2}}\;,
\label{eq:MCSVAF}
\ee
\nomenclature[TTex]{$\T^{\e}_\xi$}{``Tube'' of external recollision}%
where $\T^{\e}_\xi$ is the region spanned by a ball of radius $R^\e$
with center moving on the
curve of parametric equation $\left(\D(s)\right)_{s\in (0,t-\e t / R^\e)}\;,$ defined by
\be
\D(s) := \frac{t}{t-s}\Big[(\xi(s)-\xi(t))+ (\xi(t)-x_1) + \sum_{k=1}^r P^kV_k(t^k-s) -  {\sum_{k \leq r}}^* \e\,\o^k\Big]
\label{eq:eqTUBE}
\ee
\nomenclature[De]{$\D$}{Axis of $\T^{\e}_\xi$}%
for $s \in (t^{r+1},t^{r})$, with the sum ${\sum}^*$ running over all the nodes of type $1$.
\end{lem}
The above result holds for any choice of the variables defining $\z^{\EE,i}(s)$.

\medskip
\ni {\bf Proof of Lemma \ref{lem:SPAN}.} 
Given $r \in \{0,1,\cdots,n^i\}$, at time $s \in (t^{r+1},t^{r})$ (remind $t^0 \equiv t$) the virtual trajectory 
reads:
\bea
\xi^{\EE,i}(s) && = x_1 - v_1(t-t^1)-\eta^2(t^1-t^2)-\cdots - \eta^{r+1}(t^r-s)+ {\sum_{k \leq r}}^* \e\,\o^k \nn\\
&& = x_1-v_1(t-s) - (\eta^2-\eta^1)(t^1-s)-\cdots - (\eta^{r+1}-\eta^{r})(t^r-s)
+ {\sum_{k \leq r}}^* \e\,\o^k\nn\\
&& = x_1-v_1(t-s) - \sum_{k=1}^r P^kV_k(t^k-s) +  {\sum_{k \leq r}}^* \e\,\o^k\;.
\label{eq:expexpxieis}
\eea

By assumption, the velocity of the target is bounded as $|\eta| \leq \e^{-\theta_3/2}$
and the same is true for the bullet (see \eqref{eq:boundeta^k}).
Furthermore, both the trajectories of bullet and target may have at most $n$ jumps of entity $\e$ in position, where
$n \leq \e^{-3/4} \log\e^{-\theta_2}$ as fixed by Proposition \ref{basicprop}.
Namely, the displacements are bounded by 
\bea
\max\left(|\xi(s)-\xi(t)|,|\xi^{\EE,i}(s)-x_1|\right) &&\leq |\eta| (t-s) + n \,\e \nn\\
&& \leq \e^{-\theta_3/2} (t-s) + \e^{1/4} \log\e^{-\theta_2}\;.
\label{eq:bounddisp}
\eea
From this, we deduce two remarks on the overlap condition $\{\inf_{s \in (0,t)} |\xi(s) - \xi^{\EE,i}(s)| < \e\}$.

\medskip
{\bf (1)} Time $s$ realizing the condition can {\em not} be too close to $t$.
In fact, since by hypothesis $ |\xi(t)-x_1| \geq \e^{\theta}$,
\bea
 |\xi(s) - \xi^{\EE,i}(s)| && \geq  |\xi(t)-x_1| - |\xi(s)-\xi(t)| -|\xi^{\EE,i}(s)-x_1|\nn\\
 && \geq \e^\theta - 2\e^{-\theta_3/2} (t-s) -2 \e^{1/4} \log\e^{-\theta_2}\;,
\eea
which implies, through simple algebra,
\be
(t-s) > \e^{\theta + \theta_3 / 2} / 2 - \e^{1/4+\theta_3/2}\log\e^{-\theta_2} - \e^{1+\theta_3/2}/2\;,
\ee
and hence
\be
(t-s) >\e^{\theta + \theta_3 / 2} /4 \label{eq:constsen}
\ee
if $\theta < 1/4$ and $\e$ is small enough.

\medskip
{\bf (2)} The condition implies that bullet and target are initially (i.e. at time $t$) not too far from each other,
i.e.
\bea
|\xi(t)-x_1| && \leq  |\xi(s) - \xi^{\EE,i}(s)| + |\xi(s)-\xi(t)| + |\xi^{\EE,i}(s)-x_1| \nn\\
&& < \e + 2\e^{-\theta_3/2} (t-s) + 2\e^{1/4} \log\e^{-\theta_2}\nn\\
&& \leq  3\,t\, \e^{-\theta_3/2}\;. \label{eq:LDmr}
\eea

\medskip
Inserting now \eqref{eq:expexpxieis} and using \eqref{eq:defReps}, \eqref{eq:constsen}, the overlap condition assumes the form
\bea
\inf_{s \in (0,t-\e t / R^\e)}\Big|(\xi(s)-\xi(t))+ (\xi(t)-x_1) + v_1(t-s) + \sum_{k=1}^r P^kV_k(t^k-s) -  
{\sum_{k \leq r}}^* \e\,\o^k\Big| < \e \;.
\nn\\\label{eq:conditionvorC}
\eea
Alternatively, using the position variable
\be
X := -v_1 t
\ee
and definition \eqref{eq:eqTUBE}, one has:
\be
\inf_{s \in (0,t-\e t / R^\e)} \frac{t-s}{t}|X - \D(s) |  < \e\;.
\ee
Thus, taking into account the above Remarks {\bf (1)} and {\bf (2)}, conditions
\be
\inf_{s \in (0,t-\e t / R^\e)} |X - \D(s) |  < R^\e
\ee
and \eqref{eq:LDmr} have to be both satisfied.

We conclude that
\bea
&& \int dv_1 \mathbbm{1}_{\mbox{$\{\inf_{s \in (0,t)} |\xi(s) - \xi^{\EE,i}(s)| < \e\}$}}
\, \mathbbm{1}_{\HH^{i}_{1}} \\
&& 
\leq 
\frac{1}{t^3} \left(\int dX \,\mathbbm{1}_{\mbox{$\{\inf_{s \in (0,t-\e t / R^\e)} |X - \D(s) |  < R^\e\}$}}\right)
\, \mathbbm{1}_{|\xi(t)-x_1| \leq  3t \e^{-\theta_3/2}}\;.\nn
\eea
But $\D$ does {\em not} depend on $X$. Therefore, the integral in $dX$ is nothing but the
volume of the region $\T^{\e}_\xi$. \qed

\subsubsection {Volume of $\T^{\e}_\xi$} \label{sec:VTube}

The parametric curve $\D$ inherits its features from the trajectories of the bullet
$\xi^{\EE,i}$ and of the target $\xi$, namely:

-- $\D(s)$ is piecewise smooth, with singularity points in the set $T \cup \{t^1,t^2,\cdots,t^{n^{i}}\}$;

-- at most $n$ singular points $\t_1^*, \t_2^*,\cdots$ are jumps of entity $\e$;

-- all the singular points are finite jumps in the velocity $\D'(s)$.

\begin{figure}[htbp] 
\centering
\includegraphics[width=3in]{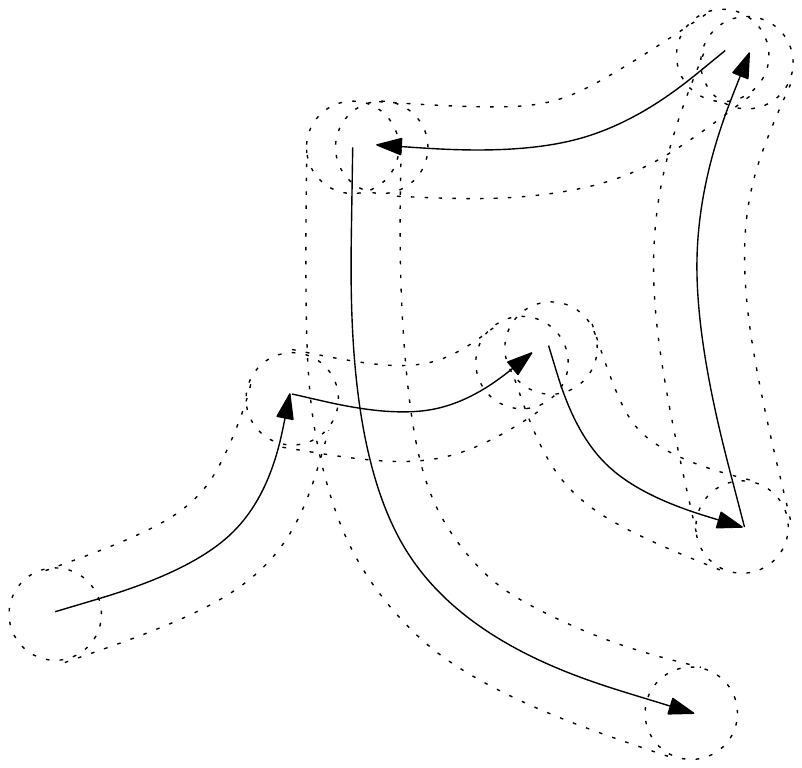} 
\caption{Curve $\D$. The boundary of the region $\T^{\e}_\xi$ is in dotted lines.}
\label{fig:Tube}
\end{figure}
\ni See e.g. Figure \ref{fig:Tube}.

Let $L^\e$ be the length of the curve. If there were no jumps in position, $|\T^{\e}_\xi|$ would be
bounded by $(4\pi /3)(R^\e)^3 + \p (R^\e)^2 L^\e$. In fact, the volume of a tube with a cuspid in $s$
is bounded by the volume of the smooth tube where we put $\D'(s^+)/ |\D'(s^+)|= \D'(s^-)/ |\D'(s^-)|$.
Moreover, observe that $n$ jumps in position produce an error in $|\T^{\e}_\xi|$ which is at most
$(4\pi /3)(R^\e)^3 n$. Therefore,
\be
|\T^{\e}_\xi| \leq \p (R^\e)^2 L^\e + (4\pi /3)(R^\e)^3 (n+1)\;.
\label{eq:boundtubepre}
\ee

Let us give a bound on $L^\e$. Denoting by $\bar\D$ the continuous parametric curve obtained from $\D$ by disregarding
the positional jumps of entity $\e$, we can write
\be
L^\e = \int_0^{t-\e t / R^\e} ds \, |\bar\D'(s)|\;.
\ee
If $s \in (t^{r+1},t^{r})$ and outside singularity points, one has
\be
\bar\D'(s) = \frac{\bar\D(s)}{t-s}
+ \frac{t}{t-s}\left(\eta(s)-\sum_{k=1}^r P^k V_k \right)\;.
\ee
We have now all the ingredients to provide a uniform bound on this, i.e.:

\begin{center}
\ni $(t-s) >\e^{\theta + \theta_3 / 2} /4$;

\medskip

\ni $|\eta| \leq \e^{-\theta_3/2}$;

\medskip

\ni $|\xi(t)-x_1| \leq  3t \e^{-\theta_3}$ (by \eqref{eq:MCSVAF});

\medskip

\ni $\sup_{s\in (0,t)}|\sum_{k=1}^r P^kV_k(t^k-s)| \leq  2t \e^{-\theta_3/2}$ and

\medskip

\ni $|\sum_{k=1}^r P^kV_k| \leq 2\e^{-\theta_3/2}$ (we integrate over the set $\AA$ defined by \eqref{eq:defAA}).
\end{center}

\ni We infer that
\be
|\bar\D(s)| \leq 4t\e^{-\theta - \theta_3 / 2} 
\Big[ \e^{-\theta_3/2} t + 3t \e^{-\theta_3/2} + 2t \e^{-\theta_3/2}
\Big] = 24\, t^2 \e^{-\theta - \theta_3}\;,
\ee
\bea
|\bar\D'(s)| && \leq 4\e^{-\theta - \theta_3 / 2} \, 
24\, t^2 \e^{-\theta - \theta_3}
+ 4t\e^{-\theta - \theta_3 / 2} \left( \e^{-\theta_3/2} + 2\e^{-\theta_3/2}\right)\nn\\
&& = 96\,t^2 \e^{-2\theta - 3\theta_3 / 2}+12\, t\e^{-\theta - \theta_3 }
\eea
and hence
\be
L^\e \leq  C\,  t^3\, \e^{-2\theta - 3\theta_3 / 2}\;,
\label{lenghtcurve}
\ee
where $C>0$ is a pure constant.

Collecting \eqref{eq:boundtubepre}, \eqref{lenghtcurve}, \eqref{eq:defReps} and
$n \leq \e^{-3/4} \log\e^{-\theta_2}$, we finally obtain
\be
|\T^{\e}_\xi| \leq C\, t^5 \, \e^{2-4\theta - (5/2) \theta_3}
\label{eq:boundTexfinal}
\ee
for some pure constant $C>0$.

\subsubsection {Conclusion} \label{sec:VTubeConc}

Equations \eqref{eq:AMmia}, \eqref{eq:nspccm}, \eqref{eq:MCSVAF} and \eqref{eq:boundTexfinal}
lead to
\bea
&& \sum_{\Gamma(1,n_1)} \int d\L \, dv_{1} 
\,\chi^{ov}_{\xi}(\bze^\e) \, F_{\theta_3}(1)
\leq (Dt)^{n_1} \frac{\e^{\g_1}}{2} + C\, t^2 \, \e^{2-4\theta - (5/2) \theta_3} \nn\\
&&\ \ \ \cdot \sum_{\Gamma(1,n_1)}\sum_{i=1}^{n_1+1} \frac{(D't)^{n_1-n^{i}}}{(n_1-n^{i})!}
\int_0^t dt^1\int_0^{t^1} dt^2\cdots \int_0^{t^{n^i-1}} dt^{n^i}
\int d\o^1\cdots d\o^{n^{i}} \nn\\
&&\ \ \ \cdot \int\, dV_1\cdots dV_{n^{i}}\,e^{-(\b/2)\sum_{k}a_k^2 + 
\frac{(\b/2)}{n^i+1}(\sum_{k}a_k)^2}\;.
\label{eq:SCTMIC}
\eea

It remains to compute the Gaussian integral.
This is conveniently done in the variables $a_k$. Note that $V_k \to a_k$,
defined by \eqref{eq:transak}, is a further translation. Then,
\bea
&& \int dV_1\cdots dV_{n^{i}}\,e^{-(\b/2)\sum_{k}a_k^2 + 
\frac{(\b/2)}{n^i+1}(\sum_{k}a_k)^2}\nn\\
&& =  \int da_1\cdots da_{n^{i}}\,e^{-(\b/2)\sum_{k}a_k^2 + 
\frac{(\b/2)}{n^i+1}(\sum_{k}a_k)^2}\nn\\
&& \equiv I_{n^{i}}\;.
\eea
For $n\geq 1$, denoting $S_n = \sum_{k=1}^n a_k $, $Q_n = \sum_{k=1}^n a^2_k$,
\bea
&& I_n = \int da_1\cdots da_{n}\,e^{-(\b/2)Q_n+ 
\frac{(\b/2)}{n+1}S^2_n} \nn\\
&& = \int da_1\cdots da_{n-1}\,e^{-(\b/2)Q_{n-1} + 
\frac{(\b/2)}{n+1}S^2_{n-1}}\, \int da \,
e^{-(\b/2)a^2 + \frac{(\b/2)}{n+1} a^2 + \frac{(\b/2)}{n+1} 2a \cdot S_{n-1}}\nn\\
&& = \int da_1\cdots da_{n-1}\,e^{-(\b/2)Q_{n-1} + 
\frac{(\b/2)}{n+1}S^2_{n-1}}\, \int da \,
e^{-\frac{(\b/2)n}{n+1} \left(a-\frac{S_{n-1}}{n}\right)^2} e^{+\frac{(\b/2)}{n(n+1)} S^2_{n-1}} 
\nn\\
&& = \int da_1\cdots da_{n-1}\,e^{-(\b/2)Q_{n-1} + 
\frac{(\b/2)}{n+1}S^2_{n-1}}\, \left(\frac{2\pi}{\b}\right)^{3/2}\left(\frac{n+1}{n}\right)^{3/2}
e^{+\frac{(\b/2)}{n(n+1)} S^2_{n-1}} 
\nn\\
&& = I_{n-1} \, \left(\frac{2\pi}{\b}\right)^{3/2}\left(\frac{n+1}{n}\right)^{3/2}\;.
\eea
Iterating $n$ times up to $I_0 \equiv 1$, one gets the result
\bea
&& I_n = \left(\frac{2\pi}{\b}\right)^{\frac{3}{2}n}\left(n+1\right)^{3/2}\;.
\eea

Replace this into \eqref{eq:SCTMIC}. Then 
\bea
&& \sum_{\Gamma(1,n_1)} \int d\L \, dv_{1} 
\,\chi^{ov}_{\xi}(\bze^\e) \, F_{\theta_3}(1)
\leq (Dt)^{n_1} \frac{\e^{\g_1}}{2} + C\, t^2 \, \e^{2-4\theta - (5/2) \theta_3} \nn\\
&&\ \ \ \cdot \sum_{\Gamma(1,n_1)}\sum_{i=1}^{n_1+1} \frac{(D't)^{n_1-n^{i}}}{(n_1-n^{i})!}
\frac{(D't)^{n^{i}}}{n^{i}!} (n^{i}+1)^{3/2}\nn\\
&& \leq (Dt)^{n_1} \frac{\e^{\g_1}}{2} + C\, t^2 \, \e^{2-4\theta - (5/2) \theta_3}
(2D' t)^{n_1} \sum_{\Gamma(1,n_1)} \frac{(n_1+1)^{5/2}}{n_1!}\nn\\
&& = (Dt)^{n_1} \frac{\e^{\g_1}}{2} + C\, t^2 \, \e^{2-4\theta - (5/2) \theta_3}
(2D' t)^{n_1}(n_1+1)^{5/2}\;.
\eea
In the first term, $\g_1<1$ arbitrary (from Lemma \ref{lem:MRresINT}). Restrict now 
\be
\g_1 < \min [1,2-4\theta-(5/2)\theta_3]\;.
\label{eq:g1boundGL}
\ee
Lemma \ref{lem:MR} is proved. \qed


\subsection{Completion of the proof of Theorem \ref{thm:MR}} \label{sec:concpro}
\setcounter{equation}{0}    
\def\theequation{4.6.\arabic{equation}}

\subsubsection{Proof of \eqref{eq:repid2thm}} \label{sec:proof2}

We insert $g^\e(t)$ into \eqref{eq:EKthmREP} by writing
\bea
f_{J}^{\e}(t) && = \sum_{ H \subset J} (f_{1}^{\e}(t))^{\otimes H} E_{J\setminus H}(t)\nn\\
&& = \sum_{ H \subset J} (f_{1}^{\e}(t)-g^\e(t)+g^\e(t))^{\otimes H} E_{J\setminus H}(t)\nn\\
&& = \sum_{ H \subset J} (g^{\e}(t))^{\otimes H} E^\EE_{J\setminus H}(t)\;,
\label{eq:fitgEEE}
\eea
where the Enskog error term of order $k$ is
\bea
E^\EE_K(t) && := \sum_{Q \subset K}(f_{1}^{\e}(t)-g^\e(t))^{\otimes Q} E_{K\setminus Q}(t)\nn\\
&& = \sum_{Q \subset K}(E^\EE_1(t))^{\otimes Q} E_{K\setminus Q}(t)\;. \label{eq:balmfinE}
\eea

Resorting to the respective BBGKY and Enskog tree expansions, Eq.s \eqref{eq:reds} and 
\eqref{eq:treeexpErev}--\eqref{eq:treeexpE}, the one--point Enskog error reads
\bea
E^\EE_1(t) && = \sum_{n=0}^\infty \sum _{\G (1,n)} \int d \L 
\left(\prod B^\e f^{\e}_{0,1+n}(\bze^\e(0)) - \prod B^{\EE} g^{\e}_{0,1+n}(\bze^{\EE}(0))\right)\nn\\
&& = \sum_{n=0}^\infty \sum _{\G (1,n)} \int d \L 
\prod B^\e \left(f^{\e}_{0,1+n}(\bze^\e(0)) -g^{\e}_{0,1+n}(\bze^{\e}(0))\right)\nn\\
&&+ \sum_{n=0}^\infty \sum _{\G (1,n)} \int d \L 
\left(\prod B^\e g^{\e}_{0,1+n}(\bze^\e(0)) - \prod B^{\EE} g^{\e}_{0,1+n}(\bze^{\EE}(0))\right)\;,\nn\\
\label{eq:2contCE}
\eea
where $g^{\e}_{0,1+n} = f_0^{\otimes (1+n)}$.

Applying Hypothesis \ref{hyp:cefe0} first and then Hypotheses \ref{hyp:f0} and \ref{hyp:bound},
the rate of convergence of the initial data is 
\bea
&& |f^{\e}_{0,1+n}(\bze^\e(0)) -g^{\e}_{0,1+n}(\bze^{\e}(0))| \nn\\
&& \leq |(f^{\e}_{0,1})^{\otimes (1+n)}(\bze^\e(0)) -g^{\e}_{0,1+n}(\bze^{\e}(0))| + \e^{\g_0} (2z)^{1+n}
e^{-(\b/2)\sum_{i\in S(1)}(\eta^{\e}_1(0))^2} \nn\\
&& \leq 2\,\e^{\g_0} (2z)^{1+n} e^{-(\b/2)\sum_{i\in S(1)}(\eta^{\e}_1(0))^2}
\eea
for $\e$ small and $1+n < \e^{-\a_0}$ (same estimate with no $\e^{\g_0}$ for larger $n$). 
Using this and the estimates of Lemma \ref{lem:STE} and its proof:
\bea
&& \big|\sum_{n=0}^\infty \sum _{\G (1,n)} \int d \L 
\prod B^\e \left(f^{\e}_{0,1+n}(\bze^\e(0)) -g^{\e}_{0,1+n}(\bze^{\e}(0))\right)\Big|\nn\\
&& \leq 2\e^{\g_0} \sum_{n=0}^\infty \sum _{\G (1,n)} \int d \L \prod |B^\e|
(2 z)^{1+n}\   e^{-(\b/2)\sum_{i\in S(1)}(\eta^{\e}_1(0))^2}\nn\\
&&\ \ \  + 2 \sum_{n \geq \e^{-\a_0}}  \sum _{\G (1,n)} \int d \L 
\prod |B^\e| (2 z)^{1+n}\   e^{-(\b/2)\sum_{i\in S(1)}(\eta^{\e}_1(0))^2}\nn\\
&& \leq 3\,\e^{\g_0}\, \bar C \, e^{-(\b/4)v_1^2}
\eea
where $\bar C = \bar C(2z,\b) > 0$ and $t < \bar t$. 

The last term in \eqref{eq:2contCE} is due to the differences among the IBF and the EBF.
Since, in absence of internal recollisions of the IBF and of internal overlaps of the EBF, the two flows coincide,
it holds that
\bea
&&\sum_{n=0}^\infty \sum _{\G (1,n)} \int d \L 
\left(\prod B^\e g^{\e}_{0,1+n}(\bze^\e(0)) - \prod B^{\EE} g^{\e}_{0,1+n}(\bze^{\EE}(0))\right)\nn\\
&& = \sum_{n=0}^\infty \sum _{\G (1,n)} \int d \L 
\left(\prod B^\e \chi^{int} g^{\e}_{0,1+n}(\bze^\e(0)) - \prod B^{\EE} 
\chi^{i.o.} g^{\e}_{0,1+n}(\bze^{\EE}(0))\right)\nn\\
\eea
where $\chi^{int}=\chi^{int}(\bze^\e)$ and $\chi^{i.o.}=\chi^{i.o.}(\bze^\EE)$ are defined
by \eqref{eq:defChiir} and \eqref{eq:defchiio} respectively.
We use, in order, $|g^{\e}_{0,1+n}| \leq2\, (2z)^{1+n} e^{-(\b/2)\sum_{i\in S(1)}(\eta^{\e}_1(0))^2}$, 
Corollary \ref{cor:Btrick}, Lemma \ref{lem:MRresINT} and Eq. \eqref{eq:MRresINTbiss}
to deduce that
\bea
&& \sum_{n=0}^\infty \sum _{\G (1,n)} \int dv_1d \L 
\left(\prod |B^\e| \chi^{int} g^{\e}_{0,1+n}(\bze^\e(0)) + \prod |B^{\EE}| 
\chi^{i.o.} g^{\e}_{0,1+n}(\bze^{\EE}(0))\right)\nn\\
&& \leq \bar C' \,\e^{\theta_1} + \e^{\g_1 - \theta_1}\, 2\, z\,\sum_{n\geq 0 } (2zDt)^n
\eea
for any $\g_1 \in (0,1)$, arbitrary $\theta_1$ and $\e$ small enough. Up to constants
depending on $z,\b$, this is smaller than $\e^{\g_1/2}$ when $t < t^*$.

We conclude that $\int dv |E^{\EE}_1(t)| \leq \e^{\g k}$ for any $\g < \min [\g_0, \g_1 / 2]$
and in particular for the $\g$ appearing in \eqref{eq:EKthm} (remind \eqref{eq:intrg1}).

The final result follows readily from \eqref{eq:balmfinE} and \eqref{eq:EKthm}. \qed

\subsubsection{Proof of \eqref{eq:repid3thm}} \label{sec:proof3}

From Equation \eqref{eq:fitgEEE} one gets
\bea
f_{J}^{\e}(t) &&= \sum_{H \subset J} (f(t))^{\otimes H} E^\BB_{J\setminus H}(t)\;,\nn\\
E^\BB_K(t) &&:= \sum_{Q \subset K}(g^{\e}(t)-f(t))^{\otimes Q} E^{\EE}_{K\setminus Q}(t)\;.
\label{eq:fineeee}
\eea

We resort once again to the tree expansions, Eq.s 
\eqref{eq:treeexpErev}--\eqref{eq:treeexpE} and \eqref{fj}--\eqref{eq:TBf}.
By $g^{\e}_{0,1+n} = f_0^{\otimes (1+n)} = f_{0,1+n}$ and \eqref{eq:equalityB}, we obtain
\bea
g^{\e}(t)-f(t) && = \sum_{n=0}^\infty \sum _{\G (1,n)} \int d \L 
\left(\prod B^{\EE} g^{\e}_{0,1+n}(\bze^{\EE}(0)) - \prod B f_{0,1+n}(\bze(0))\right)\nn\\
&& = \sum_{n=0}^\infty \sum _{\G (1,n)} \int d \L \prod B
\Big[f_{0,1+n}(\bze^{\EE}(0)) - f_{0,1+n}(\bze(0))\Big]\;.
\label{eq:get-ft}
\eea

Using the Lipschitz--regularity assumption on $f_0$, Eq. \eqref{eq:betBbetEE} and
\be
|\xi^\EE_i(0)-\xi_i(0)| \leq n\e
\ee
(as follows from the figure at page \pageref{fig:PSflows}, (iii)-(iv)), one finds
\bea
\Big|f_{0,1+n}(\bze^{\EE}(0)) - f_{0,1+n}(\bze(0))\Big| \leq 
2^{1+n} \left(\max(L, 2z)\right)^{1+n} 
e^{-(\b/2)\sum_{i\in S(1)}(\eta^{\e}_1(0))^2}\, (n\e)\nn\\
\eea
for $\e$ small and $n < \e^{-\a}$ with arbitrary $\a < 1$ (same estimate with no $(n\e)$ for larger $n$).
Inserting into \eqref{eq:get-ft} and by further application of the estimates of Lemma \ref{lem:STE}, one proves
\be
\int dv \,|g^{\e}(t)-f(t)| \leq  \bar C'' \e^{1-\a} \label{eq:diffEB}
\ee
for suitable $\bar C''=\bar C''(z,\b,L)$ and $t<t^*$.
Since $\a$ is here arbitrary, we conclude that this is smaller than $\e^{\g k}$ with $\g$ as in \eqref{eq:EKthm}.

Equation \eqref{eq:repid3thm} follows from \eqref{eq:fineeee} and \eqref{eq:repid2thm}. \qed

\subsection{Convergence of high order fluctuations} \label{sec:proofcorconv}
\setcounter{equation}{0}    
\def\theequation{4.7.\arabic{equation}}

We argue, in this section, on Theorem \ref{cor:MR}. Note preliminarily that, if
the test functions $\varphi_1,\cdots,\varphi_j$ have disjoint supports, then the result
follows immediately from Theorem \ref{thm:MR}. Indeed a simple algebra (see the remark
after \eqref{eq:befRds} below) leads to the identity 
\be
\EEE^\e\left[\prod_{i= 1}^j\Big(F_i(t) - \EEE^\BB[\varphi_i(t)]\Big)\right] 
= \int_{\RRR^{6j}} d\bz_j \, \varphi(z_1)\cdots \varphi(z_j)\,E^\BB_j(\bz_j,t)
\label{eq:ODSB}
\ee
and hence to the result by observing that no $\d-$overlap occurs in the integrand
of the r.h.s. for $\e$ small (so \eqref{eq:repid3thm} can be applied). 
Whenever the $\varphi_i$'s have supports which are not disjoint,
the estimate of the l.h.s. of \eqref{eq:ODSB} complicates considerably.

In the present section, we will work with the extended version of the correlation error $E_k$ over 
$\RRR^{6k}$, defined by
\be
f^{\e}_{J}  =\sum_{K \subset J}   
\left(f^{\e}_{1}\right)^{\otimes K}
\bar E_{J \setminus K}\;,
\ee
where $\bar E_K: \RRR^{6k} \to \RRR$ and \eqref{eq:WenEXC} is used. Since no confusion arises, we shall denote
$\bar E_k = E_k$.
\nomenclature[EzKKba]{$\bar E_{\KK}$}{Extension of $E_{\KK}$ to the whole space}%

\subsubsection{Proof of Theorem \ref{cor:MR}} \label{sec:eq:proofcorconvFLU}

Let us replace, for the moment, $\EEE^\BB[\varphi_i]$ by $\EEE^\e[F_i]$.
Then we compute the fluctuation of order $j$, namely Equation \eqref{eq:FAV}, i.e.
\be
\EEE^\e\left[\prod_{i= 1}^j\Big(F_i(t) - \EEE^\e[F_i(t)]\Big)\right] = \sum_{L \subset J} \left( (-1)^{l}\prod_{i\in L} \EEE^\e[F_i(t)] \right) \EEE^\e\Big[\prod_{i\in J \setminus L} F_i(t)\Big]\;.
\label{eq:fluctEeps}
\ee
We use, again, $l = |L|, j = |J|$ and so on.
From \eqref{def:obs} and by symmetry of the state, 
\bea
&& \EEE^\e[F_i(t)] = \e^2 \sum_{n \geq 0} \frac{1}{n!} \int_{\MM_n} dz_{1}\cdots dz_{n} 
W^{\e}_{n}(\bz_{n},t) \sum_{j=1}^n \varphi_i(z_j)\nn\\
&& \ \ \ \ \ \ \ \ \ \ \ \ = \e^2 \int dz\, \varphi_i(z)\, \r^\e_1(z,t)\nn\\
&& \ \ \ \ \ \ \ \ \ \ \ \ = \int dz\, \varphi_i(z)\, f^\e_1(z,t)\;,
\eea
where we introduced the correlation function \eqref{eq:defcftris}
and its rescaled version \eqref{eq:defrcfhs}. Similarly, for 
$K = \{1,2,\cdots,k\}$,
\bea
&& \EEE^\e\Big[\prod_{i\in K} F_i(t)\Big] = \e^{2k} \,\EEE^\e\Big[ \sum_{j_1,\cdots,j_k} \varphi_1(z_{j_1})
\cdots \varphi_k(z_{j_k})\Big] \nn\\
&& \ \ \ \ \ \ \ \ \ \ \ \ \ \ \ \ \ \ = \e^{2k} \,\EEE^\e\Big[ \sum_{m=1}^k 
\sum_{\substack {P_1, \cdots, P_m  \\ \cup_q P_q=  K\\  |P_q| \geq1\\ P_q \cap P_h= \emptyset, q\neq h}}  
\sum_{\substack{j_1,\cdots,j_m \\ j_q \neq j_h, q \neq h}}  \prod_{q=1}^m \prod_{i \in P_q} \varphi_i (z_{j_q}) \Big] \nn\\
&& \ \ \ \ \ \ \ \ \ \ \ \ \ \ \ \ \ \ = \sum_{m=1}^{k} \e^{2k-2m}
\sum_{\substack {P_1, \cdots, P_m  \\ \cup_q P_q=  K\\  |P_q| \geq1\\ P_q \cap P_h= \emptyset, q\neq h}}  
\int dz_1 \cdots dz_m\, \prod_{q=1}^m \prod_{i \in P_q} \varphi_i (z_q)\, f^\e_m(z_1,\cdots,z_m,t)\;.\nn\\
\label{eq:opiut}
\eea
Observe that, in the evaluation of this integral, the observables $\{\varphi_i\}_{i \in P_q}$ are {\em contracted} in 
the variable $z_q$. We insert the two previous expressions into \eqref{eq:fluctEeps}.
Setting $$\Phi_{P_q} = \prod_{i \in P_q} \varphi_i (z_q)\;,$$ 
we find
\bea
&& \EEE^\e\left[\prod_{i= 1}^j\Big(F_i(t) - \EEE^\e[F_i(t)]\Big)\right] \nn\\
&& = \sum_{L \subset J} \left( (-1)^{l}\prod_{i\in L} \int \varphi_i f^\e_1 \right)
\sum_{m=1}^{j-l} \e^{2j-2l-2m}
\sum_{\substack {P_1, \cdots, P_m  \\ \cup_q P_q=  J \setminus L\\  |P_q| \geq1\\ P_q \cap P_h= \emptyset, q\neq h}}  
\int d\bz_M \, \prod_{q=1}^m \Phi_{P_q}\, f^\e_m(\bz_M,t)\;,\nn\\
\label{eq:mnbvc}
\eea
where $M = \{1,\cdots, m\}$.
Denoting $S = L \cup \{P_i\;; \ |P_i| = 1\}\;,$ 
\bea
&& \EEE^\e\left[\prod_{i= 1}^j\Big(F_i(t) - \EEE^\e[F_i(t)]\Big)\right] 
= \sum_{S \subset J} \sum_{m=1}^{j-s} \e^{2j-2m-2s}
\sum_{\substack {P_1, \cdots, P_m  \\ \cup_q P_q=  J \setminus S\\  |P_q| \geq 2\\ P_q \cap P_h= \emptyset, q\neq h}}  \nn\\
&&\cdot \int d\bz_{M} \, d\bz'_S\,\left(\prod_{i \in S} \varphi_i(z'_{i})\right)\, \left(\prod_{q=1}^m \Phi_{P_q}\right)
\sum_{L \subset S} (-1)^l (f_1^{\e})^{\otimes l}(\bz'_L) f^{\e}_{m+s-l}(\bz_M,\bz'_{S \setminus L})\;.\nn\\
\label{eq:fluctCF}
\eea

Let us write this expression in terms of correlation errors.
For any $S \subset K$, the following algebraic identities hold:
\bea
&& \sum_{L \subset S}(-1)^l (f_1^{\e})^{\otimes L}f^{\e}_{K \setminus L}\nn\\
&&  = 
\sum_{L \subset S}(-1)^l (f_1^{\e})^{\otimes L} \sum_{L' \subset K \setminus S} \d_{L',\emptyset}
(f_1^{\e})^{\otimes L'} f^{\e}_{K \setminus (L \cup L')}
\nn\\
&& = \sum_{L \subset S}(-1)^l (f_1^{\e})^{\otimes L} \sum_{L' \subset K \setminus S} \left( \sum_{L'' \subset L'}(-1)^{l''}\right)
(f_1^{\e})^{\otimes L'} f^{\e}_{K \setminus (L \cup L')}\nn\\
&& = \sum_{L \subset K \setminus S} (f_1^{\e})^{\otimes L} \sum_{L' \subset K \setminus L} (-1)^{l'}
(f_1^{\e})^{\otimes L'} f^{\e}_{K \setminus (L \cup L')}\nn\\
&& = \sum_{L \subset K \setminus S} (f_1^{\e})^{\otimes L} E_{K \setminus L}
\label{eq:AIF}
\eea
where in the last step we used the definition of correlation error, Eq. \eqref{correrr1}, extended in the whole space.
Notice that in the fourth line we just renamed $L\cup L'' \to L'$, $L' \setminus L'' \to L$. Inserting 
\eqref{eq:AIF} into \eqref{eq:fluctCF}, we obtain
\bea
&& \EEE^\e\left[\prod_{i= 1}^j\Big(F_i(t) - \EEE^\e[F_i(t)]\Big)\right] 
= \sum_{S \subset J} \sum_{m=1}^{j-s} \e^{2j-2m-2s}
\sum_{\substack {P_1, \cdots, P_m  \\ \cup_q P_q=  J \setminus S\\  |P_q| \geq 2\\ P_q \cap P_h= \emptyset, q\neq h}}  \nn\\
&&\cdot \int d\bz_{M} \, d\bz'_S\,\left(\prod_{i \in S} \varphi_i(z'_{i})\right)\, \left(\prod_{q=1}^m \Phi_{P_q}\right)
\sum_{L \subset M} (f_1^{\e})^{\otimes l}(\bz'_L) E_{m+s-l}(\bz_M,\bz'_{S \setminus L})\;.\nn\\
\label{eq:befRds}
\eea

\medskip
\ni {\bf Remark (observables with disjoint support)} \label{rem:ODS} Assume that $\supp\varphi_i
\cap \supp\varphi_j = \emptyset$ for $i \neq j$. Then the above algebra becomes trivial,
because no contractions are possible (it must be $z_{j_r} \neq z_{j_s}$ for $r \neq s$)
and the only surviving term in \eqref{eq:opiut} is
$m=k$. Eq.s \eqref{eq:mnbvc} and \eqref{correrr1} lead immediately to 
\be
\EEE^\e\left[\prod_{i= 1}^j\Big(F_i(t) - \EEE^\e[F_i(t)]\Big)\right] 
= \int_{\RRR^{6j}} d\bz_j \, \varphi(z_1)\cdots \varphi(z_j)\,E_j(\bz_j,t)
\label{eq:ODS}
\ee
and the same computation with $f^\e_1$ replaced by $f$ leads to \eqref{eq:ODSB}.

\medskip
Observe that, actually, $2m \leq j-s$. Hence, by Proposition \ref{cor:EKthmOBS} (to be proved below)
\bea
\Big|\EEE^\e\left[\prod_{i= 1}^j\Big(F_i(t) - \EEE^\e[F_i(t)]\Big)\right] \Big|
&& \leq G^j \sum_{S \subset J} \sum_{m=1}^{j-s} \e^{j-s} 
\sum_{\substack {P_1, \cdots, P_m  \\ \cup_q P_q=  J \setminus S\\  |P_q| \geq 2\\ P_q \cap P_h= \emptyset, q\neq h}}  
\sum_{L \subset M}\, 2^l\,  \e^{\g (m+s-l)} \nn\\
&& \leq \e^{\g j} (8GC)^j j! \nn\\
&& \leq \e^{\g' j}
\eea
for any $j < \e^{-\a'}$, $t<t^*$, $\g' < \g - \a'$ and $\e$ small enough (having bounded the sum over partitions as in \eqref{R1}).

On the other hand, $|\EEE^\e[F_i(t)] - \EEE^\BB[\varphi_i(t)]| = |\int \varphi_i E_1^\BB| \leq G \e^\g$ by Theorem 
\ref{thm:MR}. Therefore, slightly decreasing $\a'$, 
we conclude that
\bea
&&\sup_{j < \e^{-\a'}}\,
\Big|\EEE^\e\left[\prod_{i= 1}^j\Big(F_i(t) - \EEE^\BB[\varphi_i(t)]\Big)\right]\Big| \nn\\
&& \leq \sup_{j < \e^{-\a'}}\,\sum_{L \subset J } \Big|\EEE^\e\left[\prod_{i \in L}\Big(F_i(t) - \EEE^\e[F_i(t)]\Big)\right] \Big|
\; (G \e^\g)^{j-l} \nn\\
&& \leq \sup_{j < \e^{-\a'}}\,\sum_{L \subset J } \e^{\g' l} (G \e^\g)^{j-l}\;,
\eea
which goes to zero as a power of $\e$. Theorem \ref{cor:MR} is proved. \qed

\medskip
In the proof of Theorem \ref{cor:MR} we had to estimate the quantity 
\be
\int_{\RRR^{6k}} d\bz_k \, \varphi(z_1)\cdots \varphi(z_k)\,E_k(\bz_k,t)\;.
\label{eq:pfdr}
\ee
Note that, by Theorem, \ref{thm:MR} we control $E_k$ only in the region 
$$\MM^x_k(\d) = \{\bx_k \in \RRR^{3k},\ \ |x_i-x_j| > \d,\ i\neq j\}\;.$$
Suppose now that $\bz_k = (\bz_{Q'},\bz_{K \setminus Q'})$, where 
\be
\bx_{Q'}\in \MM^x_{q'}(\d) \cap \Big\{\bx_{Q'} \in \RRR^{3q'},\ \ 
\min_{\substack{i\in Q' \\ j \in K \setminus Q'}} |x_i-x_j| > \d \Big\}
\label{eq:choicexQ'}
\ee
and $\bx_{K \setminus Q'}\in \RRR^{3(k-q')}$ is some configuration lying in a small
measure set with overlaps at distance $\d$.
Then we cannot estimate brutally $|E_k|$ by $(const.)^k$, but we need to recover a small error
$\e^{\g q'}$, relative to the non--overlapping configurations.
That is, we need a natural improvement of Theorem \ref{thm:MR} including the 
case in which $\d-$overlaps are admitted for a subset of $\bx_k$. This is expressed by the following
corollary, whose proof is deferred to Appendix E.

\begin{cor} \label{cor:MRpre}
Under the assumptions of Theorem \ref{thm:MR}, let $Q' \subset K$.
Then there exists a positive constant $C_2 = C_2(z,\b)$ such that, for any $t< t^*$ and $\e$ small enough, 
 \be
\int_{\RRR^{3q'}} d\bv_{Q'} |E_K(t)| \leq C_2^k\,\left(\e^{\g k} +  \e^{\g'_1 q'}\, \e^{-\a_1 k}\right)
\ \ \ \ \ \ \ \ \ \mbox{$\forall$ $k < \e^{-\a}$}\;,  \label{eq:EKthmEXT}
\ee
with $\g'_1 = \min[\g_0,\g_1/2]$, $\a_1 = \theta_1 + 3\a$,
$\bx_{K \setminus Q'} \in \RRR^{3(k-q')}$ and $\bx_{Q'}$ as in \eqref{eq:choicexQ'}.
\end{cor}
The parameters $\g,\g_0,\theta_1, \a, \d $ and $\g_1$ are listed in sections \ref{sec:REF}--\ref{subsub:MRE}.

By using the above corollary we achieve next the proof of Proposition \ref{cor:EKthmOBS}. 

\subsubsection{Proof of Proposition \ref{cor:EKthmOBS}} \label{sec:eq:EKthmOBS}

Let us fix $\d = \e^\theta$ with $\theta$ in the interval
\be
\theta\in (3/14,1/4)\;.
\ee
Furthermore, let
\be
1=\sum_{Q\subset K} \, \chi_{Q}^{\d} \,\, \bar \chi _{K\setminus Q,K}^{\d}\;,
\ee
where $\chi^{\d}_{Q} =1$ if and only if any particle with index in $Q$ ``$\d-$overlaps''
with a different particle in $Q$, and $ \bar \chi _{K\setminus Q,K} ^{0}=1$ if and only if all the 
particles in $K\setminus Q$ lie at distance strictly larger than $\d$ from any other particle
in $K$.

Inserting the partition of unity, \eqref{eq:pfdr} becomes 
$$
\sum_{Q\subset K} \int d\bz_Q \, \chi_{Q}^{\d} \, \left(\prod_{i\in Q}\varphi_i(z_i)\right)
 \,  \int d\bz_{K \setminus Q}\,\left(\prod_{i\in K \setminus Q}\varphi_i(z_i)\right)\,
 \bar \chi _{K\setminus Q,K}^{\d}\,E_K(\bz_k,t) \;.
 $$
For $Q = \emptyset$ we can apply the main theorem, while, for $|Q|\geq 2$, we resort
to Corollary \ref{cor:MRpre}.
Taking the supremum over velocities of the test functions, one obtains
\bea
&& \Big| \int_{\RRR^{6k}} d\bz_k \, \varphi(z_1)\cdots \varphi(z_k)\,E_K(\bz_k,t) \Big| \nn\\
&&\leq G^k \e^{\g k} + \sum_{\substack{Q\subset K \\ |Q| > 1}} G^{k-q} 
C_2^k\,\left(\e^{\g k} +  \e^{\min[\g_0,\g_1/2] (k-q)}\, \e^{-\theta_1k-3\a k}\right)
\int d\bz_Q \, \chi_{Q}^{\d} \, \left(\prod_{i\in Q}|\varphi_i(z_i)|\right)\nn\\
&& \leq G^k \e^{\g k}+G^{k-1}C_2^k  \sum_{\substack{Q\subset K \\ |Q| > 1}} 
\,\left(\e^{\g k} +  \e^{\min[\g_0,\g_1/2] (k-q)}\, \e^{-\theta_1k-3\a k}\right)
 \int dz_{i_*} \,\varphi_{i_*} \,\int d\bz_{Q \setminus \{i_*\}}\chi_{Q}^{\d}\nn\\
&&  \leq G^k \e^{\g k}+(GC_2)^k  \sum_{\substack{Q\subset K \\ |Q| > 1}}
\,\left(\e^{\g k} +  \e^{\min[\g_0,\g_1/2] (k-q)}\, \e^{-\theta_1k-3\a k}\right)
\, (q-1)! (4\p\d^3 /3)^{q-1}\nn\\
&& \leq G^k \e^{\g k}+(GC_2 4\pi/3)^k \,\sum_{\substack{Q\subset K \\ |Q| > 1}}\,
\left(\e^{\g k} +  \e^{\min[\g_0,\g_1/2] (k-q)}\, \e^{-\theta_1k-3\a k}\right)
\e^{(3\theta- \a')(q-1)}\;,\nn\\
\label{eq:TestF}
\eea
where $i_*$ is an arbitrary element in $Q$.
In the last step, we used $k < \e^{-\a'}$. 

We choose 
\be
\a' < 7\theta-3/2\;.
\ee
Then, by \eqref{eq:g1boundGL}, 
we find
$3 \theta - \a'  -\min[\g_0,\g_1/2] > 3\theta - 7 \theta
+ 3/2 - 1/2 = -4\theta + 1>0$. In particular, using again \eqref{eq:g1boundGL}
and reminding that $\theta_3 = 1/5$, $q \geq 2$,
$$\left(3 \theta - \a'  -\min[\g_0,\g_1/2]\right) q -3\theta + \a'
\geq 6\theta - 2\a' - 2 +4\theta + (1/2) - 3\theta + \a' = 7\theta - \a' - 3/2 >0\;.$$
Therefore, the $q-$dependent factors in \eqref{eq:TestF} are very small.
The final result follows then from Eq. \eqref{eq:intrg1} for $\e$ small enough.
\qed

\subsection{Concluding remarks} \label{sec:concrem}
\setcounter{equation}{0}    
\def\theequation{4.8.\arabic{equation}}

\subsubsection{Truncated functions}

In this paper we studied the kinetic theory of expansion \eqref{correrr} for a dilute
gas of hard spheres. Similar cumulant expansions within the framework of kinetic
theory have been considered in \cite{GG12,BGSR15,LM15,LMN16}.
Moreover, they are very familiar in statistical mechanics. 
The standard example is given by the Ursell functions in the classical analysis of the equilibrium state 
in a gas at low (finite) density, chapter 4.4 of \cite{Ru69}. Typically, one expands in truncated functions 
\be
f^{\e}_J = \sum_{1\leq m \leq j}
\sum_{\substack {P_1, \cdots, P_m  \\ \cup_q P_q=  J\\ P_q \cap P_h= \emptyset, q\neq h}} 
\prod_i f^{\e, T}_{P_i}
\label{eq:TF}
\ee
and focuses on the decay properties of $f^{\e, T}_{j}$ and on related physical quantities.
In connection to this, combinatorial methods have been intensively studied under the 
generic name of ``cluster expansion'', see e.g. \cite{Ko06}.

Note that \eqref{eq:TF} defines implicitly the truncated functions, as we did for the correlation
errors in Eq.s \eqref{eq:exampleDEF}-\eqref{correrr}. 
The difference is that in \eqref{eq:TF} the sum runs over {\em all} partitions of $j$ elements. 
A direct comparison with \eqref{correrr} yields:
\be
E_J = \sum_{1\leq m \leq j/2}
\sum_{\substack {P_1, \cdots, P_m  \\ \cup_q P_q=  J\\ P_q \cap P_h= \emptyset, q\neq h
\\ |P_q| \geq 2}}  \prod_i f^{\e, T}_{P_i}\;.
\label{eq:EjTF}
\ee
In other words, $E_J$ is a ``partially truncated correlation function'' with respect to 
clusters of size at least $2$. 

The interpretation of \eqref{eq:EjTF} is now quite transparent.
The $f^{\e,T}_j$ measures events of $j$ {\em maximally
correlated} particles with at least $j-1$ recollisions connecting all the particles.
The $E_j$ measures events of $j$ {\em minimally correlated} particles 
with at least $\lceil j/2 \rceil$ recollisions, i.e. just one per particle. If $\e^{\g_1}$
is the size of one single recollision, then we expect $|f^{\e,T}_j| = ~ O (\e^{\g_1(j-1)})$
and $|E_j| = O (\e^{\g_1 (j/2)})$.

During the revision of a first version of the present paper\footnote{{\em arXiv:1405.4676}, 2014.}, 
it appeared a preprint by Bodineau, Gallagher, Saint--Raymond with a derivation  
of the linearized Boltzmann equation for the hard sphere gas at equilibrium, global in time \cite{BGSR15}.
Here a similar notion of cumulant expansion is introduced and the control of truncated
functions seems to be crucial to reach arbitrary times.
The combinatorial problem and the estimates of multiple recollisions are however different in this context.

\subsubsection{Time of validity}  \label{subsec:time}

Note that we did not optimize the time interval $(0,t^*)$ for which the main theorem holds.
In fact, in Section \ref{sec:proof5} we used $t^* < (eC_1)^{-1}$ which can be strictly smaller
than the value $\bar t$, appearing in Proposition \ref{prop:STE} and ensuring Lanford's 
validity result. It is easy to extend our result up to $\bar t$ by paying the price of worst 
values of $\g,\a$. It is enough to notice that in Eq. \eqref{eq:almconc} we disregarded the 
truncation on $n$, i.e. $\sum_{n=0}^{\log\e^{-\theta_2 k}}$ (see Lemma \ref{prop3}).
Substituting $t^*$ by $\bar t$ in \eqref{eq:almconc} and using
$$\sum_{n=0}^{\log\e^{-\theta_2 k}}(e C_1 \bar t)^n \leq \e^{-\theta_2 \log(\bar C_1)\, k}$$
for $\bar C_1 > \max(1,e C_1 \bar t)$, one obtains that condition \eqref{eq:intrg1} is replaced by
\be
\g < \min[\g_0,\g_1/2] - \theta_1-3\a - \theta_2 \log(\bar C_1)\;.
\ee
The final result follows for a different choice of the cutoff parameters.

\subsubsection{Canonical ensemble}  \label{subsec:canonical}

The definition of ``state'' used in this paper includes the canonical 
$W^{\e}_{0,n} = 0$ for all $n \neq N$, $N \sim \e^{-2}$.
Let us focus now on the BBGKY solution, Eq. \eqref{eq:reds}, in the case of a state of this form.
Even if we ignore the dynamical correlations (see Remark page \pageref{rem:PF}) and assume
$W^{\e}_{0,N} = (w^{\e})^{\otimes N}$, the formula does not exhibit complete factorization.
The reason is twofold:

1. we work with r.c.f. $f^{\e}_j = \e^{2j} N (N-1) \cdots (N-j+1)\, (w^{\e})^{\otimes j}$;

2. an additional correlation is present, given by the constraint  $\sum n_i \leq N-j$. 

\ni The main advantage of using a grand canonical formalism is to get rid of these extra correlations.

Observe that the above effects have nothing to do with the dynamics and are uniquely determined
by the special structure of the initial data. Actually our main result does cover a canonical state 
obeying the assumptions. However we have not verified, in a canonical case, Hypothesis \ref{hyp:cefe0},
for which a more elaborate expansion than \eqref{eq:expID} seems to be necessary.

\subsubsection{Spatial domain}  \label{subsec:BOX}

Our results have been established in the whole space $\RRR^3$. A
natural question is how to extend the analysis to the case of a region $\L\subset \RRR^3$
with prescribed boundary conditions. We discuss the major points in what follows.

Assumptions on the boundary conditions ensuring existence and 
uniqueness of the $n-$particle flow have been studied in previous literature, e.g. 
\cite{Ale75,MPPP76,CIP94}. Once the flow is well defined, the setting and the hierarchical formulas of
sections \ref{sec:AS}-\ref{sec:hs} can be easily adapted, see for instance \cite{BLLS80,S81,Sp83}.
Note that, in the case of a bounded domain, all the sums over $n$ (number of particles)
become finite, both in the definition of correlation functions and in the related tree expansions
(see also the states considered in Appendix A).
Indeed, due to the hard sphere exclusion, $W^\e_n = 0$ for $n> N_{cp} = \,$ close--packing number.

However this does not produce any change in the combinatorics of Step 1.
The graph expansion (Eq. \eqref{eq:AL45}) is applied, as written, to \eqref{eq:papertp} even when
$n > N_{cp}$ (i.e., to zero terms). This produces non--zero error terms with overlapping trees and 
total number of created particles larger than $N_{cp}$. 
Such terms are the ``close--packing correlation'' which is therefore automatically taken into account by our method.
Since $N_{cp} \sim \e^{-3}$ and
$\a$ is certainly much smaller than $3$, this correlation is just a part of the first error term in Lemma 
\ref{prop3}, related to the cutoff $\theta_2$ (truncation on the number of creations in a collection of trees).

An extra difficulty comes from the geometrical estimates of recollisions, Step 3 of the proof.
The case of a vessel of arbitrary geometry with reflecting and/or diffusive boundary conditions eludes our techniques. 
On the other hand, the analysis of this paper can be easily adapted to some simple situation as the case of a gas contained in a parallelepiped 
with periodic or reflecting boundary conditions. Let us discuss this point
 in some more detail\footnote{A discussion 
similar to the one that follows appears in \cite{EP10, BGSR13}.}.
 
Consider the gas confined in
$
\Lam_0= (0,L_1) \times (0,L_2) \times (0,L_3)
$, $L_i>0$
and assume periodic boundary conditions for the free flow.
After a moment of thought, one realizes that
the overlap condition appearing in Lemma \ref{lem:MR}, i.e. $\inf_s |\xi(s) - \xi^{\e}_k(s)|<\e$, 
can be represented as follows:
\be
\mathbbm{1}(\inf_s |\xi(s) - \xi^{\e}_k(s)|<\e) \leq \sum_{\bm m \in \Z^3} \mathbbm{1}(\inf_s |\xi(\bm m; s) - 
\xi^{\e}_k(\RRR^3; s)|<\e)
\ee
where $\bm m = (m_1,m_2,m_3)$,
\be
\xi(\bm m; s) = \xi(\RRR^3; s)+(m_1L_1,m_2L_2,m_3L_3)
\label{eq:PCflow}
\ee
and $\xi(\RRR^3; s), \xi^{\e}_k(\RRR^3; s)$ are the trajectories in $\RRR^3$ computed with {\em no} 
boundary conditions.
In other words, if the bullet $k$ hits the target $\xi$ in the torus $\L_0$, then the bullet $k$ moving
in the whole space hits some periodic copy of $\xi$, also moving in the whole space.

Since the time is finite and the velocities are bounded, it follows that there is no serious modification
to the estimates of this paper. Reflecting boundary conditions are treated in the same way, but the periodic
translations in \eqref{eq:PCflow} are replaced by reflections.


\begin{appendices}
\renewcommand\thetable{\thesection\arabic{table}}
  \renewcommand\thefigure{\thesection\arabic{figure}}
  
 \section*{Appendices}

 \addcontentsline{toc}{subsection}{A \ \ \ Chaotic states of hard spheres}
 \subsection*{A \ \ \ Chaotic states of hard spheres} \label{sec:initial}
 \setcounter{equation}{0}    
 \def\theequation{A.\arabic{equation}}

We consider here the most natural construction of hard sphere states which factorize in the Boltzmann--Grad limit,
and show that they satisfy the hypotheses stated in Section~\ref{sec:AS2}.

Let $\bW^{\e}_{0}$ be the grand canonical state over $\MM$ with system of densities 
\be
\frac{1}{n!}W^{\e}_{0,n}(\bz_n)=\frac{1}{\ZZ_\e} \frac{e^{-\m_{\e}}\m_{\e}^n}{n!}
f^{\otimes n}_0(\bz_n) \;,
\label{eq:exampleWn0}
\ee
where $\e^2\m_\e = 1$,
\be
\ZZ_\e=\sum_{n\geq 0} \frac{e^{-\m_{\e}}\m_{\e}^n}{n!}\ZZ_{n}^{can}\;,
\label{eq:APPAa2}
\ee
and where the ``canonical'' normalization constant is
\be
\ZZ_{n}^{can} = \int_{\MM_n}d\bz_n \,f^{\otimes n}_0(\bz_n)
= \int_{\RRR^{6n}} d\bz_n \,f^{\otimes n}_0(\bz_n)\prod_{1\leq i < k \leq n}\bar\chi^0_{i,k}\;,
\ee
($\ZZ_0^{can}=1$) with $ \bar \chi^0_{i,k}$ the indicator function of the set $\{ |x_i -x_k| > \e \}$.
The function $f_0$ can be any probability density over $\RRR^6$ satisfying $f_0(x,v) \leq (h(x)/2)
e^{-(\b/2)v^2}$, for some $h \in L^1(\RRR^3;\RRR^+)$ with $\esssup_x h(x) = z$, and $z,\b>0$.

\medskip
\ni {\bf Remarks.} 

\ni -- The state introduced is a ``maximally factorized state'' in the sense that the only correlations are due to the
hard sphere exclusion. A Gibbs state in equilibrium statistical mechanics is of this form. 

\ni -- The probability of finding 
$n$ particles is $p_n = \ZZ^{can}_n\ZZ_{\e}^{-1} (1/n!)e^{-\m_{\e}}\m_{\e}^n$
and the distribution of the $n$ particles $(\ZZ^{can}_n)^{-1}f_0^{\otimes n}$. The factor multiplying the 
Poissonian distribution measures the probability of having $n$ non--overlapping spheres. 

\ni -- The asymptotic behaviour of the normalization constants can be proved to be
$\ZZ_{n}^{can} \sim e^{-Cn^2\e^3}$ ($n >> \e^{-2}$, $C>0$) 
and $\ZZ_{\e} \sim e^{-C\e^{-1}}$ (see e.g. \cite {PSS13}).

\bigskip

\ni{\bf Proposition A.1}\;\; {\it The state of the system defined by \eqref{eq:exampleWn0} admits r.c.f. satisfying 
Hypotheses \ref{hyp:bound}, \ref{hyp:cefe0} and \ref{hyp:f0}.} 

\bigskip

\ni {\bf Proof.} For this particular state, it is convenient to check \eqref{eq:sbo}--\eqref{eq:E'K} first.

By \eqref{eq:defcftris} and \eqref{eq:defrcfhs},
the rescaled correlation functions are
\be
f_{0,j}^{\e}(\bz_j) = \frac{F^\e(\bz_j)}{\ZZ_\e} f^{\otimes j}_0(\bz_j)\;,
\ee
where
\be
F^\e(\bz_j)=\sum_{n\geq 0}\, \frac{e^{-\m_{\e}}\m_{\e}^n}{n!}\, F^{j+n}_{can}(\bz_j)
\ee
and
\be
F^{j+n}_{can}(\bz_j) = \int_{\RRR^{6n}} d\bz_{j,n} 
f^{\otimes n}_0(\bz_{j,n})\left(\prod_{i=1}^j\prod_{k=j+1}^{j+n}\ \bar\chi^0_{i,k}\right)
\left(\,\prod_{j+1\leq i < k \leq {j+n}}\bar\chi^0_{i,k}\right) \label{eq:defFjncan}
\ee
($F^{j}_{can}(\bz_j)=1$).

For any  $j,n \geq 1$,
we rewrite $F^{j+n}_{can}(\bz_j)$ by using
\be
\prod_{i=1}^j\prod_{k=j+1}^{j+n}\ \bar\chi^0 _{i,k}=\prod_{i=1}^j (1- \chi^0_{i,J^c})
\label{eq:expID}
\ee
where $J^c = \{j+1,\cdots,j+n\}$ and
\bea
&& \chi^0_{i,J^c}=\left(1-\prod_{k=j+1}^{j+n} \bar \chi^0_{i,k}\right)\nn\\
&&\ \ \ \ \ \  =\mathbbm{1}_{\mbox{$ \{ {\bf z}_{j+n}\ |\  \exists\, k \in J^c\ $ such that $\ |x_i-x_k| \leq \e \}$}}\;.\nn
\eea

Expanding the product in \eqref{eq:expID}, we find
\be
\prod_{i=1}^j\prod_{k=j+1}^{j+n}\ \bar\chi^0_{i,k}
= \sum_{K \subset J}(-1)^k \chi^0_{K,J^c}\;, \label{eq:expDI}
\ee
with $$\chi^0_{K,J^c}=\prod_{i\in K}  \chi^0_{i,J^c}\;.$$

Inserting \eqref{eq:expDI} into \eqref{eq:defFjncan}, we arrive to
\be
f_{0,j}^{\e}(\bz_j) = \sum_{L \subset J}f^{\otimes L}_0(\bz_L) {E}^{\BB,0}_{J \setminus L}(\bz_{J \setminus L})\;,
\ee
where ${E}^{\BB,0}_\emptyset = 1$ and, for $k\geq 1$, 
\bea
{E}^{\BB,0}_K(\bz_k) = (-f_0)^{\otimes k}(\bz_k) \frac{1}{\ZZ_\e}
\sum_{n\geq 1} \frac{e^{-\m_{\e}}\m_{\e}^n}{n!} \int d\bz_{k,n}\,f_0^{\otimes n}(\bz_{k,n})
\ \chi^0_{K,K^c} \prod_{k+1\leq i < h \leq {k+n}}\bar\chi^0_{i,h}\;.\nn\\
\label{eq:E'0Kexex}
\eea

Let $a$ be the maximum number
of three--dimensional hard spheres that can be simultaneously overlapped by a single one, and $q = q(\bx_k)$
the minimum number of different spheres in $K^c$ necessary to satisfy the condition $\chi^0_{K,K^c}=1$
({\em any} sphere in $K$ is overlapped by at least one sphere in $K^c$).
Then $k/a \leq q \leq k$ and
\be
\chi^0_{K,K^c} \leq  \sum_{\substack{Q \subset K^c \\ |Q| = q}}
\chi^0_{Q,K}\;.
\ee
It follows that
\bea
&& |{E}^{\BB,0}_K(\bz_k)| \leq \ \left(z/2\right)^{k} e^{-(\b/2)\sum_{i \in K} v_i^2}\, \frac{1}{\ZZ_\e}
\, \sum_{n\geq q}\, \frac{e^{-\m_{\e}}\m_{\e}^n}{n!} \nn\\
&&\ \ \ \ \ \ \ \ \ \ \ \ \ \ \ \sum_{\substack{Q \subset K^c \\ |Q| = q}}\,\int d\bz_{k,n}\,f_0^{\otimes n}(\bz_{k,n})
\,\chi^0_{Q,K}\, \prod_{k+1\leq i < h \leq {k+n}}\bar\chi^0_{i,h}\;.
\eea

Note now that $\chi^0_{Q,K} = \prod_{i\in Q}\chi^0_{i,K}$ and, for all $i\in Q$,
\be
\int dz_{i}\,f_0(z_{i})\, \chi^0_{i,K} \leq (z/2) (2\p/\b)^{3/2} \,k\, B\,\e^{3}
\ee
where $B$ is the volume of the unit ball.
The remaining $n-q$ integration variables reconstruct $\ZZ^{can}_{n-q}$, so that we get
\be
|{E}^{\BB,0}_K(\bz_k)| \leq \ \left(z/2\right)^{k} 
e^{-(\b/2)\sum_{i \in K} v_i^2} \frac{1}{\ZZ_\e} \ \sum_{n\geq q} \,
\frac{e^{-\m_{\e}}\m_{\e}^n}{n!}\,
\binom{n}{q}\,k^q\,(C \e^3)^q \, \ZZ^{can}_{n-q}\;.
\ee
Here and below we indicate by $C$ a positive constant, possibly changing from line to line and depending
on $z,\b,a,B$.

Using $\frac{1}{n!}\binom{n}{q}k^q \leq \frac{(ke)^q}{q^q(n-q)! }\leq C^q / (n-q)!$ and reminding \eqref{eq:APPAa2},
we deduce
\bea
|{E}^{\BB,0}_K(\bz_k)| && \leq \ \left(z/2\right)^{k} 
e^{-(\b/2)\sum_{i \in K} v_i^2} (C \e)^q \nn\\
&& \leq z^{k} 
e^{-(\b/2)\sum_{i \in K} v_i^2}\, C^k\, \e^{k/a}\;.
\eea
This implies the estimate in \eqref{eq:E'K} by choosing $\g'_0 < 1/a$ and $\e$ small enough\footnote{The bad 
value of $\g'_0$ is due to the 
uniform estimate in $\MM_k(\e)$, which includes situations similar to close--packing. If the mutual
distance between the particles in $K$ is order 1, then $q=k$ and the above computation gives
$\g'_0 < 1$.}.

Hypotheses \ref{hyp:bound} and \ref{hyp:f0} follow immediately.

Finally, observe that Hypothesis \ref{hyp:cefe0} and 
\eqref{eq:sbo}--\eqref{eq:E'K} are equivalent. 
Indeed, starting from \eqref{eq:sbo}, 
setting $ f_0^{\otimes H} = (f^\e _{0,1}-E_1^{\BB,0})^{\otimes H} $ and expanding, one finds formula
\eqref{eq:repid1} with 
\be
E^0_{K} = \sum_{Q \subset K} (-1)^q (E_1^{\BB,0})^{\otimes Q} E_{K \setminus Q}^{\BB,0}\;,
\ee
hence \eqref{eq:E'K} implies $|E_K^0| \leq 2^k \e^{\g'_0 k}z^k\ e^{-(\b/2)\sum_{i \in K} v_i^2} < 
\e^{\g_0 k}z^k\ e^{-(\b/2)\sum_{i \in K} v_i^2}$ for any $\g_0<\g'_0$ (and $\e$ small).
The proof of the inverse statement is similar (one finds $\g'_0<\g_0$).
\qed

\bigskip
We conclude this appendix with the proof of the properties presented in Section \ref{subsec: AS2''}.

\bigskip
\ni {\bf Proof of Property 1.}
Hypothesis \ref{hyp:cefe0} is obtained from Property 1 in the case $S = \JJ = J$.

Let us show that Hypothesis \ref{hyp:cefe0} implies \eqref{eq:repid1intest} for a generic partition
of the set $S$.
Inverting \eqref{eq:repid1int} we find
\be
E^0_{\KK} =\sum_{\QQ \subset \KK}  (-1)^{|\QQ|} 
\left(\prod_{S \in \QQ}f^{\e}_{0,S} \right) 
f^{\e}_{0,K\setminus Q}\;. \label{eq:defE0PPK}
\ee
We use the notation $K = \cup_{i\in \KK}S_i$, $Q = \cup_{i\in \QQ}S_i$ etc..
By using \eqref{eq:repid1}, it follows that
\be
E^0_{\KK} =\sum_{\QQ \subset \KK} (-1)^{|\QQ|} 
\sum_ {\substack {L_1, \cdots, L_{|\QQ|} \\ L_r \subset S_{i_r}}} \prod_{r=1}^{|\QQ|} E_{L_r}^0 
\sum_{L_0 \subset K \setminus Q} E^0_{L_0}
 \left(f^{\e}_{0,1}\right)^{\otimes L^c}\;,
\label{eq:repid1int3}
\ee
where $i_1, \cdots, i_{|\QQ|}$ are the indices of the clusters in $\QQ$, and 
$L^c = K \setminus \cup_{r=0}^{|\QQ|}L_r$.
Note that the first sum is over subsets of clusters, while the other sums run over subsets of indices of particles.
Setting  $L = \cup_{r=0}^{|\QQ|}L_r$ we notice that, for given $\QQ$ and $L$, one has $L_r = L \cap S_{i_r}$
and $L_0 = L \cap (K \setminus Q)$. Therefore we rewrite \eqref{eq:repid1int3} as
\be
E^0_{\KK} =\sum_{L \subset K} \left(f^{\e}_{0,1}\right)^{\otimes K \setminus L}
\sum_{\QQ \subset \KK} (-1)^{|\QQ|} E^0_{L \cap (K \setminus Q)}
\prod_{r=1}^{|\QQ|} E_{L \cap S_{i_r}}^0 \;.
\label{eq:repid1int4}
\ee

Now observe that, in the above sum, $L$ must be such that $|L\cap S_{i} |> 0$ for all $i \in \KK$.
Otherwise if $L\cap S_{i} =\emptyset$ for some $i$, setting $S^*=S_{i}$, since $E_{L \cap S_{i}}^0=1$,
\bea
&& \sum_{\substack{\QQ \subset \KK \\ S^* \in \QQ}} 
(-1)^{|\QQ|} E^0_{L \cap (K \setminus Q)}
\prod_{r=1}^{|\QQ|} E_{L \cap S_{i_r}}^0
+\sum_{\substack{\QQ \subset \KK \\ S^* \notin \QQ}} 
(-1)^{|\QQ|} E^0_{L \cap (K \setminus Q)}
\prod_{r=1}^{|\QQ|} E_{L \cap S_{i_r}}^0 \nn\\
&& = - \sum_{\QQ \subset \KK \setminus \{S^*\}} 
(-1)^{|\QQ|} E^0_{L \cap (K \setminus Q)}
\prod_{r=1}^{|\QQ|} E_{L \cap S_{i_r}}^0
+ \sum_{\QQ \subset \KK \setminus \{S^*\}} 
(-1)^{|\QQ|} E^0_{L \cap (K \setminus Q)}
\prod_{r=1}^{|\QQ|} E_{L \cap S_{i_r}}^0 \nn\\
&& = 0\;.
\eea
As a consequence, using \eqref{eq:assbound} and \eqref{eq:EK} in \eqref{eq:repid1int4},
we deduce
\bea
|E^0_{\KK}| && \leq 2^{|\KK|}\,z^k\ e^{-(\b/2)\sum_{i \in K} v_i^2}
\sum_{\substack{L \subset K \\ L \cap S_i \neq \emptyset\ \forall i}}\e^{\g_0 |L|}\nn\\
&& \leq 4^k \e^{\g_0 |\KK|}\, z^k\ e^{-(\b/2)\sum_{i \in K} v_i^2}\;,
\eea
so that \eqref{eq:repid1intest} follows by reducing the values of $\g_0,\a_0$.
\qed

\bigskip
\ni {\bf Proof of Property 2.}
We rewrite \eqref{expexten} as
\be
f^{\e}_{0,S} = \sum_{\HH \subset \JJ}   
\bar \chi_{\HH,\JJ}^0\left(\prod_{i \in \HH}\bar \chi_{S_i}^0 f^{\e}_{0,S_i}\right) 
\bar \chi_{J \setminus H}^0 E^0_{\JJ\setminus \HH}
 \label{eq:appAp2}
\ee
where $\bar \chi^0_{\HH,\JJ}=1$  if and only if all the particles in $S_i$
do not overlap with any other particle in $S_k$ for any choice of $i\in\HH, k \in \JJ$, $k\neq i$.

By virtue of  Lemma \ref{lem:CEoG} we expand
\be
\label{expscript}
\bar \chi_{\HH,\JJ}^0 =\sum_{ \QQ \subset \HH} R(\QQ,\JJ\setminus \HH)
\ee
and then we get
\be
|R(\QQ,\JJ\setminus \HH)| \leq C^{|\QQ|}\, |\QQ|! \,\chi^0_{\QQ,\QQ \cup (\JJ \setminus \HH)}\;.
\label{eq:boundappA}
\ee

Inserting \eqref {expscript} in \eqref {eq:appAp2} we obtain \eqref {eq:repid2intext} with
\be
\bar E^0_{\KK} = \sum_{\substack {\HH_1,\HH_2  \\ \HH_1 \cup \HH_2=\KK\\ \HH_1 \cap \HH_2=\emptyset}} 
  R(\HH_1,\HH_2) \left(\prod_{i \in \HH_1} \,\bar \chi_{S_i}^0 \, f^{\e}_{0,S_i}\right)
\left(\bar\chi^0_{H_2} E^0_{\HH_2}\right)\;.
\ee
The bound \eqref{boundEtilde} follows from \eqref{eq:boundappA}. \qed

 \addcontentsline{toc}{subsection}{B \ \ \ Graph expansion}
 \subsection*{B \ \ \ Graph expansion} \label{sec:B}
 \setcounter{equation}{0}    
 \def\theequation{B.\arabic{equation}}

{\bf Proof of  Lemma \ref{lem:CEoG}.} By addition/subtraction we find
\begin{equation}
\label{step0}
\bar \chi _{L,L\cup L_0} =1-\sum
_{\substack {L_1,L_2  \\ L_1 \cup L_2=L\\ L_1 \cap L_2=\emptyset \\ l_1 \geq1}} 
\chi _{L_1,L  \cup L_0} \bar \chi _{L_2,L  \cup L_0 }=1-
\sum_{\substack {L_1,L_2  \\ L_1 \cup L_2=L\\ L_1 \cap L_2=\emptyset \\ l_1 \geq1}} 
\chi _{L_1,L_1  \cup L_0} \bar \chi _{L_2,L_0  \cup L_1 \cup L_2}\;.
\end{equation}
Note that $l_1 = |L_1| >0$ and $\chi _{L_1,L  \cup L_0}=\chi _{L_1,L_1  \cup L_0}$, 
because any vertex in $L_2$ is not connected. Iterating once,
\begin{equation}
\label{1step}
\bar \chi _{L,L\cup L_0} =
1-\sum
_{\substack {L_1 \subset L  \\ l_1 \geq1}} \chi _{L_1,L_1 \cup L_0}+
\sum
_{\substack {L_1,L_2,L_3  \\ L_1 \cup L_2\cup L_3 =L\\ L_i \cap L_j=\emptyset,  i\neq j \\ l_1 \geq1, l_2 \geq 1}} 
\chi _{L_1,L_0\cup L_1} 
\chi_{L_2,L_0\cup L_1 \cup L_2} 
\bar \chi _{L_3,L_0  \cup L_1\cup L_2 \cup L_3} \;.
\end{equation}
Then, successive iterations yield the following expansion:
\be
\bar \chi _{L,L\cup L_0} =\sum_{r=0}^{|L|} (-1)^r 
\sum_{\substack {L_1, \cdots, L_r  \\ \cup_i L_i \subset L \\  l_i \geq1\\ L_i \cap L_j= \emptyset, i\neq j }}   
\chi_{L_1,L_0\cup L_1 } \cdots \chi _{L_r, L_0  \cup L_1 \cdots \cup L_r }\;,
\label{2step}
\ee
where the $r=0$ term has to be interpreted as $1$. I.e.
\be
\bar \chi _{L,L\cup L_0} =\sum_{Q\subset L} R(Q,L_0)\;,
\ee
with
\be
R(Q,L_0):=
\sum_{r=1}^{q} (-1)^r 
\sum_{\substack {L_1, \cdots, L_r  \\ \cup_i L_i=Q \\  l_i \geq1\\ L_i \cap L_j= \emptyset, i\neq j}}   
\chi_{L_1,L_0\cup L_1 } \cdots
\chi _{L_r, L_0  \cup L_1 \dots \cup L_r }\;,
\label{3step}
\ee
and $R(\emptyset,L_0)=1$.

From this expression follows
\bea
\label{R1}
|R(Q, L_0) | &&\leq  \sum_{r=1}^{q}
\sum_{\substack {L_1, \cdots, L_r  \\ \cup_i L_i=Q \\  l_i \geq1\\ L_i \cap L_j= \emptyset, i\neq j}}   
\chi_{Q, Q\cup L_0 }\nn\\
&& \leq   \chi_{Q, Q\cup L_0 }
 \sum_{r=1}^{q}  \sum_{\substack {l_1, \cdots, l_r  \\ l_i \geq 1 }}  \frac {q !} {l_1! \cdots l_r!} \nn\\ 
&&\leq \chi_{Q, Q\cup L_0 }\,q!\,C^q \;.
\eea
\qed

\medskip

 \addcontentsline{toc}{subsection}{C \ \ \ Reduction to energy functionals}
 \subsection*{C \ \ \ Reduction to energy functionals} \label{sec:C}
 \setcounter{equation}{0}    
 \def\theequation{C.\arabic{equation}}

{\bf Proof of  Lemma \ref{prop3}.} 

\bigskip
\ni {\bf (a)} 
We first use the bound \eqref{eq:boundRov} and the assumptions on the initial state
(see \eqref{eq:insBisfce}) to estimate $E_K$ as given by \eqref{eq:errorEexp}.
Notice that \eqref{eq:insBisfce} can be applied for $k + n < \e^{-\a_0}$, which is
ensured by $k < \e^{-\a}$, $n \leq  \log\e^{-\theta_2k}$ for arbitrary positive $\theta_2$ 
and $\a < \a_0$, as soon as $\e$ is small enough.
We deduce:
\bea
&& \int d\bv_K |E_K(t)| \nn\\
&& \leq z^k\,C^k\sum_{\substack{L_0,Q, B \\Ê\subset \ K \\ \mbox{{\scriptsize disjoint}}}} q!\,b!
\sum_{n=0}^{\log\e^{-\theta_2 k}}  z^n
\sum_{\G(k,n)} \int d\bv_k d\L \prod |B^\e| \chi_{L_0 }^{rec}\,  \chi^{ov}_{Q, K} \, \chi^{0}_{B, K}\,
\e^{\g_0(k-q-l_0-b)} e^{-(\b/2) \HH_K}  \nn\\
&&+z^k\,C^k \sum_{\substack{L_0,Q \\Ê\subset \ K \\ \mbox{{\scriptsize disjoint}}}} q!\,(k-q-l_0)!
\sum_{n > \log\e^{-\theta_2 k}}\  z^n
\ \sum_{\G(k,n)}  \int d\bv_k d\L \prod |B^\e| \, e^{-(\b/2) \HH_K} \;.
\label{eq:proof1}
\eea
The symbol $C$ is always used for pure positive constants. Note that, in the error produced by the truncation
on $n$, the last line of \eqref{eq:errorEexp} has been estimated simply by $z^{k+n}C^k e^{-(\b/2) \HH_K}
(k-q-l_0)!$, as follows from \eqref{boundEtilde}, \eqref{eq:defE0PPK} and Hypothesis \ref{hyp:bound}.

Proceeding exactly as in the proof of Lemma \ref{lem:STE} (case $a=1$), the last term in \eqref{eq:proof1} 
is bounded, for $t  < \bar t$ (see \eqref{shtime}), by
\bea
&& C^k \,k!\, (4\p/\b)^{\frac{3}{2}k}\,(C(z,\b)e)^{k}  \sum_{n > \log\e^{-\theta_2 k}}(\bar t \, C(z,\b)e)^n \nn\\
&& \leq (C')^k \,k^k\,\e^{\theta_2 \log(\bar t C(z,\b)e)^{-1} k }\nn\\
&& \leq (C')^k \,\e^{\theta_2 \log(\bar t C(z,\b)e)^{-1} k -\a k}
\leq \e^{\g k}/4\;, \label{eq:proofkjhg}
\eea
for a suitable $C'= C'(z,\b)>0$. In the last line we used $k < \e^{-\a}$,
\be
\g < \theta_2 \log(\bar t C(z,\b)e)^{-1}-\a 
\label{eq:limgamt*}
\ee
and $\e$ small enough.

Since $ \chi^{ov}_{Q, K} \, \chi^{0}_{B, K} \leq \chi^{ov}_{Q\cup B, K}$
(overlap at time zero implies overlap in $[0,t]$), 
renaming $Q \cup B \to Q$, Eq. \eqref{eq:proof1} yields
\bea
&&\int d\bv_K |E_K(t)| \\
&&\leq z^kC^kk!\,\sum_{\substack{L_0,Q\\Ê\subset \ K \\ \mbox{{\scriptsize disjoint}}}}
\sum_{n=0}^{\log\e^{-\theta_2 k}}  z^n
\sum_{\G(k,n)} \int d\bv_k d\L \prod |B^\e| \chi_{L_0 }^{rec} \chi^{ov}_{Q, K} 
\e^{\g_0(k-q-l_0)} e^{-(\b/2) \HH_K} + \frac{\e^{\g k}}{4}\;.\nn
\eea

\bigskip
\ni {\bf (b)} 
Next we truncate the integration domain to the sphere of energy smaller than $2\e^{-\theta_3}$,
for arbitrary $\theta_3 > 0$. The corresponding error is bounded by
\bea
&& C^k\,k!\,\sum_{\substack{L_0,Q\\Ê\subset \ K \\ \mbox{{\scriptsize disjoint}}}}
\sum_{n = 0}^{\log\e^{-\theta_2 k}} z^{k+n} \sum_{\G(k,n)} \int d\bv_k d\L \prod |B^\e|\,
e^{-(\b/2) \HH_K} \, \mathbbm{1}_{\HH_K > \e^{-\theta_3}}\nn\\
&& \leq e^{-(\b/4) \e^{-\theta_3}}\,C^k\,k!\,\sum_{\substack{L_0,Q\\Ê\subset \ K \\ \mbox{{\scriptsize disjoint}}}}
\sum_{n = 0}^{\log\e^{-\theta_2 k}} z^{k+n} \sum_{\G(k,n)} \int d\bv_k d\L \prod |B^\e|\, e^{-(\b/4) \HH_K} \nn\\
&& \leq e^{-(\b/4) \e^{-\theta_3}}\,C^k\,k!\,4^k\,(8\p/\b)^{3 k/2} \,(eC(z,\b/2))^k\, \sum_{n=0}^{\log\e^{-\theta_2 k}} 
(eC(z,\b/2) \bar t)^n\nn\\
&& \leq (C'')^k\, e^{-(\b/4) \e^{-\theta_3}}\,k^k 
\, (C'')^{\log\e^{-\theta_2 k}}\label{eq:trEint}
\eea
for a suitable $C''= C''(z,\b)>1$. From second to third line
we repeated the proof of Lemma \ref{lem:STE} with $a=1$ and $\b \to \b/2$.
Note that \eqref{eq:trEint} is in turn bounded, for 
$k< \e^{-\a}$, by $(C'')^k\, e^{-(\b/4) \e^{-\theta_3}+\e^{-\a}\log\e^{-\a}+\e^{-\a}\log\e^{-\theta_2}\log C''}$, which is smaller
than $\e^{\g k}/4$ if $\theta_3>\a$ and $\e$ is small enough.

Remembering \eqref{eq:Fth2}, it follows that
\bea
&&\int d\bv_K |E_K(t)| \\
&&\leq z^kC^kk!\,\sum_{\substack{L_0,Q\\Ê\subset \ K \\ \mbox{{\scriptsize disjoint}}}}
\sum_{n=0}^{\log\e^{-\theta_2 k}}  z^n
\sum_{\G(k,n)} \int d\bv_k d\L \prod |B^\e| \chi_{L_0 }^{rec} \chi^{ov}_{Q, K} 
\e^{\g_0(k-q-l_0)} F_{\theta_3}(K) + \frac{2\e^{\g k}}{4}\;.\nn
\eea

\bigskip
\ni {\bf (c)} 
Finally, we introduce a truncation of the cross--section factors $\prod |B^\e|$, by
applying Corollary \ref{cor:Btrick}, i.e. Eq. \eqref{eq:Btrick}.
Note that, with respect to that result, the only difference here is that the integral is computed by using 
the mixed flow \eqref{eq:mixIBF} instead of the IBF. This causes of course no modification, except for the expression
of the characteristic function in \eqref{eq:Btrick}. 
According to \eqref{eq:Btrick}, the error produced is therefore $C^k k! (\bar C')^k \e^{\theta_1 k}
\leq (C''')^k k^k \e^{\theta_1 k}$ for a suitable $C'''= C'''(z,\b)>0$ and arbitrary $\theta_1>0$,
which is, in turn, smaller than $\e^{\g k}/4$ for $k < \e^{-\a} $, $\g < \theta_1-\a$ 
and $\e$ small enough. 

We conclude that, for any $t<\bar t$,
\bea
&&\int d\bv_K |E_K(t)| \leq \frac{3\e^{\g k}}{4} \\
&&+ \,z^kC^kk!\,\e^{-\theta_1 k}\,\sum_{\substack{L_0,Q\\Ê\subset \ K \\ \mbox{{\scriptsize disjoint}}}}
\sum_{n=0}^{\log\e^{-\theta_2 k}}  z^n
\sum_{\G(k,n)} \int d\bv_k d\L  
\,\mathbbm{1}_{L_0} \,
\tilde{\mathbbm{1}}_{K \setminus L_0}\,
\chi_{L_0 }^{rec} \chi^{ov}_{Q, K} 
\e^{\g_0(k-q-l_0)} F_{\theta_3}(K)\;,\nn
\eea
where the characteristic functions $\mathbbm{1}$ are those defined after \eqref{eq:defBemix}. \qed

\medskip
\ni {\bf Remark (choice of parameters)} \label{rem:choicepar}
If we choose the parameters as in \eqref{par2},
then \eqref{par1}-\eqref{eq:ag0} ensures that all the conditions in the proof above 
are satisfied. In fact, in part (a) of the proof we just need to check \eqref{eq:limgamt*} which 
reads $\g < (1/2) - \a$ and follows from 
$\g < a(\g_0) - 3 \a < 1/4 - 3\a $. In part (b), the condition $\a < \theta_3 = 1/5$ follows
from $\a < (1/3) a(\g_0) < 1/12 $. Finally in part (c) the condition $\g < \theta_1 - \a = a(\g_0)-\a$ is guaranteed
by $\g < a(\g_0) - 3\a$.

\addcontentsline{toc}{subsection}{D \ \ \ Estimate of internal recollisions}
 \subsection*{D \ \ \ Estimate of internal recollisions} \label{sec:D}
 \setcounter{equation}{0}    
 \def\theequation{D.\arabic{equation}}

\ni {\bf Proof of Lemma \ref{lem:MRresINT}.} 
It is convenient to use the Enskog backwards flow $\bze^\EE$
introduced in Section \ref{sec:EBF}.
For any given value of the variables $(x_1,\G(1,n_1),\bv_{n_1+1},\bo_{n_1},\bt_{n_1})$, if the IBF $\bze^\e$
delivers an internal recollision, then the EBF $\bze^\EE$ delivers an internal overlap (two particles
of the flow having a distance smaller than $\e$). 
That is,
\be
\chi^{int} \leq \chi^{i.o.}(\bze^\EE)\;,
\ee
where
\be
\chi^{i.o.}=\chi^{i.o.}(\bze^\EE)=1 \label{eq:defchiio}
\ee
if and only if the EBF associated to the $1-$particle 
tree exhibits at least one overlap between two particles.
Therefore in what follows we shall focus on the proof of the estimate
\be
\sum_{\Gamma(1,n_1)} \int dv_{1} d\L
\,\chi^{i.o.}\, e^{-(\b/2)\sum_{i\in S(1)}v_i^2} 
\leq  \frac{\e^{\g_1}}{2}(Dt)^{n_1}\;. \label{eq:MRresINTbiss}
\ee
Remind that $d\L = d\L(\bt_{n_1},\bo_{n_1},\bv_{1,1+n_1})$ and $\Gamma(1,n_1)=(k_1,\cdots,k_{n_1})$.

We start with
\be
\chi^{i.o.} \leq \sum_{s=2}^{n_1} \,\sum_{h=k_s,s+1}\,\sum_{\substack{i=1,\cdots,s\\i\neq k_s,s+1}} \,
\chi^{i.o.}_{(i,h), s}\;, \label{eq:coor1jhg}
\ee
where $\chi^{i.o.}_{(i,h), s} = \chi^{i.o.}_{(i,h), s}(\bze^\EE) = 1$ if and only if:

\ni (i) going backwards in time, the first overlap between particles $i$ and $h$ takes place at a time $\tau \in (0, t_s]$;

\ni (ii) particles $i$ and $h$ move freely in $(\tau,t_s)$;

\ni (iii) at time $t_s$
\be
 \eta^{\EE}_h(t^-_s)\neq  \eta^{\EE}_{k_s}(t^+_s)\;.
 \label{eq:change'}
\ee

\begin{figure}[htbp] 
   \centering
   \includegraphics[width=4.5in]{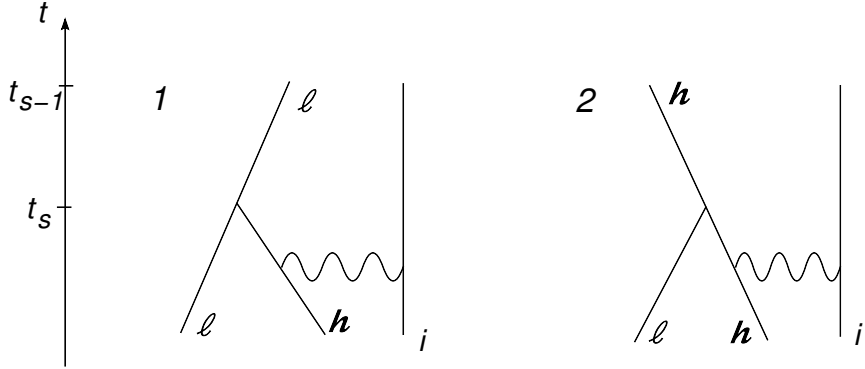} 
   \caption{Case $1$: $h = s+1$, $k_s = \ell$. Case $2$: $h = k_s$, $\ell = s+1$.
   }
   \label{fig:intrecf}
\end{figure}

\ni Notice that particle $h$ is involved in the creation process at time $t_s$.
See Figure \ref{fig:intrecf} below for a scheme of the possible situations and observe
that, by virtue of (iii), we are excluding case $2$ for incoming collision configurations at the creation time $t_s$.

From \eqref{eq:MRresINTbiss}--\eqref{eq:coor1jhg} one gets
\bea
&& \sum_{\Gamma(1,n_1)} \int dv_{1} d\L
 \,\chi^{i.o.}\, e^{-(\b/2)\sum_{i\in S(1)}v_i^2} \\
&& \leq
\sum_{\Gamma(1,n_1)}\,\sum_{s=2}^{n_1} \,\sum_{h=k_s,s+1}\,\sum_{\substack{i=1,\cdots,s\\i\neq k_s}} \,
 \int dv_{1} d\L\, \chi^{i.o.}_{(i,h), s}\,
 e^{-(\b/2)\sum_{i\in S(1)}v_i^2}\;. \nn
\eea
Note that $\chi^{i.o.}_{(i,h), s}$ depends actually only on $\bze^\EE_{1+s}$, hence we can immediately
integrate out the node variables $$t_{s+1},\cdots,t_{n_1},\o_{s+1}\cdots,\o_{n_1},v_{s+2},\cdots,v_{1+n_1}$$
and sum over the tree variables $k_{s+1},\cdots,k_{n_1}$. 
Applying \eqref{eq:nnnfact}, 
$$\sum_{k_{s+1},\cdots,k_{n_1}}\int d\bt_{s,n_1-s} = (s+1)(s+2)\cdots (n_1) t^{n_1-s}/(n_1-s)!
\leq e^{n_1}t^{n_1-s}\;,$$
thus we infer that 
\bea
&& \sum_{\Gamma(1,n_1)} \int dv_{1} d\L
 \,\chi^{i.o.}\, e^{-(\b/2)\sum_{i\in S(1)}v_i^2}
 \leq e^{n_1}\,\sum_{s=2}^{n_1} \,(D't)^{n_1-s}\,
 \sum_{\Gamma(1,s)}\,\sum_{h=k_s,s+1}\,\sum_{\substack{i=1,\cdots,s\\i\neq k_s}} \,
\nn\\
&& \cdot
 \int dv_{1} d\L(\bt_s,\bo_s,\bv_{1,1+s})\,\chi^{i.o.}_{(i,h), s}\,
 e^{-(\b/2)\sum_{i = 1}^{1+s}v_i^2}\;, \label{eq:irtspoint}
\eea
where $D' = 4\pi\, (2\pi/\b)^{3/2}$ and,
in the last line, we are left with integrals associated to $1-$particle, $s-$collision trees.

If $\chi^{i.o.}_{(i,h), s} = 1$, then there are two possibilities: either $h = s+1$ ($h$ is created
at $t_s$) or $k_s = h$ ($h$ is the progenitor of $s+1$), see Figure \ref{fig:intrecf}.
Let us resort to the notation of virtual trajectories, 
to deal with both cases simultaneously (Definition \ref{def:VT}, applied to $\bar\bze=\bze^\EE$).
We set
$$
W=\eta^\EE_h (t_s^-)-\eta^\EE_i(t_s)\;, \quad\quad\quad W_0= \eta^{\EE,h}(t_s^+) - \eta^\EE_i(t_s)
$$
and
$$
Y=\xi^\EE_h (t_s^-)-\xi^\EE_i(t_s)\;, \quad\quad\quad Y_0=\xi^{\EE,h} (t_{s-1}^-)-\xi^\EE_i(t_{s-1})\;.
$$
We remind that $t^+, t^-$ denote the limit from the future (post--collision) or from the past 
(pre--collision) respectively.
Note that \eqref{eq:change'} is, in this notation, 
\be
\label{change}
 \eta^{\EE,h}(t^-_s)\neq  \eta^{\EE,h}(t^+_s)\;,
\ee
namely the virtual trajectory of particle $h$ changes velocity at time $t_s$.

The overlap--condition implies
\be
\inf_{\t\in (0,t_s)} | Y-W\t| \leq \e\;. \label{eq:YWirc}
\ee
Put $\hat W=\frac W {|W|}$ if $W\neq 0$ and $W = (1,0,0)$ otherwise. 
Eq. \eqref{eq:YWirc} implies in turn
$$
|Y\wedge \hat W| \leq \e\;,
$$
i.e.
\be
|(Y_0-W_0 t_{s-1}) \wedge \hat W + (W_0 \wedge \hat W) t_s | \leq 2\,\e\;,
\label{eq:conIRF}
\ee
where the factor $2$ takes into account the jump in position in the virtual trajectory of particle
$h$ at time $t_s$, case $1$.
Therefore, we may bound the last line in \eqref{eq:irtspoint} by replacing $\chi^{i.o.}_{(i,h), s}$ with
the indicator function of the events \eqref{eq:conIRF} and $W\neq W_0$
(which takes into account \eqref{change}).

By definition of the Enskog flow, $Y_0$ and $W_0$ do not depend on $t_s$ (since they concern the previous history).
Moreover, the velocities in $(0,t_{s})$, which we denote
\be
(\eta^-_1,\cdots,\eta^-_{s+1}) = (\eta^\EE_1(t^-_s),\cdots,\eta^\EE_{s+1}(t^-_{s}))\;,
\label{eq:precollIR}
\ee
are also independent of the times $t_1,\cdots,t_s$: they depend only on previous velocities and 
impact vectors. In particular, $W$ does not depend on $t_s$, so that in \eqref{eq:conIRF} a {\em linear}
relation in $t_s$ appears. On the other hand,
the integral in $t_s$ over the condition \eqref{eq:conIRF} is bounded by $\min(t,4\e|W_0\wedge \hat W|^{-1})$.
Hence, for an arbitrary $\g_1\in (0,1)$,
\bea
&&\sum_{\Gamma(1,s)}\,\sum_{h=k_s,s+1}\,\sum_{\substack{i=1,\cdots,s\\i\neq k_s}} \,
 \int dv_{1} d\L(\bt_{s},\bo_s,\bv_{1,1+s})\, 
 \,\chi^{i.o.}_{(i,h), s}\,e^{-(\b/2)\sum_{i = 1}^{1+s}v_i^2}\label{eq:iztrtrew}\\
 && \leq 
(4/t)^{\g_1} t \,\e^{\g_1} \sum_{\Gamma(1,s)}\,\sum_{h=k_s,s+1}\,\sum_{\substack{i=1,\cdots,s\\i\neq k_s}} \,
\int dv_{1} d\L'(\bt_{s-1},\bo_s,\bv_{1,1+s})\, \frac{1}{|W_0\wedge \hat W|^{\g_1}}
 \, e^{-(\b/2)\sum_{i = 1}^{1+s}v_i^2}\;,\nn
 \eea
where $d\L'(\bt_{s-1},\bo_s,\bv_{1,1+s})$ is the measure $d\L(\bt_{s},\bo_s,\bv_{1,1+s})$ deprived
of $dt_s$ and multiplied, in case $2$ of Figure \ref{fig:intrecf}, 
by the characteristic function of $\o_s \cdot (v_{1+s}-\eta^{\EE,h}(t^+_s))>0$
(coming from the condition $W \neq W_0$).

It remains to prove that the integral of the singular function $|W_0\wedge \hat W|^{-\g_1}$ converges. To do so, let us
first express $W_0$ in terms of the pre--collisional variables \eqref{eq:precollIR}. Applying the elastic 
collision rule, Eq. \eqref{eq:collpp}, one finds
\bea
W_0 &&=  \left(\eta^{\EE,h}(t_s^+)- \eta^{\EE,h}(t_s^-)\right)+W \nn\\
&& = P_s W_\ell + W\;,\nn
\eea
where
\be
W_\ell = \eta^-_\ell - \eta^-_h\nn
\ee
and
\be \label{eq:opPs}
P_s X := \begin{cases}
\displaystyle  P_{\o_s}^{\perp} X := X - \o_s (\o_s\cdot X)\ \ \ \ \ \ \ \ \ \mbox{case 1, outgoing collision}\\
\displaystyle  X \ \ \ \ \ \ \ \ \ \ \ \ \ \ \ \ \ \ \ \ \ \ \ \ \ \ \ \ \ \ \ \ \ \ \ \ \ \ \mbox{case 1, incoming collision}\\
\displaystyle  P_{\o_s}^{\parallel} X := \o_s (\o_s\cdot X)\ \ \ \ \ \ \ \ \ \ \ \ \ \ \ \mbox{case 2, outgoing collision}\\
\displaystyle  0 \ \ \ \ \ \ \ \ \ \ \ \ \ \ \ \ \ \ \ \ \ \ \ \ \ \ \ \ \ \ \ \ \ \ \ \ \ \ \ \mbox{case 2, incoming collision}
\end{cases}\;.
\ee
Cases $1,2$ are those in Figure \ref{fig:intrecf}, while we remind that the incoming / outgoing collisions
are depicted in Figure \ref{fig:creations} on page \pageref{fig:creations}
(here corresponding respectively to the negative / positive
sign of the scalar product $\o_s \cdot (v_{1+s}-\eta^{\EE,h}(t^+_s))$). 
Moreover, the ``case''  depends only
on the structure of the chosen tree $\G(1,s)$.
It follows that
\be
\frac{1}{|W_0\wedge \hat W|} = \frac{1}{|P_s W_\ell\wedge \hat W|}
\label{eq:opPsBLA}
\ee
which we may insert into \eqref{eq:iztrtrew}.

Next, we change variables according to $v_1,v_2,\cdots,v_{s+1} \to \eta^-_1,\cdots,\eta^-_{s+1}$. 
This is an invertible and measure--preserving transformation, for any fixed value of $\o_1,\cdots,\o_s$,
(since the single hard--sphere collision \eqref{eq:collpp} is so). Moreover, by the conservation of energy
at collisions, $e^{-(\b/2)\sum_{i = 1}^{1+s}v_i^2}= e^{-(\b/2)\sum_{i = 1}^{1+s}(\eta^-_i)^2}$. 
From \eqref{eq:irtspoint}, \eqref{eq:iztrtrew} and \eqref{eq:opPsBLA}, we thus obtain
\bea
&&  \sum_{\Gamma(1,n_1)} \int dv_{1} d\L
 \,\chi^{i.o.}\, e^{-(\b/2)\sum_{i\in S(1)}v_i^2}\leq e^{n_1}\,\sum_{s=2}^{n_1} \,(D't)^{n_1-s}\,
(4/t)^{\g_1} t \,\e^{\g_1} \,s! \, 2s \label{eq:D14}\\
 &&  \ \cdot \frac{t^{s-1}}{(s-1)!}\,
\int d\bo_s \int d\bet^-_{s+1} \left(\frac{e^{-(\b/2)\sum_{i = 1}^{1+s}(\eta^-_i)^2}}{|P_{\o_s}^{\perp} W_1\wedge \hat W|^{\g_1}}
+ \frac{e^{-(\b/2)\sum_{i = 1}^{1+s}(\eta^-_i)^2}}{|W_1\wedge \hat W|^{\g_1}}
+ \frac{e^{-(\b/2)\sum_{i = 1}^{1+s}(\eta^-_i)^2}}{|P_{\o_s}^{\parallel} W_1\wedge \hat W|^{\g_1}}\right)\;,\nn
 \eea
where we renamed $1,2,3$ particles $\ell,h,i$ respectively (hence $W_1 = \eta^-_1-\eta^-_2, W = \eta^-_2-\eta^-_3$).

Let us now give a bound of the explicit integral $\int d\bet^-_{s+1} \frac{ e^{-(\b/2)\sum_{i = 1}^{1+s}(\eta^-_i)^2}}{|\tilde P_sW_1
\wedge \hat W|^{\g_1}}$,
where $\tilde P_sW_1 = P_{\o_s}^{\perp} W_1, W_1$ or $P_{\o_s}^{\parallel} W_1$.
Since $W^2+W_1^2 \leq 2(\eta^-_1)^2+4(\eta^-_2)^2+2(\eta^-_3)^2$, applying the 
translations $(\eta^-_1,\eta^-_2) \to (W_1 = \eta^-_1-\eta^-_2,
W=\eta^-_2-\eta^-_3)$, we find
\bea
&& \int d\bet^-_{s+1} \frac{ e^{-(\b/2)\sum_{i = 1}^{1+s}(\eta^-_i)^2}}{|\tilde P_sW_1\wedge \hat W|^{\g_1}}
\leq \int d\bet^-_{s+1} e^{-(\b/2)\sum_{\substack{i > 3}}(\eta^-_i)^2} \,e^{-(\b/4)(\eta^-_3)^2}\,
\frac{e^{-(\b/8)(W_1^2+W^2)} }{|\tilde P_sW_1\wedge \hat W|^{\g_1}}\nn\\
&& \ \ \ \ \ \ \ \ \ \ \ \ \ \ \ \ = \int d\bet^-_{2,s-1} 
\,e^{-(\b/2)\sum_{\substack{i > 3}}(\eta^-_i)^2} \,e^{-(\b/4)(\eta^-_3)^2}\,
\int dW_1 dW\frac{e^{-(\b/8)(W_1^2+W^2)} }{|\tilde P_sW_1\wedge \hat W|^{\g_1}}\nn\\
&& \ \ \ \ \ \ \ \ \ \ \ \ \ \ \ \ \leq C^s_\b \int dW_1 \frac{e^{-(\b/8)W_1^2} }{|\tilde P_sW_1|^{\g_1}}\nn\\
&& \ \ \ \ \ \ \ \ \ \ \ \ \ \ \ \ \leq C^s_\b\, C_{\b,\g_1},
\label{eq:D15}
\eea
for suitable constants $C_\b,C_{\b,\g_1}>0$ and for any $\g_1<1$ (with $C_{\b,\g_1}$ diverging in the
case $\tilde P_sW_1=P_{\o_s}^{\parallel} W_1$ as $\g_1 \to 1$).

Inserting \eqref{eq:D15} into Eq. \eqref{eq:D14} and performing the sums, we obtain the final result. \qed

\addcontentsline{toc}{subsection}{E \ \ \ Proof of Corollary \ref{cor:MRpre}}
 \subsection*{E \ \ \ Proof of Corollary \ref{cor:MRpre}} \label{sec:E}
 \setcounter{equation}{0}    
 \def\theequation{E.\arabic{equation}}
 
The result follows from minor modifications in the proof of Theorem \ref{thm:MR}.

First of all, by Property 2 on page \pageref{boundEtilde},
case $S = \JJ = J$, applied to the state with r.c.f.~$f^\e_j$ and correlation errors $\bar E_k \equiv E_k$,
\bea
|E_K| && \leq \sum_{H \subset K}  
\left(C^{h}\, h!\,  \chi^0_{H,K}\, \left(f^{\e}_{1}\right)^{\otimes H}\right) \, 
\left(\bar\chi^0_{K \setminus H} |E_{K \setminus H}|\right)\nn\\
&& =  \sum_{H \subset K \setminus Q'}  
\left(C^{h}\, h!\,  \chi^0_{H,K}\, \left(f^{\e}_{1}\right)^{\otimes H}\right) \, 
\left(\bar\chi^0_{K \setminus (Q' \cup H)} |E_{K \setminus H}|\right)\;.
\eea
Remind that $\chi^0_{H,K}=1$ if and only if any particle with index in $H$ overlaps
with a different particle in $K$, which implies $H \subset K \setminus Q'$. Moreover, $\bar\chi^0_{K \setminus H}=1$
if and only if all particles in $K \setminus H$ do not overlap among themselves, which implies 
$\bar\chi^0_{K \setminus H} = \bar\chi^0_{K \setminus (Q' \cup H)}$. In particular,
 \bea
\int_{\RRR^{3q'}} d\bv_{Q'} |E_K(t)|  \leq 
 \sum_{H \subset K \setminus Q'}  
\left(C^{h}\, h!\,  \chi^0_{H,K}\, \left(f^{\e}_{1}\right)^{\otimes H}\right) \, 
\bar\chi^0_{K \setminus (Q' \cup H)} 
\int_{\RRR^{3q'}} d\bv_{Q'}  |E_{K \setminus H}| \nn\\
\label{eq:BDOUZTD}
\eea
and we are allowed to replace expression \eqref{eq:errorEexp} into \eqref{eq:BDOUZTD}.

The estimate of the integral on the r.h.s. differs from that of the main theorem from the fact that
we integrate only with respect to a subset of velocities $Q' \subset K$. Furthermore, we know
that particles in $Q'$ are at distance larger than $\d$ from any other particle in $K$, but we
have no information on the relative distance of particles in $K \setminus Q'$.
Observe that the proof of pages \pageref{sec:REF}--\pageref{eq:g1boundGL} applies unchanged, except for the 
following modifications.
\begin{enumerate}
\item In Lemma \ref{prop3}, one integrates only over $d\bv_{Q'}$. However the integral 
over velocities is not used in the proof (see Appendix C). Therefore one gets the same result apart from an overall
$(const.)^k$. This produces the first term in \eqref{eq:EKthmEXT}.
\item In Proposition \ref{basicprop}, one integrates only over $d\bv_{Q'}$ and $\e^{\g_1 \frac{q + l_0}{2}}=
\e^{\g_1 \frac{|Q \cup L_0|}{2}}$ 
has to be replaced by $\e^{\g_1 \frac{|(Q \cup L_0)\cap Q'|}{2}}$. Indeed in the proof of the proposition, Section
4.4.2.c, when the bullet $\a_i$ is outside $Q'$, Lemma \ref{lem:MR} cannot be applied. Instead of estimate
\eqref{eq:qwertz}, one uses then the simple estimate
\be
\sum_{\G_{\a_i}}  \,\int d \L_{\a_i} \, \chi^{(\a_i,\b_i)}\,F_{\theta_3}(\a_i) \leq (D't)^{n_{\a_i}}\;.
\ee
\item In \eqref{eq:almconc}, one integrates only over $d\bv_{Q'}$ and, by virtue of the previous two points,
one gets $\e^{\min[\g_0,\g_1/2] q'}$ instead of $\e^{\min[\g_0,\g_1/2] k}$.
This produces the second term in \eqref{eq:EKthmEXT}.
\end{enumerate}
\qed

\end{appendices}

\bigskip
\bigskip

\ni {\bf Acknowledgments.}
We would like to thank Raffaele Esposito and Herbert Spohn for valuable discussions and suggestions.
S. Simonella has been supported by Indam--COFUND Marie Curie fellowship 2012, call 10
and by the German Research Foundation, DFG grant 269134396.


\addcontentsline{toc}{section}{Notation index}
\renewcommand{\nomname}{Notation index}
\printnomenclature

\end{document}